\documentclass[journa,twocolumnl]{IEEEtran}

\ifCLASSINFOpdf

\else

\fi
\usepackage{amsfonts}
\usepackage{amsthm}
\usepackage{amsmath}
\usepackage{graphicx}
\usepackage{multirow}

\begin{document}

\title{Sparse Optimization Problem with $s$-difference Regularization }

\author{Yuli~Sun, Xiang~Tan, Xiao~Li, Lin~Lei, Gangyao~Kuang$^*$

\small{\emph{College of Electronic science and Technology, National University of Defense Technology, Changsha 410073, China} }}



\maketitle

\begin{abstract}

In this paper, a $s$-difference type regularization for sparse recovery problem is proposed, which is the difference of the normal penalty function $R\left( {\bf{x}} \right)$ and its corresponding $s$-truncated function $R\left( {{{\bf{x}}^s}} \right)$. First, we show the equivalent conditions between the ${\ell _0}$ constrained problem and the unconstrained $s$-difference penalty regularized problem. Next, we choose the forward-backward splitting (FBS) method to solve the non-convex regularizes function and further derive some closed-form solutions for the proximal mapping of the $s$-difference regularization with some commonly used $R\left( {\bf{x}} \right)$, which makes the FBS easy and fast. We also show that any cluster point of the sequence generated by the proposed algorithm converges to a stationary point. Numerical experiments demonstrate the efficiency of the proposed $s$-difference regularization in comparison with some other existing penalty functions.
\end{abstract}

\begin{IEEEkeywords}
Sparse constrained,  Forward-Backward Splitting, proximal operator, difference of convex, truncated function.
\end{IEEEkeywords}

\IEEEpeerreviewmaketitle

\section{Introduction}

\subsection{Background}
In recent years, sparse optimization problems have drawn lots of attention in many applications such as compressive sensing, machine learning, image processing and medical imaging. Signal and image processing problems are usually expressed as
\begin{equation}
\label{eqn_1}
A\left( {\bf{x}} \right) + {\bf{n}} = {\bf{b}}
\end{equation}
where $A$  is the linear or non-linear operator, $\bf{b}$  is the observation data, and $\bf{n}$ represents the observation noise or error. Since problem (\ref{eqn_1}) is often ill-posed and the error $\bf{n}$ is unknown, solving (\ref{eqn_1}) is difficulty. To overcome this ill-posed problem, we need to make some constraints to narrow the solution space, such as the prior sparsity of the signals. Then the problem can be formulated as
\begin{equation}
\label{eqn_2}
\mathop {\min }\limits_{\bf{x}} \phi \left( {\bf{x}} \right) + P\left( {\bf{x}} \right)
\end{equation}
where the loss function $\phi \left( {\bf{x}} \right)$ is the data fidelity term related to (\ref{eqn_1}), for example, the least square (LS) loss function $\left\| {A\left( {\bf{x}} \right) - {\bf{b}}} \right\|_2^2$ or the least-absolute (LA) loss function $\left\| {A\left( {\bf{x}} \right) - {\bf{b}}} \right\|_1^{}$; $P\left( {\bf{x}} \right)$ is the regularizes function to penalize the sparsity of ${\bf{x}}$. Intuitively, $P\left( {\bf{x}} \right)$ should be selected as the ${\ell _0}$-norm ${\left\| {\bf{x}} \right\|_0}$, represents the number of nonzero elements in ${\bf{x}}$. However, minimizing the ${\ell _0}$-norm is equivalent to finding the sparsest solution, which is known to be NP-hard problem. A favorite and popular approach is using the ${\ell _1}$-norm convex approximation, i.e., ${\left\| {\bf{x}} \right\|_1}$ to replace the ${\ell _0}$ [1]. This ${\ell _1}$ model has been widely used in many different applications, such as radar systems [2-3], communications [4], computed tomography (CT) [5] and magnetic resonant imaging (MRI) [6]. It has been proved that the signal ${\bf{x}}$ can be recovered by this ${\ell _1}$ model under some assumption of the operator $A$, such as the restricted isometry property (RIP) of $A$ when the operator is a sensing matrix [1]. However, the ${\ell _1}$-norm regularization tends to underestimate high-amplitude components of ${\bf{x}}$ as it penalizes the amplitude uniformly, unlike ${\ell _0}$-norm in which all nonzero entries have equal contributions. This may lead to reconstruction failures with the least measurements [7-8], and brings undesirable blocky images [9-10]. It is quite well-known that the when it promotes sparsity, the ${\ell _1}$-norm does not provide a performance close to that of the ${\ell _0}$-norm, and lots of theoretical and experimental results in CS and low-rank matrix recovery suggest that better approximations of the ${\ell _0}$-normand matrix rank give rise to better performances.

Recently, researchers began to investigate various non-convex regularizes to replace the ${\ell _1}$-norm regularization and gain some better reconstructions. In particular, the ${\ell _p}$(quasi)-norm with $p \in \left( {0,1} \right)$ [11-16], can be regarded as a interpolation between the ${\ell _0}$ and ${\ell _1}$, and a continuation strategy to approximate the ${\ell _0}$ as $p \to 0$. The optimization strategies include half thresholding [14, 17-20] and iterative reweighting [11-12, 15]. Other non-convex regularizations and algorithms have also been designed to outperform ${\ell _1}$-norm regularization and seek better reconstruction: capped ${\ell _1}$-norm [21-23], transformed ${\ell _1}$-norm [24-26], sorted ${\ell _1}$-norm [27-28], the difference of the ${\ell _1}$ and ${\ell _2}$-norms (${\ell _{1 - 2}}$) [29-31], the log-sum penalty (LSP) [8], smoothly clipped absolute deviation (SCAD) [32-33], minimax-concave penalty (MCP) [34-36].

On the other hand, there are some approaches which do not approximate the ${\ell _0}$-norm, such as the iterative hard thresholding (IHT) algorithm [37-38], which operate directly on the ${\ell _0}$ regularized cost function or the $s$-sparse constrained optimization problem. Moreover, there are some acceleration methods for the IHT: accelerated IHT (AIHT) [39], proximal IHT (PIHT) [40], extrapolated proximal IHT (EPIHT) [41] and accelerated proximal IHT [42]. Meanwhile, there are some researchers transformed the ${\ell _0}$-norm problem into an equivalent difference of two convex functions, and then using the difference of convex algorithm (DCA) and the proximal gradient technique to solve the subproblem [43-44].

To address these nonconvex regularization problems, many iterative algorithms are investigated by researchers, such as the DCA [45-48] (or Convex-ConCave Procedure (CCCP) [49], or the Multi-Stage (MS) convex relaxation [22]), and its accelerate versions: Boosted Difference of Convex function Algorithms (BDCA) [50] and proximal Difference-of-Convex Algorithm with extrapolation (pDCAe) [51], the alternating direction method of multipliers (ADMM) [52], split Bregman iteration (SBI) [53], General Iterative Shrinkage and Thresholding (GIST) [54], nonmonotone accelerated proximal gradient (nmAPG) [55], which is an extension of the APG [56].
\subsection{Contributions}
In many applications, the non-convex ${\ell _0}$-norm based regularization has its advantages over the convex ${\ell _1}$-norm , such as image restoration [41, 53, 57-58], bioluminescence [59], CT [9-10], MRI reconstruction [60-61]. Thus, in this paper, we are interested in the following ${\ell _0}$ constrained problem
\begin{equation}
\label{eqn_3}
\mathop {\min }\limits_{\bf{x}} \phi \left( {\bf{x}} \right)\  {\rm{subject \ to}}\   {\left\| {\bf{x}} \right\|_0} \le s
\end{equation}
i.e. this $s$-sparse problem tries to find the solution minimizing $\phi \left( {\bf{x}} \right)$ under the constraint that the number of non-zero coefficients below a certain value, where $s \in \left\{ {1,2, \cdots ,N} \right\}$.

This paper can be viewed as a natural complement and extension of Gotoh et al. framework [43]. First, we rewrite the ${\ell _0}$ constrained problem (\ref{eqn_3}) as difference of two functions, one of which is the convex or nonconvex function $R\left( {\bf{x}} \right)$ and the other is the corresponding truncated function $R\left( {{{\bf{x}}^s}} \right)$. Then, we consider the unconstrained minimization problem by using this $s$-difference $R\left( {\bf{x}} \right) - R\left( {{{\bf{x}}^s}} \right)$ type regularizations. Second, we propose fast approaches to deal with this non-convex regularizes function, which is based on a proximal operator corresponding to $R\left( {\bf{x}} \right) - R\left( {{{\bf{x}}^s}} \right)$. Moreover, we derive some cheap closed-form solutions for the proximal mapping of $R\left( {\bf{x}} \right) - R\left( {{{\bf{x}}^s}} \right)$  with some commonly used $R\left( {\bf{x}} \right)$, such as ${\left\| {\bf{x}} \right\|_1}$, ${\left\| {\bf{x}} \right\|_2}$, ${\left\| {\bf{x}} \right\|_1} - a{\left\| {\bf{x}} \right\|_2}$, LSP, MCP and so on.  Third, we prove the convergence performance of the proposed algorithm, and show that any cluster point of the sequence generated by the proposed algorithm converges to a stationary point. We also show a link between the proposed algorithm with some related regularizations and algorithms. Finally, we evaluate the effectiveness of the proposed algorithm via numerical experiments. The reconstruction results demonstrate that the proposed difference penalty function with closed-form solutions is more accurate than the ${\ell _1}$-norm and other non-convex regularization based methods, and faster than the DCA based algorithms.
\subsection{Outline and notation}
The rest of this paper is structured as follows. In section 2, we define the constrained sparse optimization. In section 3, we propose the reconstruction algorithm by using the proximal operator with closed-form solutions. In section 4, we provide some theorems to demonstrate the convergence of the proposed algorithm. Section 5 presents the numerical results. In the end, we provide our conclusion in section 6.

Here, we define our notation. For a vector $\textbf{x}\in\mathbb{R}^{N}$, it can be written as ${\bf{x}} = \left( {{x_1},{x_2}, \cdots ,{x_N}} \right)$, and its ${\ell _p}$-norm is defined as ${\left\| {\bf{x}} \right\|_p} = {\left( {\sum\nolimits_n {{{\left| {{x_n}} \right|}^p}} } \right)^{{\raise0.5ex\hbox{$\scriptstyle 1$}
\kern-0.1em/\kern-0.15em
\lower0.25ex\hbox{$\scriptstyle p$}}}}$. Especially, the ${\ell _\infty }$-norm of ${\bf{x}}$ is defined as ${\max _n}\left| {{x_n}} \right|$. Given a matrix ${\bf{A}} \in {\mathbb{R}^{M \times N}}$, the transpose of ${\bf{A}}$ is denoted by ${{\bf{A}}^T}$, the maximum eigenvalue of ${{\bf{A}}^T}{\bf{A}}$ is defined as $\left\| {\bf{A}} \right\|_2^2$. Some of the arguments in this paper use sub-vectors. The letters $\Gamma $, $\Lambda $ denote sets of indices that enumerate the elements in the vector ${\bf{x}}$. By using this sets as subscripts, ${{\bf{x}}_\Gamma }$ represents the vector that setting all elements of ${\bf{x}}$ to zero except those in the set $\Gamma $. The iteration count is given in square bracket, e.g., ${{\bf{x}}^{[k]}}$. $\left\langle { \cdot , \cdot } \right\rangle $ denotes the inner product, ${\rm{sign}}\left(  \cdot  \right)$ represents the sign of a quantity with ${\rm{sign}}\left( 0 \right) \in \left[ { - 1,1} \right]$. We also use the notation ${\mathbb{R}_ + } = \left\{ {x \in \mathbb{R}:x \ge 0} \right\}$, and if the function $f$ is defined as the composition $f = h\left( {g\left( x \right)} \right)$, then we write $f = h \circ g$.

Given a proper closed function $h\left( x \right):{\mathbb{R}^n} \to \mathbb{R} \cup \left\{ \infty  \right\}$, the subgradient of $h$ at $x$ is given by
\begin{equation}
\label{eqn_4}
\partial h\left( x \right) = \left\{ {v \in {\mathbb{R}^n}:h\left( u \right) - h\left( x \right) - \left\langle {v,u - x} \right\rangle  \ge 0,\forall u \in {\mathbb{R}^n}} \right\}
\end{equation}
In addition, if $h\left( x \right)$  is continuously differentiable, then the subdifferential reduces to the gradient of $h\left( x \right)$ denoted by $\nabla h\left( x \right)$.

\section{Penalty representation for $s$-sparse problem}

Inspired by Gotoh et al. work of [43], in which they expressed the ${\ell _0}$-norm constraint as a difference of convex (DC) function:
\begin{equation}
\label{eqn_5}
{\left\| {\bf{x}} \right\|_0} \le s \Leftrightarrow {\left\| {\bf{x}} \right\|_1} - {\left\| {\left| {\bf{x}} \right|} \right\|_s} = 0
\end{equation}
where $s \in \left\{ {1,2, \cdots ,N} \right\}$ and ${\left\| {\left| {\bf{x}} \right|} \right\|_s}$, which named top-$\left( {s,1} \right)$ norm, denotes the sum of top-$s$elements in absolute value. This notation is also known as the largest-$s$ norm (or called CVaR norm in [62-63]). Precisely,
\begin{equation}
\label{eqn_6}
{\left\| {\left| {\bf{x}} \right|} \right\|_s}: = \left| {{x_{{\pi _x}\left( 1 \right)}}} \right| + \left| {{x_{{\pi _x}\left( 2 \right)}}} \right| +  \cdots  + \left| {{x_{{\pi _x}\left( s \right)}}} \right|
\end{equation}
where ${x_{{\pi _x}\left( i \right)}}$ denotes the element whose absolute value is the $i$-th largest among the $N$ elements of vector ${\bf{x}} \in {\mathbb{R}^N}$, i.e., $\left| {{x_{{\pi _x}\left( 1 \right)}}} \right| \ge \left| {{x_{{\pi _x}\left( 2 \right)}}} \right| \ge  \cdots  \ge \left| {{x_{{\pi _x}\left( N \right)}}} \right|$. For convenience of description, we define the set $\Gamma _{\bf{x}}^s = \left\{ {{\pi _x}\left( 1 \right),{\pi _x}\left( 2 \right), \cdots ,{\pi _x}\left( s \right)} \right\}$, then we have $\Gamma _{\bf{x}}^1 \subseteq \Gamma _{\bf{x}}^2 \subseteq  \cdots  \subseteq \Gamma _{\bf{x}}^N$. By using $ \cdot \backslash  \cdot $  as the set difference, we have $\Gamma _{\bf{x}}^N\backslash \Gamma _{\bf{x}}^s = \left\{ {{\pi _x}\left( {s + 1} \right),{\pi _x}\left( {s + 2} \right), \cdots ,{\pi _x}\left( N \right)} \right\}$.

In this work, we consider a more general $s$-difference function $R\left( {\bf{x}} \right) - R\left( {{{\bf{x}}^s}} \right)$ instead of ${\left\| {\bf{x}} \right\|_1}$ to replace the ${\ell _0}$-norm constraint, where $R\left( {\bf{x}} \right)$ can be convex or nonconvex, separable or non-separable, and ${{\bf{x}}^s}$ is the best $s$ term approximation to ${\bf{x}}$, that is, any $s$-sparse vectors that minimize ${\left\| {{\bf{x}} - {{\bf{x}}^s}} \right\|_2}$. By using the definition of ${x_{{\pi _x}\left( i \right)}}$, we have
\begin{equation}
\label{eqn_7}
x_i^s = \left\{ {\begin{array}{*{20}{c}}
{{x_i},}&{\begin{array}{*{20}{c}}
{{\rm{if}}}&{i \in \Gamma _{\bf{x}}^s}
\end{array}}\\
{0,}&{\begin{array}{*{20}{c}}
{{\rm{if}}}&{i \in \Gamma _{\bf{x}}^N\backslash \Gamma _{\bf{x}}^s}
\end{array}}
\end{array}} \right.
\end{equation}
Let $P\left( {\bf{x}} \right) = R\left( {\bf{x}} \right) - R\left( {{\bf{x}}_{}^s} \right)$, $s \in \left\{ {1,2, \cdots ,N} \right\}$, we defined a class of penalty functions $P,R:{\mathbb{R}^N} \to {\mathbb{R}_ + }$ as follows (without loss of generality, functions $P\left( {\bf{x}} \right)$ and $R\left( {\bf{x}} \right)$ mentioned thought this paper all satisfy Property 1 ).

\newtheorem{Property}{Property}
\begin{Property}
The penalty functions $P,R:{\mathbb{R}^N} \to {\mathbb{R}_ + }$ satisfy the following properties.

(a) $R\left( {\bf{x}} \right) = R\left( { - {\bf{x}}} \right)$

(b) ${\left\| {\bf{x}} \right\|_0} \le s \Leftrightarrow P\left( {\bf{x}} \right) = 0$

(c) $P\left( {\bf{x}} \right)$ is a continuous function which can be written as the difference of two convex (DC) functions, that is, $P\left( {\bf{x}} \right) = {P_1}\left( {\bf{x}} \right) - {P_2}\left( {\bf{x}} \right)$, where ${P_1}\left( {\bf{x}} \right)$ and ${P_2}\left( {\bf{x}} \right)$ are convex functions.
\end{Property}

\newtheorem{Proposition}{Proposition}
\begin{Proposition}
The penalty functions listed on Table 1 all satisfy Property 1.
\end{Proposition}

See appendix A for the Proof of Proposition 1.

\begin{table*}[]
\newcommand{\tabincell}[2]{\begin{tabular}{@{}#1@{}}#2\end{tabular}}
	\centering
\caption{\label{table_1}Functions that satisfies Property 1}
 \begin{tabular}{c|l|l|l}
  \hline
 Function type  & $R\left( {\bf{x}} \right)$ & ${P_1}\left( {\bf{x}} \right)$ &  ${P_2}\left( {\bf{x}} \right)$ \\ \hline
 \multirow{2}{*}{\tabincell{c}{Convex,\\Separable}}   & ${\left\| {\bf{x}} \right\|_1}$ &  ${\left\| {\bf{x}} \right\|_1}$ & ${\left\| {{{\bf{x}}^s}} \right\|_1}$ \\ \cline{2-4}
 &$\left\| {\bf{x}} \right\|_2^2$ & $\left\| {\bf{x}} \right\|_2^2$ & $\left\| {{{\bf{x}}^s}} \right\|_2^2$\\ \hline
\multirow{2}{*}{\tabincell{c}{Convex,\\non-Separable}} & ${\left\| {\bf{x}} \right\|_2}$ & ${\left\| {\bf{x}} \right\|_2}$ & $\left\| {{{\bf{x}}^s}} \right\|_2^{}$ \\ \cline{2-4}
& $R\left( {\bf{x}} \right) = \left\{ {\begin{array}{*{20}{c}}
{{{\left\| {\bf{x}} \right\|_2^2} \mathord{\left/
 {\vphantom {{\left\| {\bf{x}} \right\|_2^2} {\left( {2\theta } \right)}}} \right.
 \kern-\nulldelimiterspace} {\left( {2\theta } \right)}},}&{{{\left\| {\bf{x}} \right\|}_2} \le \theta }\\
{{{{{\left\| {\bf{x}} \right\|}_2} - \theta } \mathord{\left/
 {\vphantom {{{{\left\| {\bf{x}} \right\|}_2} - \theta } 2}} \right.
 \kern-\nulldelimiterspace} 2}},&{{{\left\| {\bf{x}} \right\|}_2} > \theta }
\end{array}} \right.,\theta  > 0$ & $R\left( {\bf{x}} \right)$ & $R\left( {{{\bf{x}}^s}} \right)$ \\ \hline
\multirow{2}{*}{\tabincell{c}{Non-convex,\\Separable}}  & \tabincell{c}{$R\left( {\bf{x}} \right) = \sum\limits_{i = 1}^N {{r_i}\left( {{x_i}} \right)} $\\${r_i}\left( {{x_i}} \right) = \log \left( {1 + {{\left| {{x_i}} \right|} \mathord{\left/
 {\vphantom {{\left| {{x_i}} \right|} \theta }} \right.
 \kern-\nulldelimiterspace} \theta }} \right),\theta  > 0$} &  ${{{{\left\| {\bf{x}} \right\|}_1}} \mathord{\left/
 {\vphantom {{{{\left\| {\bf{x}} \right\|}_1}} \theta }} \right.
 \kern-\nulldelimiterspace} \theta } + \left( {{{{{\left\| {{{\bf{x}}^s}} \right\|}_1}} \mathord{\left/
 {\vphantom {{{{\left\| {{{\bf{x}}^s}} \right\|}_1}} \theta }} \right.
 \kern-\nulldelimiterspace} \theta } - R\left( {{{\bf{x}}^s}} \right)} \right)$ & ${{{{\left\| {{{\bf{x}}^s}} \right\|}_1}} \mathord{\left/
 {\vphantom {{{{\left\| {{{\bf{x}}^s}} \right\|}_1}} \theta }} \right.
 \kern-\nulldelimiterspace} \theta } + \left( {{{{{\left\| {\bf{x}} \right\|}_1}} \mathord{\left/
 {\vphantom {{{{\left\| {\bf{x}} \right\|}_1}} \theta }} \right.
 \kern-\nulldelimiterspace} \theta } - R\left( {\bf{x}} \right)} \right)$ \\ \cline{2-4}
&\tabincell{c}{$R\left( {\bf{x}} \right) = \sum\limits_{i = 1}^N {{r_i}\left( {{x_i}} \right)} $ \\ ${r_i}\left( {{x_i}} \right) = \left\{ {\begin{array}{*{20}{c}}
{\left| {{x_i}} \right| - {{x_i^2} \mathord{\left/
 {\vphantom {{x_i^2} {\left( {2\theta } \right)}}} \right.
 \kern-\nulldelimiterspace} {\left( {2\theta } \right)}},}&{\left| {{x_i}} \right| \le \theta }\\
{{\theta  \mathord{\left/
 {\vphantom {\theta  2}} \right.
 \kern-\nulldelimiterspace} 2}},&{\left| {{x_i}} \right| > \theta }
\end{array}} \right.,\theta  > 0$} & ${\left\| {\bf{x}} \right\|_1} + \left( {{{\left\| {{{\bf{x}}^s}} \right\|}_1} - R\left( {{{\bf{x}}^s}} \right)} \right)$ & ${\left\| {{{\bf{x}}^s}} \right\|_1} + \left( {{{\left\| {\bf{x}} \right\|}_1} - R\left( {\bf{x}} \right)} \right)$ \\ \hline
\multirow{3}{*}{\tabincell{c}{Non-convex,\\Non-separable}}  & ${\left\| {\bf{x}} \right\|_1} - a{\left\| {\bf{x}} \right\|_2},0 < a \le 1$ & ${\left\| {\bf{x}} \right\|_1} + a{\left\| {{{\bf{x}}^s}} \right\|_2}$ & ${\left\| {{{\bf{x}}^s}} \right\|_1} + a{\left\| {\bf{x}} \right\|_2}$ \\ \cline{2-4}
& $\log \left( {1 + {{{{\left\| {\bf{x}} \right\|}_2}} \mathord{\left/
 {\vphantom {{{{\left\| {\bf{x}} \right\|}_2}} \theta }} \right.
 \kern-\nulldelimiterspace} \theta }} \right),\theta  > 0$ & ${{{{\left\| {\bf{x}} \right\|}_2}} \mathord{\left/
 {\vphantom {{{{\left\| {\bf{x}} \right\|}_2}} \theta }} \right.
 \kern-\nulldelimiterspace} \theta } + \left( {{{{{\left\| {{{\bf{x}}^s}} \right\|}_2}} \mathord{\left/
 {\vphantom {{{{\left\| {{{\bf{x}}^s}} \right\|}_2}} \theta }} \right.
 \kern-\nulldelimiterspace} \theta } - R\left( {{{\bf{x}}^s}} \right)} \right)$ & ${{{{\left\| {{{\bf{x}}^s}} \right\|}_2}} \mathord{\left/
 {\vphantom {{{{\left\| {{{\bf{x}}^s}} \right\|}_2}} \theta }} \right.
 \kern-\nulldelimiterspace} \theta } + \left( {{{{{\left\| {\bf{x}} \right\|}_2}} \mathord{\left/
 {\vphantom {{{{\left\| {\bf{x}} \right\|}_2}} \theta }} \right.
 \kern-\nulldelimiterspace} \theta } - R\left( {\bf{x}} \right)} \right)$ \\ \cline{2-4}
 & $\left\{ {\begin{array}{*{20}{c}}
{{{\left\| {\bf{x}} \right\|}_2} - {{\left\| {\bf{x}} \right\|_2^2} \mathord{\left/
 {\vphantom {{\left\| {\bf{x}} \right\|_2^2} {\left( {2\theta } \right)}}} \right.
 \kern-\nulldelimiterspace} {\left( {2\theta } \right)}},}&{{{\left\| {\bf{x}} \right\|}_2} \le \theta }\\
{{\theta  \mathord{\left/
 {\vphantom {\theta  2}} \right.
 \kern-\nulldelimiterspace} 2}},&{{{\left\| {\bf{x}} \right\|}_2} > \theta }
\end{array}} \right.,\theta  > 0$ & ${\left\| {\bf{x}} \right\|_2} + \left( {{{\left\| {{{\bf{x}}^s}} \right\|}_2} - R\left( {{{\bf{x}}^s}} \right)} \right)$ & ${\left\| {{{\bf{x}}^s}} \right\|_2} + \left( {{{\left\| {\bf{x}} \right\|}_2} - R\left( {\bf{x}} \right)} \right)$   \\ \hline
\end{tabular}
\end{table*}

Remark 1. For the separable $R\left( {\bf{x}} \right) = \sum\limits_{i = 1}^N {r\left( {{x_i}} \right)} $, and $r\left( x \right)$ is continuous, symmetrical and strictly increasing on ${\mathbb{R}_ + }$, if $r\left( x \right)$ is convex, then $R\left( {\bf{x}} \right)$ satisfies Property 1; if $r\left( x \right)$ is nonconvex, while it can be written as the difference of two convex functions as $r\left( x \right) = h(x) - g(x)$, then $R\left( {\bf{x}} \right)$ also satisfies Property 1.

It is easy to see that the penalty function in Ref. [43] is a special case of $R\left( {\bf{x}} \right){\rm{ = }}{\left\| {\bf{x}} \right\|_1}$.

With the Property 1(b), we consider the following unconstrained minimization problem associated with (3):
\begin{equation}
\label{eqn_8}
\mathop {\min }\limits_{{\bf{x}} \in {\mathbb{R}^N}} \left\{ {F\left( {\bf{x}} \right) = \phi \left( {\bf{x}} \right) + \rho P\left( {\bf{x}} \right)} \right\}
\end{equation}
where $\rho  > 0$ is the penalty parameter. We make the following assumptions on the above formulation thought the paper, which are standard in image processing and many CS field.

\newtheorem{Assumption}{Assumption}
\begin{Assumption}
$\phi \left( {\bf{x}} \right)$  is continuously differentiable with Lipschitz continuous gradient, i.e., there exists $L > 0$ such that
\begin{equation}
\label{eqn_9}
{\left\| {\nabla \phi \left( {\bf{x}} \right) - \nabla \phi \left( {\bf{y}} \right)} \right\|_2} \le L{\left\| {{\bf{x}} - {\bf{y}}} \right\|_2},\forall {\bf{x}},{\bf{y}} \in {\mathbb{R}^N}
\end{equation}
\end{Assumption}

\begin{Assumption}
$F\left( {\bf{x}} \right)$ is bounded from below.
\end{Assumption}
From (\ref{eqn_8}), we can find that the difference between penalty $P\left( {\bf{x}} \right)$ and other penalty function, such as ${\ell _1}$, ${\ell _p}$, ${\ell _{1 - 2}}$ and MCP, is that there is no punishment in model (\ref{eqn_8}) when the sparsity level of ${\bf{x}}$ is under $s$, since $P\left( {\bf{x}} \right)$ is equal to zero as ${\left\| {\bf{x}} \right\|_0} \le s$. Meanwhile, the selection of the weighting parameter $\rho $ has importance influence on the performance of the reconstruction. On the one hand, $\rho $ should be big enough to give a heavy cost for constraint violation: ${\left\| {\bf{x}} \right\|_0} > s$. On the other hand, if $\rho $ is too big, the reconstruction is mostly over regularized. In light of this, we need the adjust the value of $\rho $ iteratively based on the convergence speed. The next Theorem ensures that problem (\ref{eqn_8}) is equivalent to the original $s$-sparse constraint problem (\ref{eqn_3}) as we take the limit of $\rho $, which can be proved in a similar manner to Theorem 17.1 in [71].
\newtheorem{Theorem}{Theorem}
\begin{Theorem}
Let $\left\{ {{\rho _t}} \right\}$ be an increasing sequence with ${\lim _{t \to \infty }}{\rho _t} = \infty $ and suppose that ${{\bf{x}}_t}$ is an optimal solution of (\ref{eqn_8}) with $\rho  = {\rho _t}$. Then, any limit point ${\bf{\bar x}}$ of $\left\{ {{{\bf{x}}_t}} \right\}$ is also optimal to (\ref{eqn_3}).
\end{Theorem}

See Appendix B for the proof.

In addition to Theorem 1, we have some stricter conclusions for the parameter $\rho $ under some assumptions of $P\left( {\bf{x}} \right)$ and $\phi \left( {\bf{x}} \right)$.

\begin{Proposition}
If $\phi \left( \mathbf{x} \right)$ is Lipschitz continuous with constant $\beta >0$, i.e., ${{\left\| \phi \left( \mathbf{x} \right)-\phi \left( \mathbf{y} \right) \right\|}_{2}}\le \beta {{\left\| \mathbf{x}-\mathbf{y} \right\|}_{2}},\forall \mathbf{x},\mathbf{y}\in {{\mathbb{R}}^{N}}$, and ${{\mathbf{\bar{x}}}_{\rho }}$ is an optimal solution of (\ref{eqn_8}) with some $\rho $. Suppose that there exists a constant $\eta >0$ such that $R\left( \mathbf{x} \right)-R\left( \mathbf{x}+{{\mathbf{x}}^{s}}-{{\mathbf{x}}^{s+1}} \right)\ge \eta \left\| {{\mathbf{x}}^{s+1}}-{{\mathbf{x}}^{s}} \right\|_{2}^{{}}$ for any $\mathbf{x}\in {{\mathbb{R}}^{N}}$. Then if $\rho >{\beta}/{\eta}$, ${{\mathbf{\bar{x}}}_{\rho }}$ is also optimal to (\ref{eqn_3}).
\end{Proposition}

See Appendix C for the proof.

Remark 2. Suppose that $\phi \left( \mathbf{x} \right)$ is $\beta $-Lipschitz continuous and the regularization is $P\left( \mathbf{x} \right)={{\left\| \mathbf{x} \right\|}_{1}}-{{\left\| {{\mathbf{x}}^{s}} \right\|}_{1}}$. Then if $\rho >\beta $, any optimal solution of (\ref{eqn_8}) is also optimal to (\ref{eqn_3}).

Remark 3. Suppose that $\phi \left( \mathbf{x} \right)$ is $\beta $-Lipschitz continuous. If we choose $R\left( \mathbf{x} \right)$ as $R\left( \mathbf{x} \right)={{\left\| \mathbf{x} \right\|}_{1}}-a{{\left\| \mathbf{x} \right\|}_{2}},0<a\le 1$, then any optimal solution of (\ref{eqn_8}) is also optimal to (\ref{eqn_3}) when $\rho >\frac{\beta }{1-{a}/{\left( 2\sqrt{s} \right)}}$. This can be proved by using that
\begin{equation}
\label{eqn_10}
\begin{aligned}
{\left\| {\bf{x}} \right\|_2} - {\left\| {{\bf{x}} + {{\bf{x}}^s} - {{\bf{x}}^{s + 1}}} \right\|_2} &= \frac{{\left\| {{{\bf{x}}^{s + 1}} - {{\bf{x}}^s}} \right\|_2^2}}{{{{\left\| {\bf{x}} \right\|}_2} + {{\left\| {{\bf{x}} + {{\bf{x}}^s} - {{\bf{x}}^{s + 1}}} \right\|}_2}}}\\
 &\le \frac{{\left\| {{{\bf{x}}^{s + 1}} - {{\bf{x}}^s}} \right\|_2^{}}}{{2\sqrt s }}
\end{aligned}
\end{equation}
If we choose $R\left( \mathbf{x} \right)={{\theta }_{1}}{{\left\| \mathbf{x} \right\|}_{1}}-\sum\limits_{i=1}^{N}{\log \left( 1+{\left| {{x}_{i}} \right|}/{{{\theta }_{2}}}\; \right)},{{\theta }_{1}}>{{\theta }_{2}}>0$, then the condition of $\rho $ is that $\rho >\frac{\beta }{{{\theta }_{1}}-{{\theta }_{2}}}$. Meanwhile, we can obtain similar conclusions for the $R\left( \mathbf{x} \right)$ which are the difference of ${{\left\| \mathbf{x} \right\|}_{1}}$ and MCP, or SCAD functions.

The next proposition, which is similar to Theorem 3 in [43], but with wider scope and stricter conclusion, shows another exact penalty parameters $\rho $ requirement for $\phi \left( \mathbf{x} \right)$ with Lipschitz continuous gradient $L$.

\begin{Proposition}
If Assumption 1 is satisfied and ${{\mathbf{\bar{x}}}_{\rho }}$ is an optimal solution of (\ref{eqn_8}) with some $\rho $. Suppose that there exists a constant $C>0$ such that ${{\left\| {{{\mathbf{\bar{x}}}}_{\rho }} \right\|}_{2}}\le C$ for any $\rho >0$, and there exists a constant $\eta >0$ such that $R\left( \mathbf{x} \right)-R\left( \mathbf{x}+{{\mathbf{x}}^{s}}-{{\mathbf{x}}^{s+1}} \right)\ge \eta \left\| {{\mathbf{x}}^{s+1}}-{{\mathbf{x}}^{s}} \right\|_{2}^{{}}$ for any $\mathbf{x}\in {{\mathbb{R}}^{N}}$, Then if $\rho >\frac{1}{\eta }\left( {{\left\| \nabla \phi \left( \mathbf{0} \right) \right\|}_{2}}+\left( 1+\frac{1}{2\sqrt{s+1}} \right)LC \right)$, ${{\mathbf{\bar{x}}}_{\rho }}$ is also optimal to (\ref{eqn_3}).
\end{Proposition}

See Appendix D for the proof.

Remark 4. Suppose that $\phi \left( \mathbf{x} \right)=\frac{1}{2}\left\| \mathbf{Ax}-\mathbf{b} \right\|_{2}^{2}$ and ${{\left\| {{{\mathbf{\bar{x}}}}_{\rho }} \right\|}_{2}}\le C$. If we choose $R\left( \mathbf{x} \right)$ as $R\left( \mathbf{x} \right)={{\left\| \mathbf{x} \right\|}_{1}}$, $R\left( \mathbf{x} \right)={{\left\| \mathbf{x} \right\|}_{1}}-a{{\left\| \mathbf{x} \right\|}_{2}}$ ($0<a\le 1$) and $R\left( \mathbf{x} \right)={{\theta }_{1}}{{\left\| \mathbf{x} \right\|}_{1}}-\sum\limits_{i=1}^{N}{\log \left( 1+{\left| {{x}_{i}} \right|}/{{{\theta }_{2}}}\; \right)}$ (${{\theta }_{1}}>{{\theta }_{2}}>0$), then any optimal solution of (\ref{eqn_8}) is also optimal to (\ref{eqn_3}) when $\rho >{{\left\| {{\mathbf{A}}^{T}}\mathbf{b} \right\|}_{2}}+\left( 1+\frac{1}{2\sqrt{s+1}} \right)\left\| \mathbf{A} \right\|_{2}^{2}C$, $\rho >\frac{1}{1-{a}/{\left( 2\sqrt{s} \right)}\;}\left( {{\left\| {{\mathbf{A}}^{T}}\mathbf{b} \right\|}_{2}}+\left( 1+\frac{1}{2\sqrt{s+1}} \right)\left\| \mathbf{A} \right\|_{2}^{2}C \right)$ and $\rho >\frac{1}{{{\theta }_{1}}-{{\theta }_{2}}}\left( {{\left\| {{\mathbf{A}}^{T}}\mathbf{b} \right\|}_{2}}+\left( 1+\frac{1}{2\sqrt{s+1}} \right)\left\| \mathbf{A} \right\|_{2}^{2}C \right)$, respectively.

Remark 5. Similarly to Theorem 4 in [43] by replacing penalty function ${{\left\| \mathbf{x} \right\|}_{1}}-{{\left\| \left| \mathbf{x} \right| \right\|}_{s}}$ with ordinary function $R\left( \mathbf{x} \right)-R\left( {{\mathbf{x}}^{s}} \right)$, we have the following conclusions without proof. If the conditions in Proposition 3 are satisfied, and suppose that $\phi \left( \mathbf{x} \right)=\frac{1}{2}{{\mathbf{x}}^{T}}\mathbf{Qx}+{{\mathbf{q}}^{T}}\mathbf{x}$, where $\mathbf{Q}=\left( {{q}_{ij}} \right)\in {{\mathbb{R}}^{N\times N}}$ is symmetric and $\mathbf{q}=\left( {{q}_{i}} \right)\in {{\mathbb{R}}^{N}}$, then ${{\mathbf{\bar{x}}}_{\rho }}$ is also optimal to (\ref{eqn_3}) if $\rho  > \mathop {\max }\limits_i \frac{1}{\eta }\left\{ {\left| {{q_i}} \right| + \left( {{{\left\| {{\bf{Q}}{{\bf{e}}_i}} \right\|}_2} + \frac{{\left| {{q_{ii}}} \right|}}{{2\sqrt {s + 1} }}} \right)C} \right\}$, where ${{\mathbf{e}}_{i}}$ denotes the unit vector in the $i$-th coordinate direction.

\section{Forward-Backward Splitting for the regularization of difference of two functions}
In this section, we use the FBS to solve the unconstrained minimization (8). Moreover, we derive closed-form solutions for the proximal mapping of some special regularization $s$-difference $P\left( {\bf{x}} \right)$, and this makes FBS more efficient.
\subsection{Forward-Backward Splitting and proximal operator}
Each iteration of forward-backward splitting applies the gradient descent of $\rho P\left( \mathbf{x} \right)$ followed by a proximal operator. That is
\begin{equation}
\label{eqn_11}
{{\bf{x}}^{[k + 1]}} = {\rm{pro}}{{\rm{x}}_{\beta \rho P}}\left( {{{\bf{x}}^{[k]}} - \beta \nabla \phi \left( {{{\bf{x}}^{[k]}}} \right)} \right)
\end{equation}
where $\beta >0$ is the step size, and the FBS is sometimes called the proximal gradient (PG) algorithm. The proximal operator is defined as
\begin{equation}
\label{eqn_12}
{\rm{pro}}{{\rm{x}}_{\lambda P}}\left( {\bf{y}} \right) = \arg \mathop {\min }\limits_{{\bf{x}} \in {\mathbb{R}^N}} \frac{{\left\| {{\bf{x}} - {\bf{y}}} \right\|_2^2}}{{2\lambda }} + P\left( {\bf{x}} \right)
\end{equation}
with parameter $\lambda >0$.

The equation (\ref{eqn_11}) can be broken up into a forward gradient step using the function $\phi \left( \mathbf{x} \right)$, and a backward step using the function $\rho P\left( \mathbf{x} \right)$. The proximal operator plays a central role in the analysis and solution of optimization problems. For example, the soft shrinkage operator, which is a proximal operator for ${{\ell }_{1}}$-norm regularizer, has been widely used in CS and rendering many efficient ${{\ell }_{1}}$ algorithms. The proximal operator also has been successfully used with some nonconvex regularizers, such as ${{\ell }_{p}}$, SCAD, LSP [64], and MCP [52, 65]. Usually, the closed-form solution of the proximal operator needs some special properties on $P\left( \mathbf{x} \right)$, such as convexity or separability (e.g., the ${{\ell }_{1}}$-norm, LSP, MCP, and other various separable functions in [66]), Next, we will focus on the solution of (\ref{eqn_12}) with separable and non-separable $s$-difference $P\left( \mathbf{x} \right)$.
\subsection{Closed-form solution of the proximal operator}
Denote $E\left( \mathbf{x} \right)$ as
\begin{equation}
\label{eqn_13}
E\left( {\bf{x}} \right) = \frac{{\left\| {{\bf{x}} - {\bf{y}}} \right\|_2^2}}{{2\lambda }} + P\left( {\bf{x}} \right)
\end{equation}
Let ${{\mathbf{x}}^{*}}$ be the optimal solution of (\ref{eqn_12}), i.e., ${{\bf{x}}^ * } = {\rm{pro}}{{\rm{x}}_{\lambda P}}\left( {\bf{y}} \right)$, then we have the following Proposition.

\begin{Proposition}
${{\mathbf{x}}^{*}}=\mathbf{0}$ if and only if $\mathbf{y}=\mathbf{0}$.
\end{Proposition}

\begin{IEEEproof}
Necessary condition: note that $E\left( \mathbf{x} \right)\ge 0$ for any $\mathbf{x}$, and when $\mathbf{y}=\mathbf{0}$, we have $E\left( \mathbf{0} \right)=0$. Thus if $\mathbf{y}=\mathbf{0}$, the optimal solution is ${{\mathbf{x}}^{*}}=\mathbf{0}$.
Sufficient condition: assume by contradiction that $\mathbf{y}\ne \mathbf{0}$, then we select an arbitrary non-zero dimension ${{y}_{j}}$ in $\mathbf{y}$, and construct $\mathbf{\tilde{x}}\in {{\mathbb{R}}^{N}}$ as ${\tilde x_i} = \left\{ {\begin{array}{*{20}{c}}
{0,}&{i \ne j}\\
{{{\rm{y}}_j},}&{i = j}
\end{array}} \right.$. Then we have

\begin{equation}
\label{eqn_14}
E\left( {{{\bf{x}}^ * }} \right) = E\left( {\bf{0}} \right) = \frac{1}{{2\lambda }}\sum\limits_{i = 1}^N {y_i^2}  > \frac{1}{{2\lambda }}\sum\limits_{i = 1,i \ne j}^N {y_i^2}  = E\left( {{\bf{\tilde x}}} \right)
\end{equation}
This contradicts the optimality of ${{\mathbf{x}}^{*}}$. Thus if ${{\mathbf{x}}^{*}}=\mathbf{0}$, $\mathbf{y}$ must be equal to zero.
\end{IEEEproof}

\begin{Proposition}
 For $i \in \left\{ {1,2, \cdots ,N} \right\}$, if ${{y}_i} > 0$, then we have $x_{i}^{*}\ge 0$. If ${{y}_i} < 0$, then we have $x_i^ *  \le 0$.
\end{Proposition}
\begin{IEEEproof}
We prove it by establishing contradiction. If there exits any $x_{i}^{*}<0$ when ${{y}_i} > 0$, then we select an arbitrary one and we construct $\mathbf{\tilde{x}}\in {{\mathbb{R}}^{N}}$ as ${\tilde x_j} = \left\{ {\begin{array}{*{20}{c}}
{{x}_j^ * ,}&{j \ne i}\\
{ - {x}_j^ * ,}&{j = i}
\end{array}} \right.$. We have
\begin{equation}
\label{eqn_15}
\begin{array}{c}
\left\| {{\bf{\tilde x}} - {\bf{y}}} \right\|_2^2 = \sum\limits_{j \ne i}^{} {{{\left( {{{\tilde x}_j} - {y_j}} \right)}^2}}  + {\left( {{{\tilde x}_i} - {y_i}} \right)^2}\\
 < \sum\limits_{j \ne i}^{} {{{\left( {{\rm{x}}_j^ *  - {y_j}} \right)}^2}}  + {\left( {{\rm{x}}_i^ *  - {y_i}} \right)^2} = \left\| {{{\bf{x}}^ * } - {\bf{y}}} \right\|_2^2
\end{array}
\end{equation}
The inequality follows from that $x_{i}^{*}$ has the opposite sign as ${{y}_{i}}$ and ${{y}_{i}}>0$. Since we have not changed the absolute value of ${{\tilde{x}}_{i}}$ and $R\left( \mathbf{x} \right)=R\left( -\mathbf{x} \right)$, then we have $P\left( {\mathbf{\tilde{x}}} \right)=P\left( {{\mathbf{x}}^{*}} \right)$. Combing this and (\ref{eqn_15}), we have $E\left( {\mathbf{\tilde{x}}} \right)<E\left( {{\mathbf{x}}^{*}} \right)$. This contradicts the optimality of ${{\mathbf{x}}^{*}}$ and proves that $x_{i}^{*}\ge 0$ when ${{y}_{i}}>0$. On the other hand, we can prove that $x_{i}^{*}\le 0$ when ${{y}_{i}}<0$ by using a similar method. This completes the proof.
\end{IEEEproof}
Next, we focus on the closed-form solutions of ${\rm{prox}}{_{\lambda P}}\left( {\bf{y}} \right)$ with different types of $R\left( \mathbf{x} \right)$.

\begin{Proposition}
If $R\left( \mathbf{x} \right)$ is separable, i.e., $R\left( \mathbf{x} \right)=\sum\limits_{i=1}^{N}{{{r}_{i}}\left( {{x}_{i}} \right)}$ and each ${{r}_{i}}$ is strictly increasing on ${{\mathbb{R}}_{+}}$, we have
\begin{equation}
\label{eqn_16}
x_i^ *  = \left\{ {\begin{array}{*{20}{c}}
{{y_i},}&{{\rm{if  }}\ i \in \Gamma _{\bf{y}}^s}\\
{{{\left( {{{\bf{I}}_N} + \lambda \partial R} \right)}^{ - 1}}{{\left( {\bf{y}} \right)}_i},}&{{\rm{if  }}\ i \in \Gamma _{\bf{y}}^N\backslash \Gamma _{\bf{y}}^s}
\end{array}} \right.
\end{equation}
where ${{\mathbf{I}}_{N}}$ denotes the identity operator, $\Gamma _{\bf{y}}^s = \left\{ {{\pi _y}\left( 1 \right),{\pi _y}\left( 2 \right), \cdots ,{\pi _y}\left( s \right)} \right\}$ and ${{\pi }_{y}}\left( j \right)$ is the index of the $j$-th largest amplitude of $\mathbf{y}$, i.e., $\left| {{y}_{{{\pi }_{y}}\left( 1 \right)}} \right|\ge \left| {{y}_{{{\pi }_{y}}\left( 2 \right)}} \right|\ge \cdots \ge \left| {{y}_{{{\pi }_{y}}\left( N \right)}} \right|$.
\end{Proposition}

See Appendix E for the proof.

Remark 6. Note that $x_{i}^{*}={{y}_{i}}$ if $i\in \left\{ {{\pi }_{y}}\left( 1 \right),{{\pi }_{y}}\left( 2 \right),\cdots ,{{\pi }_{y}}\left( s \right) \right\}$ in (\ref{eqn_16}). Suppose that there exits one or more components of ${{y}_{i}}$, $i\notin \left\{ {{\pi }_{y}}\left( 1 \right),{{\pi }_{y}}\left( 2 \right),\cdots ,{{\pi }_{y}}\left( s \right) \right\}$ having the same amplitude of ${{y}_{{{\pi }_{y}}\left( s \right)}}$, i.e., $\left| {{y}_{{{\pi }_{y}}\left( s-m \right)}} \right|=\cdots =\left| {{y}_{{{\pi }_{y}}\left( s \right)}} \right|=\cdots =\left| {{y}_{{{\pi }_{y}}\left( s+j \right)}} \right|$, $m\ge 0,j\ge 1$. Then there exits $C_{j+m+1}^{m+1}$ solutions of ${{\mathbf{x}}^{*}}$ as there are $C_{j+m+1}^{m+1}$ arrangements of ${{y}_{{{\pi }_{y}}\left( s-m \right)}},\cdots ,{{y}_{{{\pi }_{y}}\left( s \right)}}$.

Remark 7. If $R\left( \mathbf{x} \right)={{\left\| \mathbf{x} \right\|}_{1}}$, then the solution ${{\mathbf{x}}^{*}}$ of (\ref{eqn_12}) is
\begin{equation}
\label{eqn_17}
x_i^ *  = \left\{ {\begin{array}{*{20}{c}}
{{y_i},}&{{\rm{if  }}\ i \in \Gamma _{\bf{y}}^s}\\
{{\rm{shrink}}\left( {{y_i},\lambda } \right),}&{{\rm{if  }}\ i \in \Gamma _{\bf{y}}^N\backslash \Gamma _{\bf{y}}^s}
\end{array}} \right.
\end{equation}
where $\rm{shrink}\left( {{y}_{i}},\lambda  \right)$ denotes the soft shrinkage operator given by
\begin{equation}
\label{eqn_18}
{\rm{shrink}}\left( {{y_i},\lambda } \right) = {\rm{sign}}\left( {{y_i}} \right)\max \left\{ {\left| {{y_i}} \right| - \lambda ,0} \right\}
\end{equation}

Remark 8. If $R\left( \mathbf{x} \right)=\left\| \mathbf{x} \right\|_{2}^{2}$, then the solution ${{\mathbf{x}}^{*}}$ of (\ref{eqn_12}) is
\begin{equation}
\label{eqn_19}
x_i^ *  = \left\{ {\begin{array}{*{20}{c}}
{{y_i},}&{{\rm{if  }}\ i \in \Gamma _{\bf{y}}^s}\\
{{{{y_i}} \mathord{\left/
 {\vphantom {{{y_i}} {\left( {2\lambda  + 1} \right),}}} \right.
 \kern-\nulldelimiterspace} {\left( {2\lambda  + 1} \right),}}}&{{\rm{if  }}\ i \in \Gamma _{\bf{y}}^N\backslash \Gamma _{\bf{y}}^s}
\end{array}} \right.
\end{equation}

Remark 9. If $R\left( \mathbf{x} \right)$ is the MCP (A.3), that is ${r_i}\left( {{x_i}} \right) = \left\{ {\begin{array}{*{20}{c}}
{\left| {{x_i}} \right| - {{x_i^2} \mathord{\left/
 {\vphantom {{x_i^2} {\left( {2\theta } \right)}}} \right.
 \kern-\nulldelimiterspace} {\left( {2\theta } \right)}},}&{\left| {{x_i}} \right| \le \theta }\\
{{\theta  \mathord{\left/
 {\vphantom {\theta  {2,}}} \right.
 \kern-\nulldelimiterspace} {2,}}}&{\left| {{x_i}} \right| > \theta }
\end{array}} \right.$
($\theta>0$),then the solution ${{\mathbf{x}}^{*}}$ is: under the condition of $\theta >\lambda $, if $i \in \Gamma _{\bf{y}}^s$ or $\left| {{y}_{i}} \right|>\theta $, then $x_{i}^{*}={{y}_{i}}$; otherwise $x_{i}^{*}=\rm{sign}\left( {{y}_{i}} \right)\max \left\{ {\theta \left( \left| {{y}_{i}} \right|-\lambda  \right)}/{\left( \theta -\lambda  \right)}\;,0 \right\}$. When $\theta \le \lambda $, if $i \in \Gamma _{\bf{y}}^s$ or $\left| {{y}_{i}} \right|>\theta $, then $x_{i}^{*}={{y}_{i}}$; otherwise $x_{i}^{*}=0$. If $R\left( \mathbf{x} \right)$ is the LSP (A.2), that is ${{r}_{i}}\left( {{x}_{i}} \right)=\log \left( 1+{\left| {{x}_{i}} \right|}/{\theta }\; \right),\theta >0$, then the solution ${{\mathbf{x}}^{*}}$ is: if $i \in \Gamma _{\bf{y}}^s$, then $x_{i}^{*}={{y}_{i}}$; otherwise $x_i^ *  = {\rm{sign}}\left( {{y_i}} \right){w_i}$, and ${w_i} = \arg \mathop {\min }\limits_{{x_i} \in \Omega } \left\{ {{\textstyle{1 \over {2\lambda }}}{{\left( {{x_i} - \left| {{y_i}} \right|} \right)}^2} + \sum\nolimits_i {\log \left( {1 + {{\left| {{x_i}} \right|} \mathord{\left/
 {\vphantom {{\left| {{x_i}} \right|} \theta }} \right.
 \kern-\nulldelimiterspace} \theta }} \right)} } \right\}$, where $\Omega $ is a set composed of 3 elements or 1 element. If ${{\left( \left| {{y}_{i}} \right|-\theta  \right)}^{2}}-4\left( \lambda -\left| {{y}_{i}} \right|\theta  \right)\ge 0$, then
 \begin{equation}
\label{eqn_20}
\begin{aligned}
\Omega  = \left\{ {0,\max \left\{ {{\xi _1},0} \right\}} \right.,\left. {\max \left\{ {{\xi _2},0} \right\}} \right\}
\end{aligned}
\end{equation}
where ${\xi _1} = \frac{1}{2}\left( {\left( {\left| {{y_i}} \right| - \theta } \right) + \sqrt {{{\left( {\left| {{y_i}} \right| - \theta } \right)}^2} - 4\left( {\lambda  - \left| {{y_i}} \right|\theta } \right)} } \right)$ and ${\xi _2} = \frac{1}{2}\left( {\left( {\left| {{y_i}} \right| - \theta } \right) - \sqrt {{{\left( {\left| {{y_i}} \right| - \theta } \right)}^2} - 4\left( {\lambda  - \left| {{y_i}} \right|\theta } \right)} } \right)$. Otherwise, $\Omega =\left\{ 0 \right\}$.

Proposition 6 gives the solution of the (\ref{eqn_12}) under the conditions of $R\left( \mathbf{x} \right)$ with separable and strictly increasing properties. In fact, there are some other commonly used separable and non-convex $R\left( {\bf{x}} \right)$ also have the closed-form solution similar as (\ref{eqn_16}), such as $R\left( {\bf{x}} \right) = \left\| {\bf{x}} \right\|_p^p$ with $p = 1/2,2/3$ [14], however, these $R\left( {\bf{x}} \right)$ does not satisfy the Property 1(c), so they are not within the scope of this article. Next, we consider two special non-separable cases as the reference for other non-separable regularizations.

\begin{Proposition}
 If $R\left( \mathbf{x} \right)={{\left\| \mathbf{x} \right\|}_{2}}$, then the solution ${{\mathbf{x}}^{*}}$ of (\ref{eqn_12}) is that: when $i \in \Gamma _{\bf{y}}^s$,
  \begin{equation}
\label{eqn_21}
x_i^ *  = \frac{{\left( {{{\left\| {{{\bf{y}}^s}} \right\|}_2} + \lambda } \right)\left( {\sqrt {\left\| {{\bf{y}} - {{\bf{y}}^s}} \right\|_2^2 + {{\left( {{{\left\| {{{\bf{y}}^s}} \right\|}_2} + \lambda } \right)}^2}}  - \lambda } \right)}}{{{{\left\| {{{\bf{y}}^s}} \right\|}_2}\sqrt {\left\| {{\bf{y}} - {{\bf{y}}^s}} \right\|_2^2 + {{\left( {{{\left\| {{{\bf{y}}^s}} \right\|}_2} + \lambda } \right)}^2}} }}{y_i}
\end{equation}
when $i \in \Gamma _{\bf{y}}^N\backslash \Gamma _{\bf{y}}^s$,
\begin{equation}
\label{eqn_22}
x_i^ *  = \frac{{\sqrt {\left\| {{\bf{y}} - {{\bf{y}}^s}} \right\|_2^2 + {{\left( {{{\left\| {{{\bf{y}}^s}} \right\|}_2} + \lambda } \right)}^2}}  - \lambda }}{{\sqrt {\left\| {{\bf{y}} - {{\bf{y}}^s}} \right\|_2^2 + {{\left( {{{\left\| {{{\bf{y}}^s}} \right\|}_2} + \lambda } \right)}^2}} }}{y_i}
\end{equation}
\end{Proposition}

See Appendix F for the proof.

\begin{Proposition}
 If $R\left( \mathbf{x} \right)={{\left\| \mathbf{x} \right\|}_{1}}-a{{\left\| \mathbf{x} \right\|}_{2}}$, $0 < a \le 1$, then the solution ${{\mathbf{x}}^{*}}$ of (\ref{eqn_12}) is that:

 1) When $\left| {{y}_{\pi_{y} \left( s+1 \right)}} \right|>\lambda $, for $i \in \Gamma _{\bf{y}}^s$,

  \begin{equation}
\label{eqn_23}
x_i^ *  = \frac{{{{\left\| {{{\bf{y}}^s}} \right\|}_2} - a\lambda }}{{{{\left\| {{{\bf{y}}^s}} \right\|}_2}}}\left( {1 + \frac{{a\lambda }}{{\sqrt {\left\| {{\bf{z}} - {{\bf{z}}^s}} \right\|_2^2 + {{\left( {{{\left\| {{{\bf{y}}^s}} \right\|}_2} - a\lambda } \right)}^2}} }}} \right){y_i}
\end{equation}
for $i \in \Gamma _{\bf{y}}^N\backslash \Gamma _{\bf{y}}^s$,
 \begin{equation}
\label{eqn_24}
x_i^ *  = \left( {1 + \frac{{a\lambda }}{{\sqrt {\left\| {{\bf{z}} - {{\bf{z}}^s}} \right\|_2^2 + {{\left( {{{\left\| {{{\bf{y}}^s}} \right\|}_2} - a\lambda } \right)}^2}} }}} \right){z_i}
\end{equation}
where ${{z}_{i}}={{y}_{{{\pi }_{y}}\left( 1 \right)}}$ for $i \in \Gamma _{\bf{y}}^s$, and ${{z}_{i}}=\rm{shrink}\left( {{y}_{i}},\lambda  \right)$ for $i \in \Gamma _{\bf{y}}^N\backslash \Gamma _{\bf{y}}^s$.

2) When $\left| {{y}_{{{\pi }_{y}}\left( s+1 \right)}} \right|=\lambda $, if $a=1$, $s=1$, $\left| {{y}_{{{\pi }_{y}}\left( 1 \right)}} \right|=\lambda $, and suppose that there are $k$ components of ${y_i}$ having the same amplitude of $\lambda $, i.e., $\left| {{y_{{\pi _y}\left( {s + 1} \right)}}} \right| =  \cdots  = \left| {{y_{{\pi _y}\left( {s + k} \right)}}} \right| = \lambda  > \left| {{y_{{\pi _y}\left( {s + k + 1} \right)}}} \right|$. ${{\mathbf{x}}^{*}}$ is an optimal solution of (\ref{eqn_12}) if and only if it satisfies ${{\left\| {{\mathbf{x}}^{*}} \right\|}_{2}}=\lambda $, $x_{i}^{*}{{y}_{i}}\ge 0$, and $x_{i}^{*}=0$ when $i\in \left\{ {{\pi }_{y}}\left( k+2 \right),{{\pi }_{y}}\left( k+3 \right),\cdots ,{{\pi }_{y}}\left( N \right) \right\}$. In this case, there are infinite many solutions, equations (A.40) and (A.41) are two solution examples. When $\left| {{y}_{{{\pi }_{y}}\left( s+1 \right)}} \right|=\lambda $, and any of these conditions $a=1$, $s=1$, $\left| {{y}_{{{\pi }_{y}}\left( 1 \right)}} \right|=\lambda $ cannot be satisfied, the solution ${{\mathbf{x}}^{*}}$ is
 \begin{equation}
\label{eqn_25}
x_i^ *  = \left\{ {\begin{array}{*{20}{c}}
{{y_i},}&{i \in \Gamma _{\bf{y}}^s}\\
{0,}&{i \in \Gamma _{\bf{y}}^N\backslash \Gamma _{\bf{y}}^s}
\end{array}} \right.
\end{equation}

3) When $0\le \left| {{y}_{{{\pi }_{y}}\left( s+1 \right)}} \right|<\lambda $, the solution ${{\mathbf{x}}^{*}}$ is the same as (\ref{eqn_25}).
\end{Proposition}

We apply the similar proof framework in Ref. [29] for the fast ${{\ell }_{1-2}}$ minimization. See Appendix G for the proof.

Remark 10. When $a=0$, then $R\left( \mathbf{x} \right)={{\left\| \mathbf{x} \right\|}_{1}}-a{{\left\| \mathbf{x} \right\|}_{2}}$ reduces to $R\left( \mathbf{x} \right)={{\left\| \mathbf{x} \right\|}_{1}}$, and the corresponding solution ${{\mathbf{x}}^{*}}$ of (\ref{eqn_23}, \ref{eqn_24}, \ref{eqn_25}) reduces to (\ref{eqn_17}) as in Remark 7.

\section{Convergence analysis}
The purpose of this section is to demonstrate that the sequence of $\left\{ {{\mathbf{x}}^{[k]}} \right\}$ obtained from the FBS for (\ref{eqn_8}) is convergent.

\begin{Theorem}
If Assumption 1 and 2 are satisfied and $\beta < 1/L$, let $\left\{ {{\mathbf{x}}^{[k]}} \right\}$ be the sequence generated by the FBS for (\ref{eqn_8}), the following statements hold.

1) The sequence $\left\{ {{\mathbf{x}}^{[k]}} \right\}$ is bounded.

2) ${{\lim }_{k\to \infty }}{{\left\| {{\mathbf{x}}^{[k+1]}}-{{\mathbf{x}}^{[k]}} \right\|}_{2}}=0$.

3) Any accumulation points of $\left\{ {{\mathbf{x}}^{[k]}} \right\}$ is a stationary point of $F\left( \mathbf{x} \right)$.
\end{Theorem}

\begin{IEEEproof}
1) Rewrite (\ref{eqn_8}) and consider the following inequality

\begin{equation}
\label{eqn_26}
\begin{aligned}
&F\left( {{{\bf{x}}^{[k + 1]}}} \right) - F\left( {{{\bf{x}}^{[k]}}} \right)\\
 &= \phi \left( {{{\bf{x}}^{[k + 1]}}} \right) + \rho P\left( {{{\bf{x}}^{[k + 1]}}} \right) - \phi \left( {{{\bf{x}}^{[k]}}} \right) - \rho P\left( {{{\bf{x}}^{[k]}}} \right)\\
 &\le \left\langle {\nabla \phi \left( {{{\bf{x}}^{[k]}}} \right),{{\bf{x}}^{[k + 1]}} - {{\bf{x}}^{[k]}}} \right\rangle  + \frac{L}{2}\left\| {{{\bf{x}}^{[k + 1]}} - {{\bf{x}}^{[k]}}} \right\|_2^2\\
 & \quad + \rho P\left( {{{\bf{x}}^{[k + 1]}}} \right) - \rho P\left( {{{\bf{x}}^{[k]}}} \right)\\
 &= \rho P\left( {{{\bf{x}}^{[k + 1]}}} \right) - \rho P\left( {{{\bf{x}}^{[k]}}} \right) + \frac{L}{2}\left\| {{{\bf{x}}^{[k + 1]}} - {{\bf{x}}^{[k]}}} \right\|_2^2 \\
  & \quad + \frac{{\left\| {{{\bf{x}}^{[k + 1]}} - \left( {{{\bf{x}}^{[k]}} - \beta \nabla \phi \left( {{{\bf{x}}^{[k]}}} \right)} \right)} \right\|_2^2}}{{2\beta }} - \frac{{\left\| {\beta \nabla \phi \left( {{{\bf{x}}^{[k]}}} \right)} \right\|_2^2}}{{2\beta }} \\
  & \quad- \frac{{\left\| {{{\bf{x}}^{[k + 1]}} - {{\bf{x}}^{[k]}}} \right\|_2^2}}{{2\beta }}\\
 &= \rho \left( {E\left( {{{\bf{x}}^{[k + 1]}}} \right) - E\left( {{{\bf{x}}^{[k]}}} \right)} \right) + \left( {\frac{L}{2} - \frac{1}{{2\beta }}} \right)\left\| {{{\bf{x}}^{[k + 1]}} - {{\bf{x}}^{[k]}}} \right\|_2^2\\
 &\le \left( {\frac{L}{2} - \frac{1}{{2\beta }}} \right)\left\| {{{\bf{x}}^{[k + 1]}} - {{\bf{x}}^{[k]}}} \right\|_2^2
\end{aligned}
\end{equation}
where the $E\left( \mathbf{x} \right)$ in the third equation is the expression (\ref{eqn_13}) with $\mathbf{y}$ replaced by ${{\mathbf{x}}^{[k]}}-\beta \nabla \phi \left( {{\mathbf{x}}^{[k]}} \right)$ and set $\lambda =\beta \rho $. The first inequality comes from Assumption 1, and the second inequality is based on the fact that ${{\mathbf{x}}^{[k+1]}}$ is the optimal solution of the $E\left( \mathbf{x} \right)$. When $\beta <1/L$, we have $F\left( {{\mathbf{x}}^{[k]}} \right)\le F\left( {{\mathbf{x}}^{[0]}} \right)$ for all $k\ge 0$. Due to the level-boundedness of $F\left( \mathbf{x} \right)$ (Assumption 2), therefore the sequence $\left\{ {{\mathbf{x}}^{[k]}} \right\}$ is bounded.

2) Summing both sides of (\ref{eqn_26}) from $k=0$ to $\infty $, we can obtain
\begin{equation}
\label{eqn_27}
\left( {\frac{1}{{2\beta }} - \frac{L}{2}} \right)\sum\limits_{k = 0}^{ + \infty } {\left\| {{{\bf{x}}^{[k + 1]}} - {{\bf{x}}^{[k]}}} \right\|_2^2}  \le F\left( {\bf{0}} \right) - F\left( {{{\bf{x}}^{[k + 1]}}} \right) < \infty
\end{equation}
Since $\beta <1/L$, we can deduce that ${{\lim }_{k\to \infty }}{{\left\| {{\mathbf{x}}^{[k+1]}}-{{\mathbf{x}}^{[k]}} \right\|}_{2}}=0$ from the above relation obviously.

3) Since the sequence $\left\{ {{\mathbf{x}}^{[k]}} \right\}$ is bounded, there exists a subsequence of $\left\{ {{\mathbf{x}}^{[k]}} \right\}$, denoted as $\left\{ {{\mathbf{x}}^{[{{k}_{j}}]}} \right\}$, converging to an accumulation point ${{\mathbf{x}}^{*}}$. Considering that minimizer $\left\{ {{\mathbf{x}}^{[{{k}_{j}}+1]}} \right\}$ is a critical point of (\ref{eqn_13}) and $P\left( \mathbf{x} \right)={{P}_{1}}\left( \mathbf{x} \right)-{{P}_{2}}\left( \mathbf{x} \right)$, we have
\begin{equation}
\label{eqn_28}
\begin{aligned}
{\bf{0}} \in &\frac{{{{\bf{x}}^{[{k_j} + 1]}} - {{\bf{x}}^{[k]}} + \beta \nabla \phi \left( {{{\bf{x}}^{[k]}}} \right)}}{{\beta \rho }}\\
& + \partial {P_1}\left( {{{\bf{x}}^{[{k_j} + 1]}}} \right) - \partial {P_2}\left( {{{\bf{x}}^{[{k_j} + 1]}}} \right)
\end{aligned}
\end{equation}
Let ${{k}_{j}}\to \infty $, by using ${{\left\| {{\mathbf{x}}^{[{{k}_{j}}+1]}}-{{\mathbf{x}}^{[{{k}_{j}}]}} \right\|}_{2}}\to 0$ from the above conclusion and considering the semi-continuity of $\nabla \phi $, $\partial {{P}_{1}}$ and $\partial {{P}_{2}}$, we have that ${\bf{0}} \in \nabla \phi \left( {{{\bf{x}}^ * }} \right) + \rho \partial {P_1}\left( {{{\bf{x}}^ * }} \right) - \rho \partial {P_2}\left( {{{\bf{x}}^ * }} \right)$. Therefore, ${{\mathbf{x}}^{*}}$ is a critical point of problem (\ref{eqn_8}). This completes the proof.
\end{IEEEproof}

From the proof of Theorem 2, we have that ${{\lim }_{k\to \infty }}{{\left\| {{\mathbf{x}}^{[k+1]}}-{{\mathbf{x}}^{[k]}} \right\|}_{2}}=0$ is a necessary optimality condition of the FBS. Therefore, we can use ${{\left\| {{\mathbf{x}}^{[k+1]}}-{{\mathbf{x}}^{[k]}} \right\|}_{2}}$ as a quantity to measure the convergence performance of the sequence $\left\{ {{\mathbf{x}}^{[k]}} \right\}$ to a critical point ${{\mathbf{x}}^{*}}$.

\begin{Theorem}
If $\beta <1/L$, let $\left\{ {{\mathbf{x}}^{[k]}} \right\}$ be the sequence generated by the FBS for (\ref{eqn_8}), then for every $K\ge 1$, we have
\begin{equation}
\label{eqn_29}
\mathop {\min }\limits_{0 \le k \le K} \left\| {{{\bf{x}}^{[k + 1]}} - {{\bf{x}}^{[k]}}} \right\|_2^2 \le 2\beta \frac{{F\left( {\bf{0}} \right) - F\left( {{{\bf{x}}^ * }} \right)}}{{K\left( {1 - L\beta } \right)}}
\end{equation}
\end{Theorem}

\begin{IEEEproof}
Summing the inequality (\ref{eqn_26}) over $k=0,\cdots ,K$, we can obtain
\begin{equation}
\label{eqn_30}
\left( {\frac{1}{{2\beta }} - \frac{L}{2}} \right)\sum\limits_{k = 0}^K {\left\| {{{\bf{x}}^{[k + 1]}} - {{\bf{x}}^{[k]}}} \right\|_2^2}  \le F\left( {\bf{0}} \right) - F\left( {{{\bf{x}}^{[K + 1]}}} \right)
\end{equation}
When $\beta <{}^{1}/{}_{L}$, we have $\left\{ F\left( {{\mathbf{x}}^{[k]}} \right) \right\}$ is monotonically decreasing, which means that $F\left( {{\mathbf{x}}^{[K+1]}} \right)\ge F\left( {{\mathbf{x}}^{*}} \right)$. Substitute this into (\ref{eqn_30}), we have
\begin{equation}
\label{eqn_31}
\begin{aligned}
K\mathop {\min }\limits_{0 \le k \le K} \left\| {{{\bf{x}}^{[k + 1]}} - {{\bf{x}}^{[k]}}} \right\|_2^2 &\le 2\beta \frac{{F\left( {\bf{0}} \right) - F\left( {{{\bf{x}}^{[K + 1]}}} \right)}}{{\left( {1 - L\beta } \right)}} \\
&\le 2\beta \frac{{F\left( {\bf{0}} \right) - F\left( {{{\bf{x}}^ * }} \right)}}{{\left( {1 - L\beta } \right)}}
\end{aligned}
\end{equation}
This completes the proof.
\end{IEEEproof}
In fact, we may have a stricter conclusion for the convergence speed as $F\left( {{\mathbf{x}}^{[k+1]}} \right)-F\left( {{\mathbf{x}}^{[k]}} \right)$ can be smaller than $\left( \frac{L}{2}-\frac{1}{2\beta } \right)\left\| {{\mathbf{x}}^{[k+1]}}-{{\mathbf{x}}^{[k]}} \right\|_{2}^{2}$ in (\ref{eqn_26}).

\begin{Proposition}
 If $R\left( \mathbf{x} \right)$ is separable, i.e., $R\left( \mathbf{x} \right)=\sum\limits_{i=1}^{N}{{{r}_{i}}\left( {{x}_{i}} \right)}$， and each ${{r}_{i}}$ is strictly increasing on ${{\mathbb{R}}_{+}}$, then we have
 \begin{equation}
\label{eqn_32}
\begin{aligned}
F\left( {{{\bf{x}}^{[k + 1]}}} \right) - F\left( {{{\bf{x}}^{[k]}}} \right) \le \left( {\frac{L}{2} - \frac{1}{{2\beta }}} \right)\left\| {{{\bf{x}}^{[k + 1]}} - {{\bf{x}}^{[k]}}} \right\|_2^2 \\
+ \min \left\{ { - \frac{1}{{2\beta }}\left\| {{{\bf{x}}^{[k + 1]}} - {{\bf{x}}^{[k]}}} \right\|_2^2 + \rho {\Delta _k},0} \right\}
\end{aligned}
\end{equation}
where ${\Delta _k} = \sum\limits_{i \in {\Lambda _{k + 1}}}^{} {{r_i}\left( {x_i^{[k]}} \right)}  - \sum\limits_{i \in {\Lambda _k}}^{} {{r_i}\left( {x_i^{[k]}} \right)} $, ${\Lambda _{k + 1}} = \Gamma _{{{\bf{x}}^{[k + 1]}}}^N\backslash \Gamma _{{{\bf{x}}^{[k + 1]}}}^s$, and ${\Lambda _k} = \Gamma _{{{\bf{x}}^{[k]}}}^N\backslash \Gamma _{{{\bf{x}}^{[k]}}}^s$.
\end{Proposition}
See Appendix H for the proof.

From Proposition 9, we can find that $F\left( {{\mathbf{x}}^{[k+1]}} \right)-F\left( {{\mathbf{x}}^{[k]}} \right)\le \left( \frac{L}{2}-\frac{1}{\beta } \right)\left\| {{\mathbf{x}}^{[k+1]}}-{{\mathbf{x}}^{[k]}} \right\|_{2}^{2}$ if $\Gamma _{{{\bf{x}}^{[k + 1]}}}^s$ is the same as $\Gamma _{{{\bf{x}}^{[k]}}}^s$.

\section{Extensions}
In this section, we discuss some related algorithms for solving (\ref{eqn_8}), show a link between the DC function $P\left( \mathbf{x} \right)$ with other regularization functions, and simply extend $P\left( \mathbf{x} \right)$ to rank-constrained problem.
\subsection{Related algorithms}
Here, we discuss some related algorithms. When $\phi \left( \mathbf{x} \right)$ is convex, it is an intuitive idea that using the DCA to solve the minimization (\ref{eqn_8}). Since $P\left( \mathbf{x} \right)$ can be written as the DC functions, i.e.,$P\left( \mathbf{x} \right)={{P}_{1}}\left( \mathbf{x} \right)-{{P}_{2}}\left( \mathbf{x} \right)$, the objective function can be naturally decomposed into
 \begin{equation}
\label{eqn_33}
F\left( {\bf{x}} \right) = \phi \left( {\bf{x}} \right) + \rho P\left( {\bf{x}} \right){\rm{ = }}\left\{ {\phi \left( {\bf{x}} \right) + \rho {P_1}\left( {\bf{x}} \right)} \right\} - \rho {P_2}\left( {\bf{x}} \right)
\end{equation}
The corresponding DCA solves the minimization problem as
 \begin{equation}
\label{eqn_34}
\begin{aligned}
{{\bf{x}}^{[k + 1]}} = \arg \mathop {\min }\limits_{{\bf{x}} \in {\mathbb{R}^N}} & \left\{ {\phi \left( {\bf{x}} \right) + \rho {P_1}\left( {\bf{x}} \right) - \rho {P_2}\left( {{{\bf{x}}^{[k]}}} \right)} \right.\\
&\left. { - \rho \left\langle {{{\bf{w}}^{[k]}},{\bf{x}} - {{\bf{x}}^{[k]}}} \right\rangle } \right\}
\end{aligned}
\end{equation}
where ${{\mathbf{w}}^{[k]}}\in \partial {{P}_{2}}\left( {{\mathbf{x}}^{[k]}} \right)$. Although this problem is convex, it does not necessarily have closed-form solution and the computational cost is very expensive for large-scale problems.

On the other hand, since $\phi \left( \mathbf{x} \right)$ is continuously differentiable with $L$-Lipschitz continuous gradient, we can use the Sequential Convex Programming (SCP) [67] to solve problem (\ref{eqn_8}) by updating $\left\{ {{\mathbf{x}}^{[k]}} \right\}$ as
 \begin{equation}
\label{eqn_35}
\begin{aligned}
{{\bf{x}}^{[k + 1]}} = &\arg \mathop {\min }\limits_{{\bf{x}} \in {\mathbb{R}^N}} \left\{ {\phi \left( {{{\bf{x}}^{[k]}}} \right) + \left\langle {\nabla \phi \left( {{{\bf{x}}^{[k]}}} \right),{\bf{x}} - {{\bf{x}}^{[k]}}} \right\rangle } \right.\\
 &\quad+ \frac{L}{2}\left\| {{\bf{x}} - {{\bf{x}}^{[k]}}} \right\|_2^2 + \rho {P_1}\left( {\bf{x}} \right) - \rho {P_2}\left( {{{\bf{x}}^{[k]}}} \right)\\
&\quad \left. { - \rho \left\langle {{{\bf{w}}^{[k]}},{\bf{x}} - {{\bf{x}}^{[k]}}} \right\rangle } \right\}
\end{aligned}
\end{equation}
Meanwhile, the SCP can be thought as a variant of DCA with DC decomposition:
\begin{equation}
\label{eqn_36}
\begin{aligned}
F\left( {\bf{x}} \right) =& \left( {\rho {P_1}\left( {\bf{x}} \right) + {{L\left\| {\bf{x}} \right\|_2^2} \mathord{\left/
 {\vphantom {{L\left\| {\bf{x}} \right\|_2^2} 2}} \right.
 \kern-\nulldelimiterspace} 2}} \right)\\
 & \quad - \left( {\rho {P_2}\left( {\bf{x}} \right) + {{L\left\| {\bf{x}} \right\|_2^2} \mathord{\left/
 {\vphantom {{L\left\| {\bf{x}} \right\|_2^2} 2}} \right.
 \kern-\nulldelimiterspace} 2} - \phi \left( {\bf{x}} \right)} \right)
\end{aligned}
\end{equation}
The subproblem can be written as
\begin{equation}
\label{eqn_37}
\begin{aligned}
{{\bf{x}}^{[k + 1]}} = &\arg \mathop {\min }\limits_{{\bf{x}} \in {\mathbb{R}^N}} \left\{ {\rho {P_1}\left( {\bf{x}} \right)} \right.\\
&\left. { + \frac{L}{2}\left\| {{\bf{x}} - \left( {{{\bf{x}}^{[k]}} - \frac{1}{L}\left( {\nabla \phi \left( {{{\bf{x}}^{[k]}}} \right) - \rho {{\bf{w}}^{[k]}}} \right)} \right)} \right\|_2^2} \right\}
\end{aligned}
\end{equation}
Due to that the subproblem (\ref{eqn_37}) can be solved by using the proximal operator, Ref. [43] and [44] call this type DCA as proximal DCA (PDCA). For some simple form $P\left( \mathbf{x} \right)$, subproblem (\ref{eqn_37}) also has closed-form solution. For example, $P\left( \mathbf{x} \right)={{\left\| \mathbf{x} \right\|}_{1}}-{{\left\| {{\mathbf{x}}^{s}} \right\|}_{1}}$ and $P\left( \mathbf{x} \right)=\left\| \mathbf{x} \right\|_{2}^{2}-\left\| {{\mathbf{x}}^{s}} \right\|_{2}^{2}$. In the numerical experiment, we will compare the FBS with this PDCA and show that the FBS is more efficient than PDCA in this problem. Meanwhile, as $P\left( \mathbf{x} \right)$ is a DC function, the FBS reduces to the GIST algorithm proposed in [54].

To improve the performance of the FBS, some acceleration methods can be used in the proximal framework. Such as the Nonmonotone Accelerated proximal gradient (nmAPG) method [55], the extrapolation method in PDCA (pDCAe) [51] and the backtracking line search initialized method with Barzilai-Borwein (BB) rule [68] in GIST [54].

\subsection{Comparing with other regularization}
From the previous discussion, we have illustrated that the DC function $P\left( \mathbf{x} \right)$ can replace the ${{\ell }_{0}}$-norm constraint. And in Theorem 1 and Proposition 2, we have proved that the unconstrained problem (\ref{eqn_8}) is equal to the original sparsity constrained problem (\ref{eqn_3}) if we select proper parameter $\rho $. On the other hand, in the minimization problem (\ref{eqn_8}), $P\left( \mathbf{x} \right)$ can also be considered as a regularizes function. Then, we can investigate its performance from the aspect of sparsity metric. Figure 1 shows the contours of various regularizers.

From Fig.1, we can find that the level curves of $R\left( \mathbf{x} \right)-R\left( {{\mathbf{x}}^{s}} \right)$ approach the $x$ and $y$ axes as the values get small, hence promoting sparsity. Inspire by Sidky et al. work of [69] and Rahimi et al. work of [70], where they using toy examples to illustrate the advantages of ${{\ell }_{p}}$ and ${{{\ell }_{1}}}/{{{\ell }_{2}}}\;$, respectively, we also use a similar example to show that with some special data sets $\left( \mathbf{A},\mathbf{b} \right)$, the $R\left( \mathbf{x} \right)-R\left( {{\mathbf{x}}^{s}} \right)$ tends to select a sparser solution.

Example 1 Let $N=6$ and define

${\bf{A}}: = \left[ {\begin{array}{*{20}{c}}
1&{ - 1}&0&0&0&0\\
0&1&{ - 1}&0&0&0\\
0&1&2&1&0&0\\
2&1&1&0&1&0\\
{0.5}&{0.5}&3&0&0&{ - 1}
\end{array}} \right]$, ${\bf{b}}: = \left[ {\begin{array}{*{20}{c}}
0\\
0\\
{15}\\
{20}\\
{40}
\end{array}} \right]$

It is straightforward that any general solutions of $\mathbf{Ax}=\mathbf{b}$ have the form of $\mathbf{x}={{\left( t,t,t,15-3t,20-4t,4t-40 \right)}^{T}}$ for a scalar $t\in \mathbb{R}$. The sparest solution occurs at $t=0$ for the sparsity of $\mathbf{x}$ being 3, and some local solutions include $t=5$ for sparsity being 4 and $t=10$ for sparsity being 5. We plot the various regularize function with respect to $t$, including ${{\ell }_{1}}$, ${{\ell }_{p}}$ ($p={1}/{2}$), ${{\ell }_{1-2}}$, ${{{\ell }_{1}}}/{{{\ell }_{2}}}$, MCP ($\theta =15$) of (A.3) and the proposed $R\left( \mathbf{x} \right)-R\left( {{\mathbf{x}}^{s}} \right)$ with $R\left( \mathbf{x} \right)={{\left\| \mathbf{x} \right\|}_{1}}$, ${{\left\| \mathbf{x} \right\|}_{2}}$, ${{\left\| \mathbf{x} \right\|}_{1}}-{{\left\| \mathbf{x} \right\|}_{2}}$, ${{{\left\| \mathbf{x} \right\|}_{1}}}/{{{\left\| \mathbf{x} \right\|}_{2}}}$, MCP, and $s=3$.

\begin{figure}[!t]
\centering
{\includegraphics[width=3.5 in]{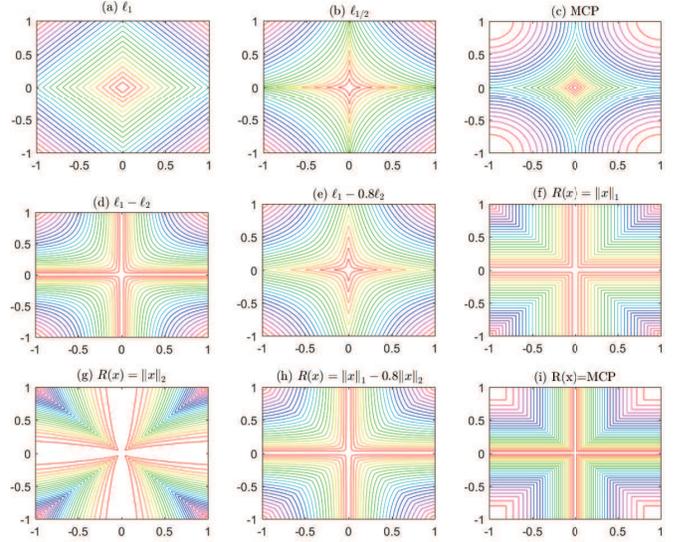}}
\caption{Level curves of different metrics. }
\end{figure}

\begin{figure*}[!t]
\centering
{\includegraphics[width=7.5 in]{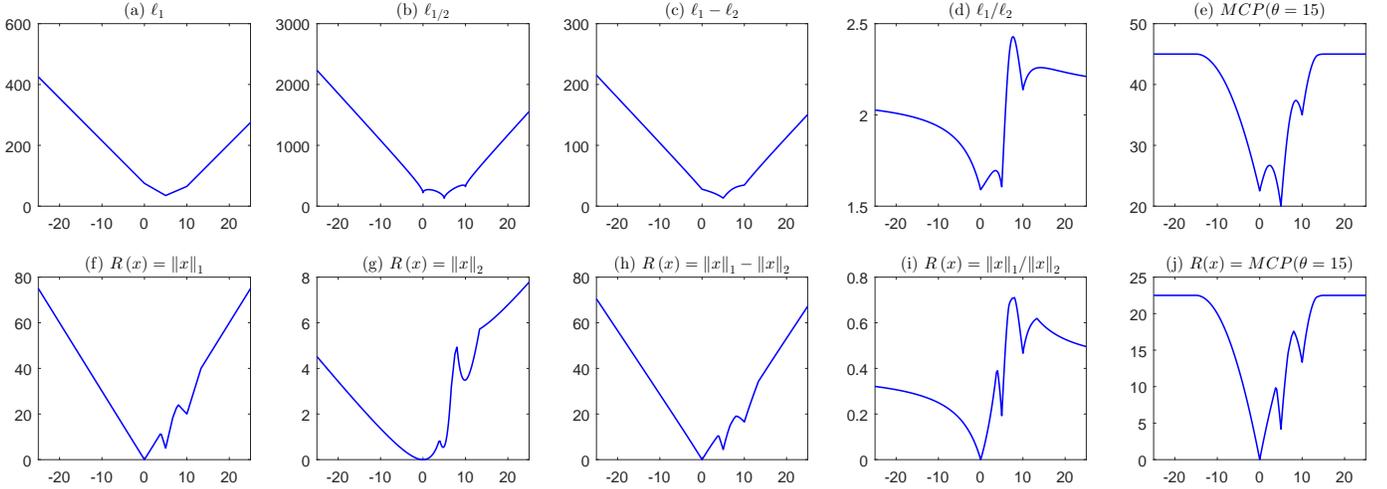}}
\caption{The objective functions of a toy example. For the top row, from the left to right, the five columns are functions of ${{\left\| \mathbf{x} \right\|}_{1}}$, ${{\left\| \mathbf{x} \right\|}_{0.5}}$, ${{\left\| \mathbf{x} \right\|}_{1}}-{{\left\| \mathbf{x} \right\|}_{2}}$, ${{{\left\| \mathbf{x} \right\|}_{1}}}/{{{\left\| \mathbf{x} \right\|}_{2}}}\;$, MCP of (A.3) with $\theta =15$, respectively. While for the bottom row, from the left to right, the five columns are functions of $R\left( \mathbf{x} \right)-R\left( {{\mathbf{x}}^{s}} \right)$ with $R\left( \mathbf{x} \right)={{\left\| \mathbf{x} \right\|}_{1}}$, ${{\left\| \mathbf{x} \right\|}_{2}}$, ${{\left\| \mathbf{x} \right\|}_{1}}-{{\left\| \mathbf{x} \right\|}_{2}}$, ${{{\left\| \mathbf{x} \right\|}_{1}}}/{{{\left\| \mathbf{x} \right\|}_{2}}}\;$, MCP, respectively.}
\end{figure*}

From Fig. 2, we can find that all these regularized functions are not differentiable at the values of $t=0,5$, and $10$, where the corresponding sparsity of $\mathbf{x}$ are all small than 6. However, only the ${{\ell _1}}/{{\ell _2}}$ and the $s$-difference $R\left( \mathbf{x} \right)-R\left( {{\mathbf{x}}^{s}} \right)$ can find the sparsest vector$\mathbf{x}$ at $t=0$ as a global minimum, where the other functions find $t=5$ as the minimum and lead to the sparsity of $\mathbf{x}$ being 4.

\subsection{Extend to rank-constrained problem}
Similar as in [43], the penalty function $P\left( \mathbf{x} \right)=R\left( \mathbf{x} \right)-R\left( \mathbf{x}_{{}}^{s} \right)$ can also be extended to rank-constrained problem based on the connection between the ${{\ell }_{0}}$-norm on ${{\mathbb{R}}^{N}}$ and the rank function for a matrix. The rank-constrained minimization problem can be formulated as
\begin{equation}
\label{eqn_38}
\mathop {\min }\limits_{\bf{w}} \phi \left( {\bf{w}} \right) \quad \rm{subject \ to} \quad {\rm{rank}}\left( {\bf{w}} \right) \le s,{\bf{w}} \in {\mathbb{R}^{M \times N}}
\end{equation}
where $s$ is a non-negative integer with $s\le q=\min \left\{ M,N \right\}$. As the rank of a matrix is equal to the number of its nonzero singular values, i.e., $\text{rank}\left( \mathbf{w} \right)={{\left\| \sigma \left( \mathbf{w} \right) \right\|}_{0}}$, where $\sigma \left( \mathbf{w} \right)$ represents the singular value vector of $\mathbf{w}$ and $\sigma _{i}^{{}}\left( \mathbf{w} \right)$ is the $i$-th largest term, then we can construct the penalty functions $P,R:\mathbb{R}_{+}^{q}\to {{\mathbb{R}}_{+}}$, $P\left( \sigma \left( \mathbf{w} \right) \right)=R\left( \sigma \left( \mathbf{w} \right) \right)-R\left( {{\sigma }^{s}}\left( \mathbf{w} \right) \right)$ that satisfies Property 1 (b) and (c), where $\sigma _{i}^{s}\left( \mathbf{w} \right)=\sigma _{i}^{{}}\left( \mathbf{w} \right)$ for $i\in \left\{ 1,2,\cdots ,s \right\}$ and $\sigma _{i}^{s}\left( \mathbf{w} \right)=0$ for else. Replace the rank constraint with the DC penalty function $P\left( \sigma \left( \mathbf{w} \right) \right)$ and consider the unconstrained problem:
\begin{equation}
\label{eqn_39}
\mathop {\min }\limits_{\bf{w}} \phi \left( {\bf{w}} \right) + \rho P\left( {\sigma \left( {\bf{w}} \right)} \right)
\end{equation}
Then we can use the FBS, DCA or ADMM algorithms to solve this rank-constrained problem.

\section{Numerical experiments}

In this section, simulations are performed to demonstrate the proposed conclusions and evaluate the performance of the  $s$-difference regularization. We mainly apply four methods in comparison with the proposed algorithm: (1) the ${\ell _1}$-norm regularization based  ${\ell _1}$-ADMM [72], (2) the  ${\ell _p}$-norm ($p = 1/2$) regularization based half thresholding [14], (3) the ${\ell _0}$-norm regularization based accelerate IHT (AIHT) [39], (4) the difference of the ${\ell _1}$ and  ${\ell _2}$-norms (${\ell _{1-2}}$) regularization based ${\ell _{1-2}}$-DCA [31]. We choose the representative $R\left( {\bf{x}} \right)$ as $R\left( {\bf{x}} \right) = {\left\| {\bf{x}} \right\|_1},{\left\| {\bf{x}} \right\|_2},{\left\| {\bf{x}} \right\|_1} - {\left\| {\bf{x}} \right\|_2}$ for comparing. All experiments are performed in MATLAB 2015b running on ASUS laptop with Intel (R) Core (TM) i7-8550U CPU, 8 GB of RAM and 64-bit Windows 10 operating system.

We focus on the following least squares problem:
\begin{equation}
\label{eqn_40}
\mathop {\min }\limits_{{\bf{x}} \in {\mathbb{R}^N}} \frac{1}{2}\left\| {{\bf{Ax}} - {\bf{b}}} \right\|_2^2 + \rho P\left( {\bf{x}} \right)
\end{equation}
and conduct experiments on simulated vector signals.

We test two types of matrices ${\bf{A}}$: the random Gaussian matrix with i.i.d. standard Gaussian entries and being normalized that each column has unit norm, and the random partial DCT matrix which is formed by randomly select rows from the full DCT matrix. For the original sparse vector ${\bf{\bar x}}$, we generate it with random index set and draw non-zero elements with standard normal distribution. The observation is ${\bf{b}} = {\bf{A\bar x}} + {\bf{n}}$, where ${\bf{n}}$ is zeros for the noiseless test, and Gaussian noise for the contaminated measurements. The initial value for all the methods is an approximated solution of the ${\ell _1}$ minimization using ADMM after $N$ iterations. The max iteration for all these methods is $5N$ except for DCA, whose max internal iteration is $5N$ and the max external iteration is $20$. The stopping condition is set to be $\frac{{{{\left\| {{{\bf{x}}^{[k]}} - {{\bf{x}}^{[k - 1]}}} \right\|}_2}}}{{\max \left\{ {{{\left\| {{{\bf{x}}^{[k]}}} \right\|}_2},1} \right\}}} < {10^{ - 5}}$.

In the first study, we look at the success rates with 100 random instances under the noise-free condition, in which we set the size of matrices ${\bf{A}}$ as $64 \times 256$. Here we consider a recovery ${{\bf{x}}^ * }$ as successful if the relative error of recovery (Rel.Err) satisfies ${{{\left\| {{{\bf{x}}^ * } - {\bf{\bar x}}} \right\|}_2}} / {{{\left\| {{\bf{\bar x}}} \right\|}_2}} \le {10^{ - 3}}$. In addition, we set sparsity parameter $s$ to the ground truth ${s_{truth}}$ for the proposed $s$-difference $P\left( {\bf{x}} \right)$. Fig 3 plots the success rates of the comparing methods for both the Gaussian matrix and the partial DCT matrix. From this, we can find that the $s$-difference regularization with $R\left( {\bf{x}} \right) = {\left\| {\bf{x}} \right\|_1}$ has the best performance for both Gaussian matrix and partial DCT matrix, the $R\left( {\bf{x}} \right) = {\left\| {\bf{x}} \right\|_1} - {\left\| {\bf{x}} \right\|_2}$ is comparable to ${\ell _{1 - 2}}$-DCA, followed by $R\left( {\bf{x}} \right) = {\left\| {\bf{x}} \right\|_2}$ and half thresholding, which outperform the ${\ell _1}$-ADMM.

\begin{figure}[!t]
\centering
{\includegraphics[width=3.5 in]{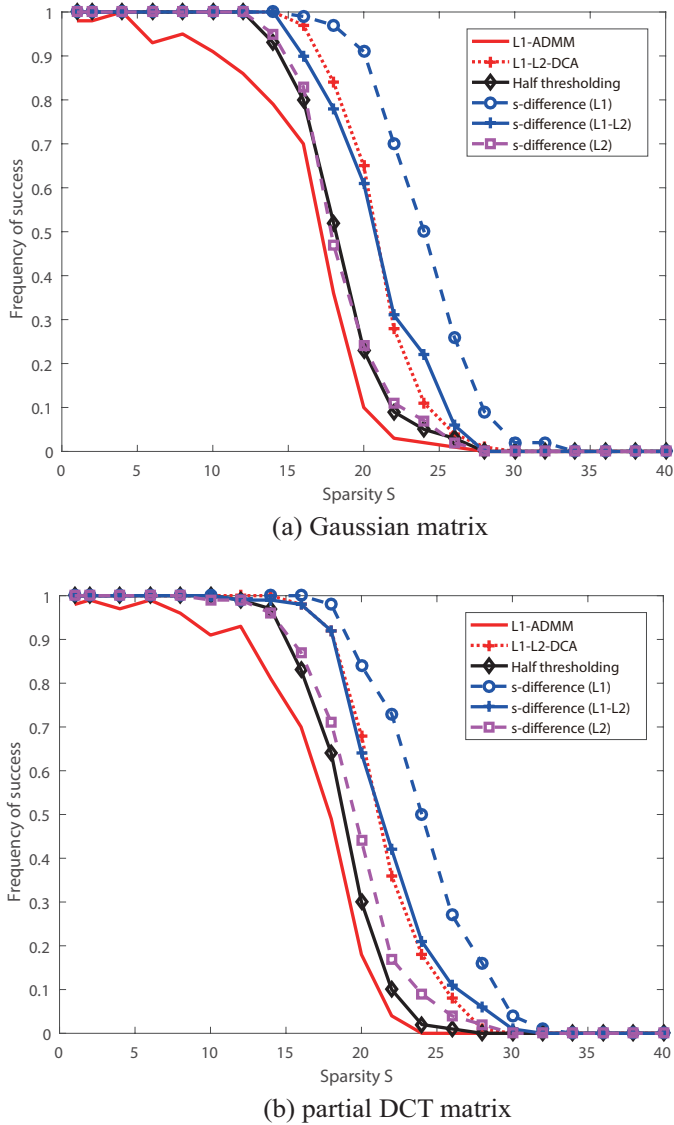}}
\caption{Success rates versus sparsity for compared methods: (a) Gaussian matrix, (b) partial DCT matrix.}
\end{figure}

In the second study, we focus on the recovery quantity of these methods under different sizes of matrix. For the noiseless case , we set $\rho = {10^{ - 1}}$ for FBS and $\rho  = {10^{ - 6}}$ for the ADMM and other types methods, set $\beta  = 10\rho $, and consider $\left( {M,N,{s_{truth}}} \right) = \left( {256i,1024i,48i} \right)$ for $i = 1,2, \cdots ,8$. Here we also set the sparsity threshold parameter to ${s_{truth}}$ for the AIHT and $s$ -difference $P\left( {\bf{x}} \right)$. For each triple $\left( {M,N,{s_{truth}}} \right)$, we generate 30 random realizations. Table 2 and 3 list the mean and standard deviation of Rel.Err for Gaussian matrix and partial DCT matrix, respectively. We also test these methods in presence of Gaussian noise as ${\bf{n}} = 0.01 * randn\left( {M,1} \right)$.  We set $ \rho  = 1$ for FBS and $\rho  = {10^{ - 3}}$ for the ADMM and other types methods, and consider $\left( {M,N,{s_{truth}}} \right) = \left( {256i,1024i,48i} \right)$ for $i = 1,2,3,4$. The recovery performance is listed in Table 4 and 5 for comparing. From Table 2 to 5, we can find that the $s$-difference $P\left( {\bf{x}} \right)$ with the ground truth sparsity threshold parameter can provide a quite competitive or slightly superior performance when comparing with AIHT and other methods under the noise-free conditions. However, under the condition of noise, AIHT performance decreases rapidly, while the $s$ -difference $P\left( {\bf{x}} \right)$ still able to provide a relatively best result.

In the third study, we focus on the accuracy and efficiency of the methods under fixed matrix ${\bf{A}}$ and the sparsity level as $\left( {M,N,{s_{truth}}} \right) = \left( {256,1024,48} \right)$. To illustrate the benefit of the closed-form solutions of proposed  $s$-difference regularization, we selectively analysis the performance of DCA, PDCA and FBS under the condition of the same regularization $P\left( {\bf{x}} \right) = {\left\| {\bf{x}} \right\|_1} - {\left\| {{{\bf{x}}^s}} \right\|_1}$. The DCA solves the minimization problem (\ref{eqn_40}) by using (\ref{eqn_34}), that is
\begin{equation}
\label{eqn_41}
\begin{aligned}
{{\bf{x}}^{[k + 1]}} = \arg \mathop {\min }\limits_{{\bf{x}} \in {^N}} \left\{ {\frac{1}{2}\left\| {{\bf{Ax}} - {\bf{b}}} \right\|_2^2 + \rho {{\left\| {\bf{x}} \right\|}_1} - \rho \left\langle {{{\bf{w}}^{[k]}},{\bf{x}}} \right\rangle } \right\}
\end{aligned}
\end{equation}
where ${{\bf{w}}^{[k]}} \in \partial {\left\| {{{\bf{x}}^{s\left[ k \right]}}} \right\|_1}$. This problem can be solved by ADMM as

\begin{equation}
\label{eqn_42}
\begin{aligned}
\mathop {\min }\limits_{{\bf{x}},{\bf{v}} \in {\mathbb{R}^N}} \left\{ {\frac{1}{2}\left\| {{\bf{Ax}} - {\bf{b}}} \right\|_2^2 + \rho {{\left\| {\bf{v}} \right\|}_1} - \rho \left\langle {{{\bf{w}}^{[k]}},{\bf{x}}} \right\rangle } \right\}\\
subject \;to \qquad  {\bf{x}} - {\bf{v}} = 0 \qquad \qquad
\end{aligned}
\end{equation}

We denote this method as DCA-ADMM for short. The PDCA solve the minimization problem (\ref{eqn_40}) by using (\ref{eqn_37}), that is
\begin{equation}
\label{eqn_43}
\begin{aligned}
&{{\bf{x}}^{[k + 1]}} = \arg \mathop {\min }\limits_{{\bf{x}} \in {\mathbb{R}^N}} \left\{ {\rho {{\left\| {\bf{x}} \right\|}_1}} \right.\\
&\left. { + \frac{L}{2}\left\| {{\bf{x}} - \left( {{{\bf{x}}^{[k]}} - \frac{1}{L}\left( {{{\bf{A}}^T}\left( {{\bf{A}}{{\bf{x}}^{[k]}} - {\bf{b}}} \right) - \rho {{\bf{w}}^{[k]}}} \right)} \right)} \right\|_2^2} \right\}
\end{aligned}
\end{equation}
and it can be solved by using soft shrinkage operator (\ref{eqn_18}). We denote this method as PDCA for short. The FBS solve the problem by using closed-form solution (\ref{eqn_17}) in Remark 7.

Figure 4 shows the convergence performance of three methods under noise-free condition with partial DCT matrix, which is measured by the Log-Rel.Err (defined as $10{\log _{10}}\left( {{\rm{Rel}}{\rm{.Err}}} \right)$) versus iteration numbers. Table 6 lists the mean of relative error, iteration number and computational time (in seconds) under the noise-free and Gaussian noise conditions as ${\bf{n}} = 0.01 * randn\left( {M,1} \right)$. From Figure 4 and Table 6, it is clear that the FBS with closed-form method leads to less error and converges faster than the DCA type methods.

\begin{figure}[!t]
\centering
{\includegraphics[width=3.5 in]{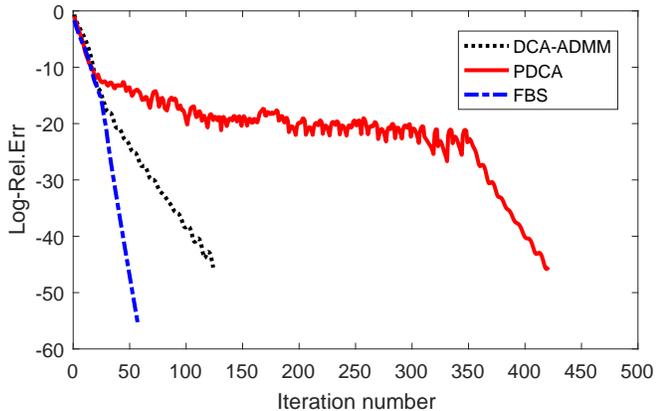}}
\caption{Convergence performance of DCA-ADMM, PDCA and FBS for solving the  $s$-difference ${\left\| {\bf{x}} \right\|_1} - {\left\| {{{\bf{x}}^s}} \right\|_1}$ regularization problem.}
\end{figure}

From the definition of  $s$-difference $P\left( {\bf{x}} \right)$, it is easy to understand that the parameter $s$ plays an important role in the algorithm. Here we focus on the problem of how the select the proper $s$ with the fixed matrix ${\bf{A}}$ and ${s_{truth}}$ as $\left( {M,N,{s_{truth}}} \right) = \left( {256,1024,48} \right)$. Figure 5 shows the performance of $s$-difference $P\left( {\bf{x}} \right) = \left( {{{\left\| {\bf{x}} \right\|}_1} - {{\left\| {\bf{x}} \right\|}_2}} \right) - \left( {{{\left\| {{{\bf{x}}^s}} \right\|}_1} - {{\left\| {{{\bf{x}}^s}} \right\|}_2}} \right)$ under different $s$ from 1 to 1000. In addition to use the FBS with closed-form solution as Proposition 8, we also consider the approximate DCA-ADMM using the similar solution of (\ref{eqn_42}) but set ${{\bf{w}}^{[k]}} \in \partial \left( {{{\left\| {{{\bf{x}}^{\left[ k \right]}}} \right\|}_2} + {{\left\| {{{\bf{x}}^{s\left[ k \right]}}} \right\|}_1} - {{\left\| {{{\bf{x}}^{s\left[ k \right]}}} \right\|}_2}} \right)$. This method is not a true DCA due to that the decomposition is not the convex function, however, this DCA-ADMM still works well as shown in Figure 5. From Figure 5, we can find that once the parameter s is less than the true sparsity ${s_{truth}}$, the performance of FBS with closed-form will drop sharply, however, the DCA-ADMM almost unaffected. This is probably because that the FBS solve the problem as the hard thresholding way when $\left| {{y_{{\pi _y}\left( {s + 1} \right)}}} \right|$ is smaller than $\lambda $ in Proposition 8, whereas the DCA-ADMM make full use of the nonconvex $P\left( {\bf{x}} \right)$ and bring better results than the ${\ell _1}$-norm methods. According to this deduction, designing an adaptive penalty parameter for FBS is quite necessary, which also is our future work. The good performance of DCA-ADMM also shows the superiority of this  $s$-difference regularization from another angle.

From Figure 5, we also have a suggestion that if we already have a preliminary range of judgements about sparsity based on prior knowledge, i.e., ${s_{truth}} \in \left( {{s_{\max }},{s_{\min }}} \right)$, then we suggest that $s$ decreases from the ${s_{\max }}$, but no less than ${s_{\min }}$, or just set $s$ be equal to ${s_{\max }}$ when the range of sparsity is not very large. Here, we also introduce an adjustment strategy to estimate the parameter $s$ when we don't know the prior sparsity range: set ${s^{[k + 1]}} = size\left( {find\left( {\left| {{{\bf{x}}^{[k]}}} \right| \ge \min \left\{ {\left| {x_{{\pi _x}\left( {{s^{[k - 1]}}} \right)}^{[k - 1]}} \right|,\varepsilon } \right\}} \right)} \right)$, where constant $\varepsilon  > 0$ is given. Some experiments show that this adjustment strategy usually can find the approximate true sparsity level ${s_{truth}}$, which means that it maybe can be used to estimate the sparsity of the unknown signal.

\begin{figure}[!t]
\centering
{\includegraphics[width=3.5 in]{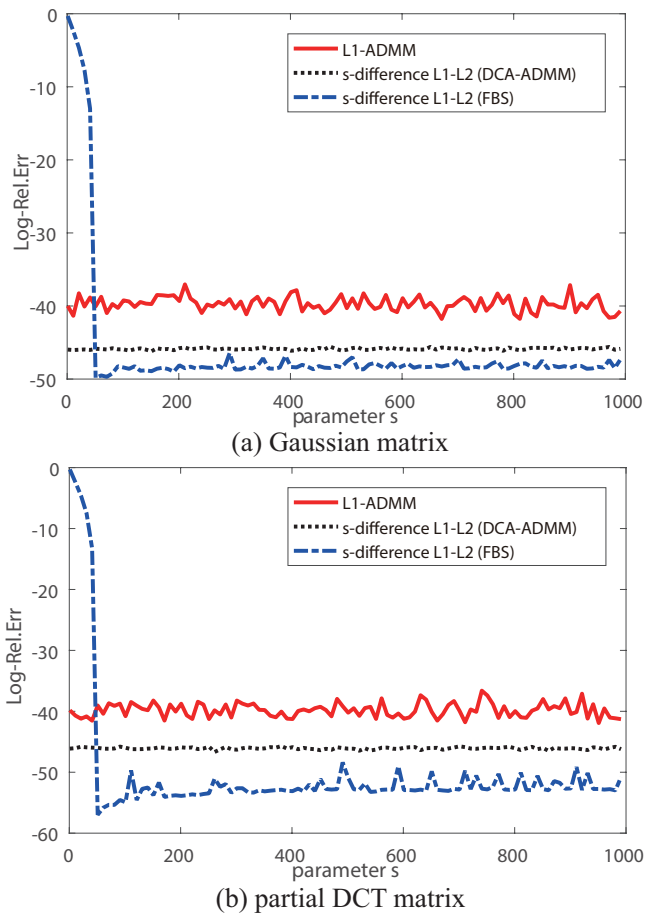}}
\caption{Recovery performance of DCA-ADMM and FBS for solving the  $s$-difference regularization problem with different parameter $s$ : (a) Gaussian matrix, (b) partial DCT matrix.}
\end{figure}

\begin{table*}
\newcommand{\tabincell}[2]{\begin{tabular}{@{}#1@{}}#2\end{tabular}}
  \centering
  \caption{\label{table_2}Mean and standard deviation of Rel.Err for different methods with Gaussian matrix under noiseless condition}
  \begin{tabular}{|c|c|c|c|c|c|c|c|c|c|}
        \hline
       $M$ & $N$ & $ {s_{true}}$  & ${\ell _1}$-ADMM & ${\ell _{1-2}}$-DCA & \tabincell{c}{Half \\thresholding} & AIHT & \tabincell{c}{$s$-difference  \\ (${\ell _1}$)} &  \tabincell{c}{$s$-difference  \\ (${\ell _{1-2}}$)} & \tabincell{c}{$s$-difference  \\ (${\ell _2}$)}   \\ \hline
       256 & 1024 &48&  \tabincell{c}{1.098e-04 \\(1.774e-08)} & \tabincell{c}{2.603e-05 \\(3.222e-11)} & \tabincell{c}{2.495e-05 \\(1.587e-11)} & \tabincell{c}{1.370e-05\\(6.481e-12)} & \tabincell{c}{1.368e-05 \\(6.382e-12)} &\tabincell{c}{\textbf{1.332e-05} \\(6.129e-12)} &\tabincell{c}{1.346e-05 \\(6.233e-12)} \\ \hline
       512 & 2048 & 96 &  \tabincell{c}{1.412e-04 \\(2.242e-08)} & \tabincell{c}{ 2.491e-05\\(1.536e-11)} &\tabincell{c}{ 2.484e-05\\(8.046e-11)} &\tabincell{c}{ \textbf{1.296e-05}\\(5.700e-12)} &\tabincell{c}{ 1.334e-05\\(5.284e-12)} &\tabincell{c}{1.356e-05 \\(4.390e-11)} &\tabincell{c}{ 1.364e-05\\(4.326e-11)}\\ \hline
       768& 3072& 144& \tabincell{c}{1.166e-04\\(2.099e-08)} & \tabincell{c}{2.652e-05\\(6.496e-12)} & \tabincell{c}{2.586e-05\\(4.274e-12)} &\tabincell{c}{1.335e-05\\(2.423e-12)} & \tabincell{c}{1.323e-05\\(2.383e-12)} & \tabincell{c}{ \textbf{1.301e-05}\\(2.312e-12)} &\tabincell{c}{1.308e-05\\(2.340e-12)} \\ \hline
       1024 & 4096&192& \tabincell{c}{1.522e-04\\(3.135e-08)} &\tabincell{c}{2.552e-05\\(7.715e-12)} &\tabincell{c}{2.495e-05\\(3.871e-12)} &\tabincell{c}{\textbf{1.243e-05}\\(1.403e-12)} &\tabincell{c}{1.284e-05\\(1.469e-12)} &\tabincell{c}{1.261e-05\\(1.403e-12)} &\tabincell{c}{1.267e-05\\(1.418e-12)} \\ \hline
       1280&5120&240&\tabincell{c}{1.239e-04\\(1.503e-08)} &\tabincell{c}{2.682e-05\\(9.676e-12)} &\tabincell{c}{2.523e-05\\(1.890e-12)} &\tabincell{c}{1.278e-05\\(7.650e-13)}
       &\tabincell{c}{1.261e-05\\(7.359e-13)} &\tabincell{c}{\textbf{1.241e-05}\\(7.275e-13)} &\tabincell{c}{1.247e-05\\(7.341e-13)}\\ \hline
       1536&6144&288&\tabincell{c}{1.038e-04\\(1.404e-08)}
       &\tabincell{c}{2.586e-05\\(8.843e-12)}
       &\tabincell{c}{2.543e-05\\(2.426e-12)}&\tabincell{c}{1.330e-05\\(1.589e-12)}
       &\tabincell{c}{1.327e-05\\(1.581e-12)}&\tabincell{c}{\textbf{1.293e-05}\\(1.509e-12)}
       &\tabincell{c}{1.298e-05\\(1.520e-12)} \\ \hline
       1792&7168&336&\tabincell{c}{1.518e-04\\(1.650e-08)}
       &\tabincell{c}{2.647e-05\\(9.289e-12)}  &\tabincell{c}{2.525e-05\\(2.889e-12)}
         &\tabincell{c}{\textbf{1.271e-05}\\(1.313e-12)}   &\tabincell{c}{1.298e-05\\(1.470e-12)}
           &\tabincell{c}{1.275e-05\\(1.394e-12)}  &\tabincell{c}{1.280e-05\\(1.405e-12)}\\ \hline
       2018&8192&384&\tabincell{c}{1.665e-04\\(1.744e-08)}
       &\tabincell{c}{2.602e-05\\(7.737e-12)}&\tabincell{c}{2.550e-05\\(2.028e-12)}
       &\tabincell{c}{1.325e-05\\(1.623e-12)}&\tabincell{c}{1.318e-05\\(1.601e-12)}
       &\tabincell{c}{\textbf{1.287e-05}\\(1.482e-12)}&\tabincell{c}{1.291e-05\\(1.493e-12)}\\ \hline
    \end{tabular}
\end{table*}

\begin{table*}
\newcommand{\tabincell}[2]{\begin{tabular}{@{}#1@{}}#2\end{tabular}}
  \centering
  \caption{\label{table_3}Mean and standard deviation of Rel.Err for different methods with  partial DCT matrix under noiseless condition}
  \begin{tabular}{|c|c|c|c|c|c|c|c|c|c|}
        \hline
       $M$ & $N$ & $ {s_{true}}$  & ${\ell _1}$-ADMM & ${\ell _{1-2}}$-DCA & \tabincell{c}{Half \\thresholding} & AIHT & \tabincell{c}{$s$-difference  \\ (${\ell _1}$)} &  \tabincell{c}{$s$-difference  \\ (${\ell _{1-2}}$)} & \tabincell{c}{$s$-difference  \\ (${\ell _2}$)}   \\ \hline
       256 & 1024 &48&  \tabincell{c}{1.357e-04\\(3.142e-08)} & \tabincell{c}{2.318e-05\\(3.280e-11)} & \tabincell{c}{1.031e-05\\(4.354e-12)} & \tabincell{c}{4.117e-06\\(2.503e-13)} & \tabincell{c}{3.059e-06\\(2.068e-13)}  &\tabincell{c}{\textbf{2.882e-06}\\(1.705e-13)}
       &\tabincell{c}{3.024e-06\\(1.846e-13)} \\ \hline
       512 & 2048 & 96 &  \tabincell{c}{6.404e-05\\(5.610e-09)} & \tabincell{c}{ 2.503e-05\\(2.257e-11)} &\tabincell{c}{1.127e-05 \\(5.751e-12)} &\tabincell{c}{ 4.226e-06\\(1.482e-13)} &\tabincell{c}{3.117e-06 \\(1.296e-13)} &\tabincell{c}{\textbf{2.975e-06}\\(1.060e-13)} &\tabincell{c}{ 3.081e-06\\(1.119e-13)}\\ \hline
       768& 3072& 144& \tabincell{c}{1.031e-04\\(2.015e-08)} & \tabincell{c}{2.446e-05\\(2.514e-11)} & \tabincell{c}{1.260e-05\\(2.958e-12)} &\tabincell{c}{4.351e-06\\(1.998e-13)} & \tabincell{c}{3.207e-06\\(1.668e-13)} & \tabincell{c}{\textbf{3.089e-06} \\(1.369e-13)} &\tabincell{c}{3.178e-06\\(1.439e-13)} \\ \hline
       1024 & 4096&192& \tabincell{c}{1.088e-04\\(1.984e-08)} &\tabincell{c}{2.480e-05\\(1.353e-11)} &\tabincell{c}{1.459e-05\\(2.107e-12)} &\tabincell{c}{4.339e-06\\(1.024e-13)} &\tabincell{c}{3.193e-06\\(8.038e-14)} &\tabincell{c}{\textbf{3.073e-06}\\(6.256e-14)} &\tabincell{c}{3.150e-06\\(6.618e-14)} \\ \hline
       1280&5120&240&\tabincell{c}{1.328e-04\\(1.753e-08)} &\tabincell{c}{2.499e-05\\(2.367e-11)} &\tabincell{c}{1.434e-05\\(1.907e-12)} &\tabincell{c}{4.403e-06\\(5.093e-13)}
       &\tabincell{c}{3.229e-06\\(4.012e-13)} &\tabincell{c}{\textbf{3.122e-06}\\(2.613e-13)} &\tabincell{c}{3.191e-06\\(2.755e-13)}\\ \hline
       1536&6144&288&\tabincell{c}{7.135e-05\\(5.683e-09)}
       &\tabincell{c}{2.509e-05\\(2.298e-11)}
       &\tabincell{c}{1.495e-05\\(2.081e-12)}&\tabincell{c}{4.199e-06\\(6.557e-14)}
       &\tabincell{c}{3.101e-06\\(7.162e-14)}&\tabincell{c}{\textbf{2.965e-06}\\(4.915e-14)}
       &\tabincell{c}{3.026e-06\\(5.044e-14))} \\ \hline
       1792&7168&336&\tabincell{c}{1.378e-04\\(1.083e-08)}
       &\tabincell{c}{2.232e-05\\(3.147e-11)}  &\tabincell{c}{1.301e-05\\(1.748e-12)}
         &\tabincell{c}{4.350e-06\\(5.552e-14)}   &\tabincell{c}{3.244e-06\\(5.568e-14)}
           &\tabincell{c}{\textbf{3.079e-06}\\(3.362e-14)}  &\tabincell{c}{3.138e-06\\(3.464e-14)}\\ \hline
       2018&8192&384&\tabincell{c}{1.300e-04\\(1.383e-08)}
       &\tabincell{c}{2.478e-05\\(1.532e-11)}&\tabincell{c}{1.506e-05\\(2.359e-12)}
       &\tabincell{c}{4.260e-06\\(3.392e-14)}&\tabincell{c}{3.162e-06\\(4.226e-14)}
       &\tabincell{c}{\textbf{3.016e-06}\\(2.550e-14)}&\tabincell{c}{3.069e-06\\(2.600e-14)}\\ \hline
    \end{tabular}
\end{table*}

\begin{table*}
\newcommand{\tabincell}[2]{\begin{tabular}{@{}#1@{}}#2\end{tabular}}
  \centering
  \caption{\label{table_4}Mean and standard deviation of Rel.Err for different methods with Gaussian matrix under Gaussian noise}
  \begin{tabular}{|c|c|c|c|c|c|c|c|c|c|}
        \hline
       $M$ & $N$ & $ {s_{true}}$  & ${\ell _1}$-ADMM & ${\ell _{1-2}}$-DCA & \tabincell{c}{Half \\thresholding} & AIHT & \tabincell{c}{$s$-difference  \\ (${\ell _1}$)} &  \tabincell{c}{$s$-difference  \\ (${\ell _{1-2}}$)} & \tabincell{c}{$s$-difference  \\ (${\ell _2}$)}   \\ \hline
       256 & 1024 &48&  \tabincell{c}{1.198e-01\\(6.182e-04)} & \tabincell{c}{1.039e-01\\(4.021e-04)} & \tabincell{c}{7.307e-02\\(1.456e-04)} & \tabincell{c}{2.094e-01\\(9.742e-04)} & \tabincell{c}{6.190e-02\\(2.614e-04))}  &\tabincell{c}{\textbf{6.034e-02}\\(2.923e-04)}
       &\tabincell{c}{6.066e-02\\(2.915e-04)} \\ \hline
       512 & 2048 & 96 &  \tabincell{c}{1.167e-01\\(1.065e-04)} & \tabincell{c}{ 1.050e-01\\(8.154e-05)} &\tabincell{c}{8.861e-02\\(9.510e-05)} &\tabincell{c}{ 2.139e-01\\(5.217e-04)} &\tabincell{c}{6.174e-02\\(2.542e-04)} &\tabincell{c}{5.929e-02\\(1.054e-04)} &\tabincell{c}{ \textbf{5.906e-02}\\(1.021e-04)}\\ \hline
       768& 3072& 144& \tabincell{c}{1.189e-01\\(1.128e-04)} & \tabincell{c}{1.088e-01\\(9.662e-05)} & \tabincell{c}{1.016e-01\\(1.297e-04)} &\tabincell{c}{2.144e-01\\(5.049e-04)} & \tabincell{c}{6.192e-02\\(2.892e-04)} & \tabincell{c}{\textbf{5.886e-02}\\(6.235e-05))} &\tabincell{c}{5.901e-02\\(6.197e-05)} \\ \hline
       1024 & 4096&192& \tabincell{c}{1.221e-01\\(6.933e-05)} &\tabincell{c}{1.132e-01\\(5.775e-05)} &\tabincell{c}{1.093e-01\\(9.770e-05)} &\tabincell{c}{2.210e-01\\(3.146e-04)} &\tabincell{c}{6.188e-02\\(1.819e-04)} &\tabincell{c}{\textbf{5.777e-02}\\(5.537e-05)} &\tabincell{c}{5.797e-02\\(5.548e-05)} \\ \hline
    \end{tabular}
\end{table*}

\begin{table*}
\newcommand{\tabincell}[2]{\begin{tabular}{@{}#1@{}}#2\end{tabular}}
  \centering
  \caption{\label{table_5}Mean and standard deviation of Rel.Err for different methods with partial DCT matrix under Gaussian noise}
  \begin{tabular}{|c|c|c|c|c|c|c|c|c|c|}
        \hline
       $M$ & $N$ & $ {s_{true}}$  & ${\ell _1}$-ADMM & ${\ell _{1-2}}$-DCA & \tabincell{c}{Half \\thresholding} & AIHT & \tabincell{c}{$s$-difference  \\ (${\ell _1}$)} &  \tabincell{c}{$s$-difference  \\ (${\ell _{1-2}}$)} & \tabincell{c}{$s$-difference  \\ (${\ell _2}$)}   \\ \hline
       256 & 1024 &48&  \tabincell{c}{7.485e-02\\(1.777e-04)} & \tabincell{c}{6.372e-02\\(1.017e-04)} & \tabincell{c}{4.190e-02\\(3.338e-05)} & \tabincell{c}{1.834e-01\\(1.216e-03)} & \tabincell{c}{4.264e-02\\(2.563e-04))}  &\tabincell{c}{\textbf{3.192e-02}\\(4.218e-05)}
       &\tabincell{c}{3.306e-02\\(3.825e-05)} \\ \hline
       512 & 2048 & 96 &  \tabincell{c}{7.503e-02\\(4.682e-05)} & \tabincell{c}{6.744e-02 \\(4.305e-05)} &\tabincell{c}{5.313e-02\\(3.419e-05)} &\tabincell{c}{1.791e-01 \\(2.688e-04)} &\tabincell{c}{4.211e-02\\(1.904e-04)} &\tabincell{c}{3.180e-02\\(3.433e-05)} &\tabincell{c}{ \textbf{3.170e-02}\\(3.092e-05)}\\ \hline
       768& 3072& 144& \tabincell{c}{7.513e-02\\(3.845e-05)} & \tabincell{c}{6.852e-02\\(2.634e-05)} & \tabincell{c}{6.195e-02\\(2.829e-05)} &\tabincell{c}{1.813e-01\\(1.902e-04)} & \tabincell{c}{4.448e-02\\(7.818e-05)} & \tabincell{c}{ \textbf{3.035e-02}\\(7.943e-06))} &\tabincell{c}{3.052e-02\\(9.563e-06)} \\ \hline
       1024 & 4096&192& \tabincell{c}{7.512e-02\\(3.957e-05)} &\tabincell{c}{6.951e-02\\(2.775e-05)} &\tabincell{c}{6.800e-02\\(2.794e-05)} &\tabincell{c}{1.796e-01\\(1.075e-04)} &\tabincell{c}{4.369e-02\\(9.062e-05)} &\tabincell{c}{\textbf{3.031e-02}\\(8.936e-06)} &\tabincell{c}{3.043e-02\\(9.242e-06)} \\ \hline
    \end{tabular}
\end{table*}

\begin{table*}
\newcommand{\tabincell}[2]{\begin{tabular}{@{}#1@{}}#2\end{tabular}}
  \centering
  \caption{\label{table_6}Mean of relative error, iteration number and computational time (sec.) under the noise-free and Gaussian noise conditions}
  \begin{tabular}{l|c c c c|c c c c}
        \hline
         \multirow{2}{*}{\tabincell{c}{Methods\\}} &  \multicolumn{2}{c}{\tabincell{c}{Noiseless condition\\Gaussian matrix}} & \multicolumn{2}{c}{\tabincell{c}{Noiseless condition\\partial DCT matrix}} & \multicolumn{2}{|c}{\tabincell{c}{Noisy  condition \\Gaussian matrix}}& \multicolumn{2}{c}{\tabincell{c}{Noisy  condition \\partial DCT matrix}} \\ \cline{2-9}
         & Rel.Err & Iter/Time&Rel.Err & Iter/Time&Rel.Err & Iter/Time&Rel.Err & Iter/Time
           \\ \hline
         ${\ell _1}$-ADMM &1.098E-04& &1.357E-04& &1.198E-01& & 7.485E-02 &   \\ \hline
         ${\left\| {\bf{x}} \right\|_1} - {\left\| {{{\bf{x}}^s}} \right\|_1}$ (DCA-ADMM) & 2.298E-05& 178/0.05&2.501E-05&170/0.05&7.182E-02&302/0.08&4.430E-02&511/0.12\\ \hline
         ${\left\| {\bf{x}} \right\|_1} - {\left\| {{{\bf{x}}^s}} \right\|_1}$ (PDCA) &3.735E-05&530/0.13&4.063E-05&460/0.12&1.179E-01&5120/1.46&1.005E-01&3559/1.08
         \\ \hline
        ${\left\| {\bf{x}} \right\|_1} - {\left\| {{{\bf{x}}^s}} \right\|_1}$ (FBS)&1.368E-05&126/0.04&3.059E-06&65/0.03&6.190E-02&195/0.06&4.264E-02&108/0.05
         \\ \hline
    \end{tabular}
\end{table*}

\section{Conclusion}
In this paper, we propose a new $s$-difference type penalty function for the sparse optimization problem, which is the difference of the normal convex or nonconvex penalty function and its corresponding $s$-truncated function. To solve this nonconvex regularization problem, we select the FBS method based on the proximal operator, which have some cheap closed-form solutions for commonly used $R\left( {\bf{x}} \right)$, such as ${\ell _1}$, ${\ell _2}$, ${\ell _{1 - 2}}$ and so on. The convergence and effectiveness of the proposed algorithm are proved and demonstrated by the theoretical proof and numerical experiments, respectively. In addition, we observed that the DCA with $s$-difference regularization gives better recovery results than the FBS using close-form solutions when the parameter $s$ is less than the true sparsity, which motivate us to find an adaptive strategy for the penalty and sparsity parameters in the future.

\appendices
\section{Proof of Proposition 1}
To prove the Proposition 1, we use the following Lemma:
\newtheorem{Lemma}{Lemma}
\begin{Lemma}
If $R:{{\mathbb{R}}^{N}}\to \mathbb{R}$ is convex, then for any $s\in \left\{ 1,2,\cdots ,N \right\}$, $R\left( {{\mathbf{x}}^{s}} \right)$ is also convex.
\end{Lemma}
\begin{IEEEproof}
 let $\mathbf{v}=diag\left\{ {{v}_{1}},{{v}_{2}},\cdots ,{{v}_{N}} \right\}$, since $R\left( \mathbf{x} \right)$ is convex, then $R\left( \mathbf{vx} \right)$ is convex. Then the $R\left( {{\mathbf{x}}^{s}} \right)$ can be written as a pointwise maximum of convex functions:
 \renewcommand{\theequation}{A.1}
 \begin{equation}
\label{eqn_A.1}
R\left( {{{\bf{x}}^s}} \right) = \mathop {\max }\limits_{\bf{v}} \left\{ {R\left( {{\bf{vx}}} \right):{v_i} \in \left\{ {0,1} \right\},{{\left\| {\bf{v}} \right\|}_1} = s} \right\}
\end{equation}
Then we have that $R\left( {{\mathbf{x}}^{s}} \right)$ is convex.

1) For the convex and separable $R(\mathbf{x})=\left\| \mathbf{x} \right\|_{p}^{p} \left( p\ge 1 \right)$, such as ${{\left\| \mathbf{x} \right\|}_{1}}$ and $\left\| \mathbf{x} \right\|_{2}^{2}$, and the convex and non-separable functions $R(\mathbf{x})={{\left\| \mathbf{x} \right\|}_{p}},\left( p>1 \right)$, such as $R(\mathbf{x})={{\left\| \mathbf{x} \right\|}_{2}}$, it is obviously that they fulfilling (a) and (b). Then by using Lemma 1, it completes the Property 1(c).

2) For the non-convex and separable functions $R\left( \mathbf{x} \right)=\sum\limits_{i=1}^{N}{{{r}_{i}}\left( {{x}_{i}} \right)}$, where ${{r}_{i}}\left( {{x}_{i}} \right)$ are equations (A.2), (A.3) and (A.4) corresponding to LSP, MCP and SCAD, respectively.
\renewcommand{\theequation}{A.2}
 \begin{equation}
\label{eqn_A.2}
{r_i}\left( {{x_i}} \right) = \log \left( {1 + {{\left| {{x_i}} \right|} \mathord{\left/
 {\vphantom {{\left| {{x_i}} \right|} \theta }} \right.
 \kern-\nulldelimiterspace} \theta }} \right),\theta  > 0
\end{equation}

\renewcommand{\theequation}{A.3}
 \begin{equation}
\label{eqn_A.3}
{r_i}\left( {{x_i}} \right) = \left\{ {\begin{array}{*{20}{c}}
{\left| {{x_i}} \right| - {{x_i^2} \mathord{\left/
 {\vphantom {{x_i^2} {\left( {2\theta } \right)}}} \right.
 \kern-\nulldelimiterspace} {\left( {2\theta } \right)}},}&{\left| {{x_i}} \right| \le \theta }\\
{{\theta  \mathord{\left/
 {\vphantom {\theta  2}} \right.
 \kern-\nulldelimiterspace} 2},}&{\left| {{x_i}} \right| > \theta }
\end{array}} \right.,\theta  > 0
\end{equation}

\renewcommand{\theequation}{A.4}
 \begin{equation}
\label{eqn_A.4}
{r_i}\left( {{x_i}} \right) = \left\{ {\begin{array}{*{20}{c}}
{\left| {{x_i}} \right|,}&{\left| {{x_i}} \right| < 1}\\
{\frac{{2\theta \left| {{x_i}} \right| - x_i^2 - 1}}{{2\left( {\theta  - 1} \right)}},}&{1 \le \left| {{x_i}} \right| < \theta }\\
{{{\left( {\theta  + 1} \right)} \mathord{\left/
 {\vphantom {{\left( {\theta  + 1} \right)} 2}} \right.
 \kern-\nulldelimiterspace} 2},}&{\left| {{x_i}} \right| \ge \theta }
\end{array}} \right.,\theta  > 2
\end{equation}

Property 1(a) and (b) is obvious. Then we need give the DC formulations for $P\left( \mathbf{x} \right)$. Take the LSP for example, we have that

\renewcommand{\theequation}{A.5}
 \begin{equation}
\label{eqn_A.5}
\begin{array}{l}
\frac{{{{\left\| {{{\bf{x}}^s}} \right\|}_1}}}{\theta } - R\left( {{{\bf{x}}^s}} \right) = \\
\mathop {\max }\limits_{\bf{v}} \left\{ {\sum\limits_{i = 1}^N {\frac{{\left| {{v_i}{x_i}} \right|}}{\theta } - \log \left( {1 + \frac{{\left| {{v_i}{x_i}} \right|}}{\theta }} \right)} :{v_i} \in \left\{ {0,1} \right\},{{\left\| {\bf{v}} \right\|}_1} = s} \right\}
\end{array}
\end{equation}

which means that ${{{\left\| {{\mathbf{x}}^{s}} \right\|}_{1}}}/{\theta }\;-R\left( {{\mathbf{x}}^{s}} \right)$ is convex as ${\left| {{v}_{i}}{{x}_{i}} \right|}/{\theta }\;-\log \left( 1+{\left| {{v}_{i}}{{x}_{i}} \right|}/{\theta }\; \right)$ is convex. Then we can rewrite $P\left( \mathbf{x} \right)$ as

\renewcommand{\theequation}{A.6}
 \begin{equation}
\label{eqn_A.6}
\begin{aligned}
P\left( {\bf{x}} \right) &= R\left( {\bf{x}} \right) - R\left( {{{\bf{x}}^s}} \right)\\
 &= \underbrace {\left\{ {{{{{\left\| {\bf{x}} \right\|}_1}} \mathord{\left/
 {\vphantom {{{{\left\| {\bf{x}} \right\|}_1}} \theta }} \right.
 \kern-\nulldelimiterspace} \theta } + \left( {{{{{\left\| {{{\bf{x}}^s}} \right\|}_1}} \mathord{\left/
 {\vphantom {{{{\left\| {{{\bf{x}}^s}} \right\|}_1}} \theta }} \right.
 \kern-\nulldelimiterspace} \theta } - R\left( {{{\bf{x}}^s}} \right)} \right)} \right\}}_{{P_1}\left( {\bf{x}} \right)}\\
 & \quad  - \underbrace {\left\{ {{{{{\left\| {{{\bf{x}}^s}} \right\|}_1}} \mathord{\left/
 {\vphantom {{{{\left\| {{{\bf{x}}^s}} \right\|}_1}} \theta }} \right.
 \kern-\nulldelimiterspace} \theta } + \left( {{{{{\left\| {\bf{x}} \right\|}_1}} \mathord{\left/
 {\vphantom {{{{\left\| {\bf{x}} \right\|}_1}} \theta }} \right.
 \kern-\nulldelimiterspace} \theta } - R\left( {\bf{x}} \right)} \right)} \right\}}_{{P_2}\left( {\bf{x}} \right)}
\end{aligned}
\end{equation}
where ${{P}_{1}}\left( \mathbf{x} \right)$ and ${{P}_{2}}\left( \mathbf{x} \right)$ are two convex functions. For MCP and SCAD, we can obtain similar formulations in the same way.

3) For the non-convex and non-separable functions, when $R\left( \mathbf{x} \right)={{\left\| \mathbf{x} \right\|}_{1}}-a{{\left\| \mathbf{x} \right\|}_{2}},0<a\le 1$, we have $R\left( \mathbf{x} \right)=R\left( -\mathbf{x} \right)$. When ${{\left\| \mathbf{x} \right\|}_{0}}\le s$, it is easy to see that $P\left( \mathbf{x} \right)=0$. When $P\left( \mathbf{x} \right)=0$, we have ${{\left\| \mathbf{x} \right\|}_{0}}\le s$; otherwise ${{\left\| \mathbf{x} \right\|}_{0}}>s$, then $\left\| \mathbf{x} \right\|_{2}^{2}\le \left\| {{\mathbf{x}}^{s}} \right\|_{2}^{2}+{{\left( {{\left\| \mathbf{x} \right\|}_{1}}-{{\left\| {{\mathbf{x}}^{s}} \right\|}_{1}} \right)}^{2}}<{{\left( \left\| {{\mathbf{x}}^{s}} \right\|_{2}^{{}}+{{\left\| \mathbf{x} \right\|}_{1}}-{{\left\| {{\mathbf{x}}^{s}} \right\|}_{1}} \right)}^{2}}$, then we have ${{\left\| \mathbf{x} \right\|}_{2}}-{{\left\| {{\mathbf{x}}^{s}} \right\|}_{2}}<{{\left\| \mathbf{x} \right\|}_{1}}-{{\left\| {{\mathbf{x}}^{s}} \right\|}_{1}}$, which means that $P\left( \mathbf{x} \right)=R\left( \mathbf{x} \right)-R\left( {{\mathbf{x}}^{s}} \right)={{\left\| \mathbf{x} \right\|}_{1}}-{{\left\| {{\mathbf{x}}^{s}} \right\|}_{1}}-a\left( {{\left\| \mathbf{x} \right\|}_{2}}-{{\left\| {{\mathbf{x}}^{s}} \right\|}_{2}} \right)>0$, and this is contradiction to $P\left( \mathbf{x} \right)=0$. Meanwhile, $P\left( \mathbf{x} \right)$ can be formulated as

\renewcommand{\theequation}{A.7}
 \begin{equation}
\label{eqn_A.7}
\begin{aligned}
P\left( {\bf{x}} \right) &= R\left( {\bf{x}} \right) - R\left( {{{\bf{x}}^s}} \right)\\
 &= \underbrace {\left\{ {{{\left\| {\bf{x}} \right\|}_1} + a{{\left\| {{{\bf{x}}^s}} \right\|}_2}} \right\}}_{{P_1}\left( {\bf{x}} \right)} - \underbrace {\left\{ {{{\left\| {{{\bf{x}}^s}} \right\|}_1} + a{{\left\| {\bf{x}} \right\|}_2}} \right\}}_{{P_2}\left( {\bf{x}} \right)}
\end{aligned}
\end{equation}

when $R\left( \mathbf{x} \right)$ is the non-separable LSP, denoted as $R\left( \mathbf{x} \right)=\log \left( 1+{{{\left\| \mathbf{x} \right\|}_{2}}}/{\theta }\; \right),\theta >0$, Property 1(a) and (b) are obvious. Note that ${{{\left\| {{\mathbf{x}}^{s}} \right\|}_{2}}}/{\theta }\;-R\left( {{\mathbf{x}}^{s}} \right)$ can be thought as a composition function $h\circ g$, where $h\left( x \right)={\left| x \right|}/{\theta }\;-\log \left( 1+{\left| x \right|}/{\theta }\; \right)$ and $g\left( \mathbf{x} \right)={{\left\| {{\mathbf{x}}^{s}} \right\|}_{2}}$, by using the above deduction, we have that ${{{\left\| {{\mathbf{x}}^{s}} \right\|}_{2}}}/{\theta }\;-R\left( {{\mathbf{x}}^{s}} \right)$ is convex. Then $P\left( \mathbf{x} \right)$ can be rewritten as

\renewcommand{\theequation}{A.8}
 \begin{equation}
\label{eqn_A.8}
\begin{aligned}
P\left( {\bf{x}} \right) &= R\left( {\bf{x}} \right) - R\left( {{{\bf{x}}^s}} \right)\\
 &= \underbrace {\left\{ {{{{{\left\| {\bf{x}} \right\|}_2}} \mathord{\left/
 {\vphantom {{{{\left\| {\bf{x}} \right\|}_2}} \theta }} \right.
 \kern-\nulldelimiterspace} \theta } + \left( {{{{{\left\| {{{\bf{x}}^s}} \right\|}_2}} \mathord{\left/
 {\vphantom {{{{\left\| {{{\bf{x}}^s}} \right\|}_2}} \theta }} \right.
 \kern-\nulldelimiterspace} \theta } - R\left( {{{\bf{x}}^s}} \right)} \right)} \right\}}_{{P_1}\left( {\bf{x}} \right)}\\
 & \quad- \underbrace {\left\{ {{{{{\left\| {{{\bf{x}}^s}} \right\|}_2}} \mathord{\left/
 {\vphantom {{{{\left\| {{{\bf{x}}^s}} \right\|}_2}} \theta }} \right.
 \kern-\nulldelimiterspace} \theta } + \left( {{{{{\left\| {\bf{x}} \right\|}_2}} \mathord{\left/
 {\vphantom {{{{\left\| {\bf{x}} \right\|}_2}} \theta }} \right.
 \kern-\nulldelimiterspace} \theta } - R\left( {\bf{x}} \right)} \right)} \right\}}_{{P_2}\left( {\bf{x}} \right)}
\end{aligned}
\end{equation}
For the non-separable type MCP and SCAD, we can obtain similar formulations in the same way.
\end{IEEEproof}

\section{Proof of Theorem  1}
\begin{IEEEproof}
This theorem can be proved in a similar manner to Theorem 17.1 in [71].
Let $\mathbf{\hat{x}}$ be an optimal solution of (\ref{eqn_3}), that is,
\renewcommand{\theequation}{A.9}
\label{eqn_A.9}
 \begin{equation}
\phi \left( {{\bf{\hat x}}} \right) \le \phi \left( {\bf{x}} \right) \quad  \rm{for \ all} \quad {\bf{x}} \quad \rm{with} \quad {\left\| {\bf{x}} \right\|_0} \le s
\end{equation}
Since ${{\mathbf{x}}_{t}}$ minimizes (\ref{eqn_8}) at $\rho ={{\rho }_{t}}$, we have that
\renewcommand{\theequation}{A.10}
\label{eqn_A.10}
 \begin{equation}
\phi \left( {{{\bf{x}}_t}} \right) + {\rho _t}P\left( {{{\bf{x}}_t}} \right) \le \phi \left( {{\bf{\hat x}}} \right) + {\rho _t}P\left( {{\bf{\hat x}}} \right) = \phi \left( {{\bf{\hat x}}} \right)
\end{equation}
By rearranging this expression, we have
\renewcommand{\theequation}{A.11}
\label{eqn_A.11}
 \begin{equation}
R\left( {{{\bf{x}}_t}} \right) - R\left( {{\bf{x}}_t^s} \right) \le \frac{1}{{{\rho _t}}}\left( {\phi \left( {{\bf{\hat x}}} \right) - \phi \left( {{{\bf{x}}_t}} \right)} \right)
\end{equation}
Suppose that $\mathbf{\bar{x}}$ is a limit point of $\left\{ {{\mathbf{x}}_{t}} \right\}$, then there exits an infinite subsequence $\mathcal{T}$ such that ${{\lim }_{t\in \mathcal{T}}}{{\mathbf{x}}_{t}}=\mathbf{\bar{x}}$. By taking the limit as $t\to \infty $, $t\in \mathcal{T}$, on both side of (A.11)
\renewcommand{\theequation}{A.12}
\label{eqn_A.12}
 \begin{equation}
0 \le R\left( {{\bf{\bar x}}} \right) - R\left( {{{{\bf{\bar x}}}^s}} \right) \le \mathop {\lim }\limits_{t \in {\cal T}} \frac{1}{{{\rho _t}}}\left( {\phi \left( {{\bf{\hat x}}} \right) - \phi \left( {{{\bf{x}}_t}} \right)} \right) = 0
\end{equation}
Therefore, we have that $R\left( {\mathbf{\bar{x}}} \right)-R\left( {{{\mathbf{\bar{x}}}}^{s}} \right)=0$, which means that $\mathbf{\bar{x}}$ is feasible to (3). Moreover, by taking the limit as $t\to \infty $ for $t\in \mathcal{T}$ on (A.10), we have that
\renewcommand{\theequation}{A.13}
\label{eqn_A.13}
 \begin{equation}
\phi \left( {{\bf{\bar x}}} \right) \le \phi \left( {{\bf{\bar x}}} \right) + \mathop {\lim }\limits_{t \in {\cal T}} {\rho _t}P\left( {{{\bf{x}}_t}} \right) \le \phi \left( {{\bf{\hat x}}} \right)
\end{equation}
Since $\mathbf{\bar{x}}$ is feasible to (\ref{eqn_3}) and $\mathbf{\hat{x}}$ is an optimal solution of (\ref{eqn_3}), then $\mathbf{\bar{x}}$ is also optimal to (\ref{eqn_3}).
\end{IEEEproof}

\section{Proof of Proposition 2}
\begin{IEEEproof}
For simplicity, we use $\mathbf{\bar{x}}$ instead of ${{\mathbf{\bar{x}}}_{\rho }}$ for an optimal solution of (8) with some $\rho $. First, we proof that ${{\left\| {\mathbf{\bar{x}}} \right\|}_{0}}\le s$. If ${{\left\| {\mathbf{\bar{x}}} \right\|}_{0}}>s$, which means that ${{\left\| {{{\mathbf{\bar{x}}}}^{(s+1)}}-{{{\mathbf{\bar{x}}}}^{s}} \right\|}_{2}}>0$. We construct a vector $\mathbf{\tilde{x}}$ as $\mathbf{\tilde{x}}=\mathbf{\bar{x}}+{{\mathbf{\bar{x}}}^{s}}-{{\mathbf{\bar{x}}}^{\left( s+1 \right)}}$, easily we have that ${{\mathbf{\tilde{x}}}^{s}}={{\mathbf{\bar{x}}}^{s}}$. When $\rho >{\beta }/{\eta }$, we have
\renewcommand{\theequation}{A.14}
\label{eqn_A.14}
 \begin{equation}
 \begin{aligned}
&F\left( {{\bf{\bar x}}} \right) - F\left( {{\bf{\tilde x}}} \right) \\
&= \phi \left( {{\bf{\bar x}}} \right) + \rho \left( {R\left( {{\bf{\bar x}}} \right) - R\left( {{{{\bf{\bar x}}}^s}} \right)} \right) - \phi \left( {{\bf{\tilde x}}} \right) - \rho \left( {R\left( {{\bf{\tilde x}}} \right) - R\left( {{{{\bf{\tilde x}}}^s}} \right)} \right)\\
 &= \phi \left( {{\bf{\bar x}}} \right) - \phi \left( {{\bf{\tilde x}}} \right) + \rho \left( {R\left( {{\bf{\bar x}}} \right) - R\left( {{\bf{\tilde x}}} \right)} \right)\\
& \ge  - \beta {\left\| {{\bf{\bar x}} - {\bf{\tilde x}}} \right\|_2} + \rho \eta {\left\| {{\bf{\bar x}} - {\bf{\tilde x}}} \right\|_2}\\
& = \left( {\rho \eta  - \beta } \right){\left\| {{{{\bf{\bar x}}}^{(s + 1)}} - {{{\bf{\bar x}}}^s}} \right\|_2} > 0
\end{aligned}
\end{equation}
This contradicts the optimality of $\mathbf{\bar{x}}$. Then we have that ${{\left\| {\mathbf{\bar{x}}} \right\|}_{0}}$ satisfies the $s$-sparse constraint of (\ref{eqn_3}). Let $\mathbf{\hat{x}}$ be an optimal solution of (\ref{eqn_3}), then we have
\renewcommand{\theequation}{A.15}
\label{eqn_A.15}
 \begin{equation}
 \begin{aligned}
\phi \left( {{\bf{\bar x}}} \right) - \phi \left( {{\bf{\hat x}}} \right) &= F\left( {{\bf{\bar x}}} \right) - \rho P\left( {{\bf{\bar x}}} \right) - F\left( {{\bf{\hat x}}} \right) + \rho P\left( {{\bf{\hat x}}} \right)\\
 &= F\left( {{\bf{\bar x}}} \right) - F\left( {{\bf{\hat x}}} \right) \le 0
\end{aligned}
\end{equation}
The inequality comes from that $\mathbf{\bar{x}}$ is the optimal solution of (\ref{eqn_8}). This means that $\mathbf{\bar{x}}$ is also optimal to (\ref{eqn_3}).
\end{IEEEproof}

\section{Proof of Proposition 3}
\begin{IEEEproof}
Similar to the previous proof of Proposition 2, we use $\mathbf{\bar{x}}$ instead of ${{\mathbf{\bar{x}}}_{\rho }}$ for an optimal solution of (\ref{eqn_8}) with some $\rho $. Assume by contradiction that ${{\left\| {\mathbf{\bar{x}}} \right\|}_{0}}>s$, which means that ${{\left\| {{\mathbf{x}}^{s+1}}-{{\mathbf{x}}^{s}} \right\|}_{2}}>0$. By constructing $\mathbf{\tilde{x}}=\mathbf{\bar{x}}+{{\mathbf{\bar{x}}}^{s}}-{{\mathbf{\bar{x}}}^{\left( s+1 \right)}}$, we have

\renewcommand{\theequation}{A.16}
\label{eqn_A.16}
 \begin{equation}
 \begin{aligned}
&F\left( {{\bf{\bar x}}} \right) - F\left( {{\bf{\tilde x}}} \right)\\
& = \phi \left( {{\bf{\bar x}}} \right) + \rho \left( {R\left( {{\bf{\bar x}}} \right) - R\left( {{{{\bf{\bar x}}}^s}} \right)} \right) - \phi \left( {{\bf{\tilde x}}} \right) - \rho \left( {R\left( {{\bf{\tilde x}}} \right) - R\left( {{{{\bf{\tilde x}}}^s}} \right)} \right)\\
 &= \phi \left( {{\bf{\bar x}}} \right) - \phi \left( {{\bf{\tilde x}}} \right) + \rho \left( {R\left( {{\bf{\bar x}}} \right) - R\left( {{\bf{\tilde x}}} \right)} \right)\\
& \ge \left\langle {\nabla \phi \left( {{\bf{\bar x}}} \right),{{{\bf{\bar x}}}^{(s + 1)}} - {{{\bf{\bar x}}}^s}} \right\rangle  - \frac{L}{2}\left\| {{{{\bf{\bar x}}}^{(s + 1)}} - {{{\bf{\bar x}}}^s}} \right\|_2^2\\
&\quad + \rho \eta {\left\| {{{{\bf{\bar x}}}^{(s + 1)}} - {{{\bf{\bar x}}}^s}} \right\|_2}\\
& \ge {\left\| {{{{\bf{\bar x}}}^{(s + 1)}} - {{{\bf{\bar x}}}^s}} \right\|_2}\left( {\rho \eta  - {{\left\| {\nabla \phi \left( {{\bf{\bar x}}} \right)} \right\|}_2} - \frac{{LC}}{{2\sqrt {s + 1} }}} \right)\\
& \ge {\left\| {{{{\bf{\bar x}}}^{(s + 1)}} - {{{\bf{\bar x}}}^s}} \right\|_2}\left( {\rho \eta  - {{\left\| {\nabla \phi \left( {\bf{0}} \right)} \right\|}_2} - \left( {1 + \frac{1}{{2\sqrt {s + 1} }}} \right)LC} \right)\\
& > 0
\end{aligned}
\end{equation}
The first inequality using Assumption 1 that
\renewcommand{\theequation}{A.17}
\label{eqn_A.17}
 \begin{equation}
 \phi \left( {\bf{y}} \right) \le \phi \left( {\bf{x}} \right) + \left\langle {\nabla \phi \left( {\bf{x}} \right),{\bf{y}} - {\bf{x}}} \right\rangle  + \frac{L}{2}\left\| {{\bf{y}} - {\bf{x}}} \right\|_2^2,\forall {\bf{x}},{\bf{y}} \in {\mathbb{R}^N}
 \end{equation}
 The third inequality follows from that
 \renewcommand{\theequation}{A.18}
\label{eqn_A.18}
 \begin{equation}
\begin{aligned}
{\left\| {\nabla \phi \left( {{\bf{\bar x}}} \right)} \right\|_2} &\le {\left\| {\nabla \phi \left( {\bf{0}} \right)} \right\|_2} + {\left\| {\nabla \phi \left( {{\bf{\bar x}}} \right) - \nabla \phi \left( {\bf{0}} \right)} \right\|_2}\\
& \le {\left\| {\nabla \phi \left( {\bf{0}} \right)} \right\|_2} + LC
\end{aligned}
 \end{equation}
 (A.16)) contradicts the optimality of $\mathbf{\bar{x}}$, then we have that ${{\left\| {\mathbf{\bar{x}}} \right\|}_{0}}$ satisfies the $s$-sparse constraint of (\ref{eqn_3}). Then we can prove that $\mathbf{\bar{x}}$ is also optimal to (\ref{eqn_3}) similar as the previous proof of Proposition 2.
\end{IEEEproof}

\section{Proof of Proposition 6}
\begin{IEEEproof}
Suppose that ${{\mathbf{x}}^{*}}$ is the optimal solution of (\ref{eqn_12}). First, we prove that if $\left| {{\text{y}}_{i}} \right|>\left| {{\text{y}}_{j}} \right|$.we have $\left| x_{i}^{*} \right|\ge \left| x_{j}^{*} \right|$; otherwise $\left| x_{i}^{*} \right|<\left| x_{j}^{*} \right|$, then we construct $\mathbf{\tilde{x}}\in {{\mathbb{R}}^{N}}$ as $\tilde{x}_{i}^{*}=\text{sign}\left( {{y}_{i}} \right)\left| x_{j}^{*} \right|$ and $\tilde{x}_{j}^{*}=\text{sign}\left( {{y}_{j}} \right)\left| x_{i}^{*} \right|$. Whether $i,j \in \Gamma _{\bf{y}}^s$ or $i,j \notin \Gamma _{\bf{y}}^s$ or $i \in \Gamma _{\bf{y}}^s,j \notin \Gamma _{\bf{y}}^s$, we always have that $R\left( {\mathbf{\tilde{x}}} \right)=R\left( {{\mathbf{x}}^{*}} \right)$ and $R\left( {{{\mathbf{\tilde{x}}}}^{s}} \right)=R\left( {{\mathbf{x}}^{*s}} \right)$. As $\left\| \mathbf{\tilde{x}}-\mathbf{y} \right\|_{2}^{2}<\left\| {{\mathbf{x}}^{*}}-\mathbf{y} \right\|_{2}^{2}$, then we can obtain $E\left( {\mathbf{\tilde{x}}} \right)<E\left( {{\mathbf{x}}^{*}} \right)$. However, this contradicts the optimality of ${{\mathbf{x}}^{*}}$.

Next, we prove that $\left| x_{{{\pi }_{y}}\left( s+1 \right)}^{*} \right|\le \left| {{y}_{{{\pi }_{y}}\left( s \right)}} \right|$. To prove this, we need to prove that $\left| x_{{{\pi }_{y}}\left( j \right)}^{*} \right|\le \left| {{y}_{{{\pi }_{y}}\left( s \right)}} \right|$ for all $j\in \left\{ s+1,s+2,\cdots ,N \right\}$. We can do this one by one, i.e., we look at $x_{{{\pi }_{y}}\left( N \right)}^{*}$ first. Easily, we have $\left| x_{{{\pi }_{y}}\left( N \right)}^{*} \right|\le \left| {{y}_{{{\pi }_{y}}\left( s \right)}} \right|$; otherwise we construct ${{\tilde{x}}_{{{\pi }_{y}}\left( N \right)}}=\text{sign}\left( {{y}_{{{\pi }_{y}}\left( N \right)}} \right)\left| {{y}_{{{\pi }_{y}}\left( s \right)}} \right|$, as ${{r}_{i}}$ is strictly increasing on ${{\mathbb{R}}_{+}}$ and symmetrical, thus we have the contradiction $E\left( {\mathbf{\tilde{x}}} \right)<E\left( {{\mathbf{x}}^{*}} \right)$, then we have $\left| x_{{{\pi }_{y}}\left( N \right)}^{*} \right|\le \left| {{y}_{{{\pi }_{y}}\left( s \right)}} \right|$. By using this deduction, we can prove that $\left| x_{{{\pi }_{y}}\left( N-1 \right)}^{*} \right|\le \left| {{y}_{{{\pi }_{y}}\left( s \right)}} \right|$ in a similar way. At last, we have $\left| x_{{{\pi }_{y}}\left( s+1 \right)}^{*} \right|\le \left| {{y}_{{{\pi }_{y}}\left( s \right)}} \right|$.

Rewrite $E\left( \mathbf{x} \right)$ as
 \renewcommand{\theequation}{A.19}
\label{eqn_A.19}
 \begin{equation}
\begin{aligned}
E\left( {\bf{x}} \right) = &\sum\limits_{j = 1}^s {\frac{1}{{2\lambda }}{{\left( {{x_{{\pi _y}\left( j \right)}} - {y_{{\pi _y}\left( j \right)}}} \right)}^2}} \\
 &+ \sum\limits_{j = s + 1}^N {\left( {\frac{1}{{2\lambda }}{{\left( {{x_{{\pi _y}\left( j \right)}} - {y_{{\pi _y}\left( j \right)}}} \right)}^2} + {r_{{\pi _y}\left( j \right)}}\left( {{x_{{\pi _y}\left( j \right)}}} \right)} \right)}
\end{aligned}
 \end{equation}

 As $\left| x_{{{\pi }_{y}}\left( s+1 \right)}^{*} \right|\le \left| {{y}_{{{\pi }_{y}}\left( s \right)}} \right|$, we have that $x_{{{\pi }_{y}}\left( j \right)}^{*}={{y}_{{{\pi }_{y}}\left( j \right)}},j=1,2,\cdots ,s$ and $x_{{{\pi }_{y}}\left( j \right)}^{*}={{\left( 1+\lambda \partial {{r}_{{{\pi }_{y}}\left( j \right)}} \right)}^{-1}}\left( {{y}_{{{\pi }_{y}}\left( j \right)}} \right),j=s+1,s+2,\cdots ,N$. This completes the proof.
\end{IEEEproof}

\section{Proof of Proposition 7}
\begin{IEEEproof}
First, we show that when $R\left( \mathbf{x} \right)={{\left\| \mathbf{x} \right\|}_{2}}$, we also have if $\left| {{\text{y}}_{i}} \right|>\left| {{\text{y}}_{j}} \right|$.we have $\left| x_{i}^{*} \right|\ge \left| x_{j}^{*} \right|$. Otherwise, we can always construct a $\mathbf{\tilde{x}}\in {{\mathbb{R}}^{N}}$, which swap the absolute value of $x_{i}^{*}$ and $x_{j}^{*}$ as the same way in the proof of Proposition 6, then we can obtain a smaller objective value. As proved in Proposition 4, ${{\mathbf{x}}^{*}}=\mathbf{0}$ if and only if $\mathbf{y}=\mathbf{0}$. Then, we only need to consider the case $\mathbf{y}\ne \mathbf{0}$.

1) If $\left| {{y}_{{{\pi }_{y}}\left( s \right)}} \right|\ne \left| {{y}_{{{\pi }_{y}}\left( s+1 \right)}} \right|$, then we have
 \renewcommand{\theequation}{A.20}
\label{eqn_A.20}
 \begin{equation}
\begin{aligned}
\left\{ {{\pi _x}\left( 1 \right),{\pi _x}\left( 2 \right), \cdots ,{\pi _x}\left( s \right)} \right\} = \left\{ {{\pi _y}\left( 1 \right),{\pi _y}\left( 2 \right), \cdots ,{\pi _y}\left( s \right)} \right\}
\end{aligned}
 \end{equation}
Easily, we have that if ${{y}_{{{\pi }_{y}}\left( s+1 \right)}}=0$, then ${{\mathbf{x}}^{*}}=\mathbf{y}$ and $E\left( {{\mathbf{x}}^{*}} \right)=0$.

When ${{y}_{{{\pi }_{y}}\left( s+1 \right)}}\ne 0$, the first-order optimality condition optimality condition of minimizing $E\left( \mathbf{x} \right)$ is that
\renewcommand{\theequation}{A.21}
\label{eqn_A.21}
 \begin{equation}
\begin{aligned}
\left\{ {\begin{array}{*{20}{c}}
{\left( {1 + \frac{\lambda }{{{{\left\| {\bf{x}} \right\|}_2}}} - \frac{\lambda }{{{{\left\| {{{\bf{x}}^s}} \right\|}_2}}}} \right){x_i} = {y_i},}&{i \in \Gamma _{\bf{y}}^s}\\
{\left( {1 + \frac{\lambda }{{{{\left\| {\bf{x}} \right\|}_2}}}} \right){x_i} = {y_i},}&{i \in \Gamma _{\bf{y}}^N\backslash \Gamma _{\bf{y}}^s}
\end{array}} \right.
\end{aligned}
 \end{equation}
By using Proposition 5, we have that $1+\frac{\lambda }{{{\left\| \mathbf{x} \right\|}_{2}}}-\frac{\lambda }{{{\left\| {{\mathbf{x}}^{s}} \right\|}_{2}}}\ge 0$ in (A.21). Using (A.21), we have
\renewcommand{\theequation}{A.22}
\label{eqn_A.22}
 \begin{equation}
\begin{aligned}
\left\{ {\begin{array}{*{20}{c}}
{\left( {1 + \frac{\lambda }{{{{\left\| {\bf{x}} \right\|}_2}}}} \right){{\left\| {{{\bf{x}}^s}} \right\|}_2} = {{\left\| {{{\bf{y}}^s}} \right\|}_2} + \lambda }\\
{{{\left\| {\bf{x}} \right\|}_2} = \frac{{\lambda {{\left\| {{\bf{x}} - {{\bf{x}}^s}} \right\|}_2}}}{{{{\left\| {{\bf{y}} - {{\bf{y}}^s}} \right\|}_2} - {{\left\| {{\bf{x}} - {{\bf{x}}^s}} \right\|}_2}}}}
\end{array}} \right.
\end{aligned}
 \end{equation}
Substitute one equation of (A.22) into another, we have
\renewcommand{\theequation}{A.23}
\label{eqn_A.23}
 \begin{equation}
\begin{aligned}
{\left\| {{{\bf{x}}^s}} \right\|_2} = \frac{{{{\left\| {{{\bf{y}}^s}} \right\|}_2} + \lambda }}{{{{\left\| {{\bf{y}} - {{\bf{y}}^s}} \right\|}_2}}}{\left\| {{\bf{x}} - {{\bf{x}}^s}} \right\|_2}
\end{aligned}
 \end{equation}
By using the equation ${{\left\| \mathbf{x} \right\|}_{2}}=\sqrt{\left\| {{\mathbf{x}}^{s}} \right\|_{2}^{2}+\left\| \mathbf{x}-{{\mathbf{x}}^{s}} \right\|_{2}^{2}}$, we have
\renewcommand{\theequation}{A.24}
\label{eqn_A.24}
 \begin{equation}
\begin{aligned}
{\left\| {\bf{x}} \right\|_2} = \sqrt {\left\| {{\bf{y}} - {{\bf{y}}^s}} \right\|_2^2 + {{\left( {{{\left\| {{{\bf{y}}^s}} \right\|}_2} + \lambda } \right)}^2}}  - \lambda
\end{aligned}
 \end{equation}
\renewcommand{\theequation}{A.25}
\label{eqn_A.25}
 \begin{equation}
 \begin{aligned}
{\left\| {{{\bf{x}}^s}} \right\|_2} = \left( {{{\left\| {{{\bf{y}}^s}} \right\|}_2} + \lambda } \right)\frac{{\sqrt {\left\| {{\bf{y}} - {{\bf{y}}^s}} \right\|_2^2 + {{\left( {{{\left\| {{{\bf{y}}^s}} \right\|}_2} + \lambda } \right)}^2}}  - \lambda }}{{\sqrt {\left\| {{\bf{y}} - {{\bf{y}}^s}} \right\|_2^2 + {{\left( {{{\left\| {{{\bf{y}}^s}} \right\|}_2} + \lambda } \right)}^2}} }}
\end{aligned}
 \end{equation}
 \renewcommand{\theequation}{A.26}
\label{eqn_A.26}
 \begin{equation}
 \begin{aligned}
\left\| {{\bf{x}} - {{\bf{x}}^s}} \right\|_2^{} = {\left\| {{\bf{y}} - {{\bf{y}}^s}} \right\|_2}\frac{{\sqrt {\left\| {{\bf{y}} - {{\bf{y}}^s}} \right\|_2^2 + {{\left( {{{\left\| {{{\bf{y}}^s}} \right\|}_2} + \lambda } \right)}^2}}  - \lambda }}{{\sqrt {\left\| {{\bf{y}} - {{\bf{y}}^s}} \right\|_2^2 + {{\left( {{{\left\| {{{\bf{y}}^s}} \right\|}_2} + \lambda } \right)}^2}} }}
\end{aligned}
 \end{equation}
 Substitute these into (A.21), then we have
 \renewcommand{\theequation}{A.27}
\label{eqn_A.27}
 \begin{equation}
 \begin{aligned}
x_i^ *  = \left\{ {\begin{array}{*{20}{c}}
{\frac{{\left( {{{\left\| {{{\bf{y}}^s}} \right\|}_2} + \lambda } \right)\left( {\sqrt {\left\| {{\bf{y}} - {{\bf{y}}^s}} \right\|_2^2 + {{\left( {{{\left\| {{{\bf{y}}^s}} \right\|}_2} + \lambda } \right)}^2}}  - \lambda } \right)}}{{{{\left\| {{{\bf{y}}^s}} \right\|}_2}\sqrt {\left\| {{\bf{y}} - {{\bf{y}}^s}} \right\|_2^2 + {{\left( {{{\left\| {{{\bf{y}}^s}} \right\|}_2} + \lambda } \right)}^2}} }}{y_i},}&{{\rm{ }}i \in \Gamma _{\bf{y}}^s}\\
{\frac{{\sqrt {\left\| {{\bf{y}} - {{\bf{y}}^s}} \right\|_2^2 + {{\left( {{{\left\| {{{\bf{y}}^s}} \right\|}_2} + \lambda } \right)}^2}}  - \lambda }}{{\sqrt {\left\| {{\bf{y}} - {{\bf{y}}^s}} \right\|_2^2 + {{\left( {{{\left\| {{{\bf{y}}^s}} \right\|}_2} + \lambda } \right)}^2}} }}{y_i},}&{{\rm{ }}i \in \Gamma _{\bf{y}}^N\backslash \Gamma _{\bf{y}}^s}
\end{array}} \right.
\end{aligned}
 \end{equation}

2) If $\left| {{y}_{{{\pi }_{y}}\left( s \right)}} \right|=\left| {{y}_{{{\pi }_{y}}\left( s+1 \right)}} \right|$, then we have a similar conclusion as Remark 6.

From the above deduction, we have the expression of ${{\mathbf{x}}^{*}}$ in (\ref{eqn_21}) and (\ref{eqn_22}) when $R\left( \mathbf{x} \right)={{\left\| \mathbf{x} \right\|}_{2}}$. This completes the proof.
 \end{IEEEproof}

\section{Proof of Proposition 8}
\begin{IEEEproof}
Similar to the previous proof of Proposition 6, we have that
 \renewcommand{\theequation}{A.28}
\label{eqn_A.28}
 \begin{equation}
 \begin{aligned}
\left| {x_i^ * } \right| \ge \left| {x_j^ * } \right| \quad \rm{if} \quad \left| {{{\rm{y}}_i}} \right| > \left| {{{\rm{y}}_j}} \right|
\end{aligned}
 \end{equation}

 As proved in Proposition 4, ${{\mathbf{x}}^{*}}=\mathbf{0}$ if and only if $\mathbf{y}=\mathbf{0}$. Then, we just consider the condition of $\mathbf{y}\ne \mathbf{0}$. Firstly, we suppose that $\left| {{y}_{{{\pi }_{y}}\left( s \right)}} \right|\ne \left| {{y}_{{{\pi }_{y}}\left( s+1 \right)}} \right|$, then we have $\left\{ {{\pi _x}\left( 1 \right),{\pi _x}\left( 2 \right), \cdots ,{\pi _x}\left( s \right)} \right\} = \left\{ {{\pi _y}\left( 1 \right),{\pi _y}\left( 2 \right), \cdots ,{\pi _y}\left( s \right)} \right\}$.

 The first-order optimality condition of minimizing $E\left( \mathbf{x} \right)$ is that
  \renewcommand{\theequation}{A.29}
\label{eqn_A.29}
 \begin{equation}
 \begin{aligned}
\left( {1 - \frac{{a\lambda }}{{{{\left\| {\bf{x}} \right\|}_2}}} + \frac{{a\lambda }}{{{{\left\| {{{\bf{x}}^s}} \right\|}_2}}}} \right){x_i} = {y_i},i \in \Gamma _{\bf{y}}^s
\end{aligned}
 \end{equation}

   \renewcommand{\theequation}{A.30}
\label{eqn_A.30}
 \begin{equation}
 \begin{aligned}
\left( {1 - \frac{{a\lambda }}{{{{\left\| {\bf{x}} \right\|}_2}}}} \right){x_i} = {y_i} - \lambda {q_i},i \in \Gamma _{\bf{y}}^N\backslash \Gamma _{\bf{y}}^s
\end{aligned}
 \end{equation}
 where $\mathbf{q}\in \partial {{\left\| \mathbf{x}-{{\mathbf{x}}^{s}} \right\|}_{1}}$ is a subgradient.

 1) First case, when $\left| {{y}_{{{\pi }_{y}}\left( s+1 \right)}} \right|>\lambda $. Easily we have $1-\frac{a\lambda }{{{\left\| {{\mathbf{x}}^{*}} \right\|}_{2}}}>0$ by using Proposition 5:  $x_{i}^{*}\left\{ \begin{matrix}
   \ge 0, & \text{if  }{{\text{y}}_{i}}>0  \\
   \le 0, & \text{if  }{{\text{y}}_{i}}<0  \\
\end{matrix} \right.$.  When ${{y}_{\pi_y \left( s+1 \right)}}>\lambda $, then ${{y}_{{{\pi }_{y}}\left( s+1 \right)}}-\lambda q>0$, so we have $1-\frac{a\lambda }{{{\left\| {{\mathbf{x}}^{*}} \right\|}_{2}}}>0$; when ${{y}_{{{\pi }_{y}}\left( s+1 \right)}}<-\lambda $, then ${{y}_{{{\pi }_{y}}\left( s+1 \right)}}-\lambda q<0$, and we also have $1-\frac{a\lambda }{{{\left\| {{\mathbf{x}}^{*}} \right\|}_{2}}}>0$.

For $i \in \Gamma _{\bf{y}}^N\backslash \Gamma _{\bf{y}}^s$, if $\left| {{y}_{i}} \right|\le \lambda $, then $x_{i}^{*}=0$. Otherwise, for this $i$, if $0<{{y}_{i}}\le \lambda $, then $x_{i}^{*}>0$ based on Proposition 5. As $1-\frac{a\lambda }{{{\left\| {{\mathbf{x}}^{*}} \right\|}_{2}}}>0$, the left-hand side (LHS) of (A.30) is positive, while the right-hand side (RHS) of (A.30) nonpositive; if $-\lambda \le {{y}_{i}}<0$, then $x_{i}^{*}<0$ based on Proposition 5. The LHS of (A.30) is negative, while the RHS of (A.30) is nonnegative; if ${{y}_{i}}=0$, we have $x_{i}^{*}=0$ based on (A.28).

For $i \in \Gamma _{\bf{y}}^N\backslash \Gamma _{\bf{y}}^s$, if any $\left| {{y}_{i}} \right|>\lambda $, then we have $x_{i}^{*}\ne 0$ based on (A.30). For this $i$, we construct a vector $\mathbf{z}\in {{\mathbb{R}}^{N}}$ as

\renewcommand{\theequation}{A.31}
\label{eqn_A.31}
 \begin{equation}
 \begin{aligned}
{z_i} = \left\{ {\begin{array}{*{20}{c}}
{{\rm{shrink}}\left( {{y_i},\lambda } \right),}&{i \in \Gamma _{\bf{y}}^N\backslash \Gamma _{\bf{y}}^s}\\
{{y_{{\pi _y}\left( 1 \right)}},}&{i \in \Gamma _{\bf{y}}^s}
\end{array}} \right.
\end{aligned}
 \end{equation}
For $i \in \Gamma _{\bf{y}}^N\backslash \Gamma _{\bf{y}}^s$, we have $\left( 1-\frac{a\lambda }{{{\left\| \mathbf{x} \right\|}_{2}}} \right){{x}_{i}}={{z}_{i}}$, then we can obtain
\renewcommand{\theequation}{A.32}
\label{eqn_A.32}
 \begin{equation}
 \begin{aligned}
\left( {1 - \frac{{a\lambda }}{{{{\left\| {\bf{x}} \right\|}_2}}}} \right){\left\| {{\bf{x}} - {{\bf{x}}^s}} \right\|_2} = {\left\| {{\bf{z}} - {{\bf{z}}^s}} \right\|_2}
\end{aligned}
 \end{equation}
For $i \in \Gamma _{\bf{y}}^s$, we have
\renewcommand{\theequation}{A.33}
\label{eqn_A.33}
 \begin{equation}
 \begin{aligned}
\left( {1 - \frac{{a\lambda }}{{{{\left\| {\bf{x}} \right\|}_2}}} + \frac{{a\lambda }}{{{{\left\| {{{\bf{x}}^s}} \right\|}_2}}}} \right){\left\| {{{\bf{x}}^s}} \right\|_2} = {\left\| {{{\bf{y}}^s}} \right\|_2}
\end{aligned}
 \end{equation}
 Substitute (A.32) into (A.33), we have
 \renewcommand{\theequation}{A.34}
\label{eqn_A.34}
 \begin{equation}
 \begin{aligned}
{\left\| {{{\bf{x}}^s}} \right\|_2} = \frac{{{{\left\| {{{\bf{y}}^s}} \right\|}_2} - a\lambda }}{{{{\left\| {{\bf{z}} - {{\bf{z}}^s}} \right\|}_2}}}{\left\| {{\bf{x}} - {{\bf{x}}^s}} \right\|_2}
\end{aligned}
 \end{equation}
 By using the equation ${{\left\| \mathbf{x} \right\|}_{2}}=\sqrt{\left\| {{\mathbf{x}}^{s}} \right\|_{2}^{2}+\left\| \mathbf{x}-{{\mathbf{x}}^{s}} \right\|_{2}^{2}}$, we have

  \renewcommand{\theequation}{A.35}
\label{eqn_A.35}
 \begin{equation}
 \begin{aligned}
{\left\| {{\bf{x}} - {{\bf{x}}^s}} \right\|_2} = {\left\| {{\bf{z}} - {{\bf{z}}^s}} \right\|_2} + \frac{{a\lambda {{\left\| {{\bf{z}} - {{\bf{z}}^s}} \right\|}_2}}}{{\sqrt {\left\| {{\bf{z}} - {{\bf{z}}^s}} \right\|_2^2 + {{\left( {{{\left\| {{{\bf{y}}^s}} \right\|}_2} - a\lambda } \right)}^2}} }}
\end{aligned}
 \end{equation}
\renewcommand{\theequation}{A.36}
\label{eqn_A.36}
 \begin{equation}
 \begin{aligned}
{\left\| {{{\bf{x}}^s}} \right\|_2} = \left( {{{\left\| {{{\bf{y}}^s}} \right\|}_2} - a\lambda } \right)\left( {1 + \frac{{a\lambda }}{{\sqrt {\left\| {{\bf{z}} - {{\bf{z}}^s}} \right\|_2^2 + {{\left( {{{\left\| {{{\bf{y}}^s}} \right\|}_2} - a\lambda } \right)}^2}} }}} \right)
\end{aligned}
 \end{equation}
\renewcommand{\theequation}{A.37}
\label{eqn_A.37}
 \begin{equation}
 \begin{aligned}
{\left\| {\bf{x}} \right\|_2} = {\left\| {{\bf{z}} - {{\bf{z}}^s}} \right\|_2}\sqrt {1 + \frac{{{{\left( {{{\left\| {{{\bf{y}}^s}} \right\|}_2} - a\lambda } \right)}^2}}}{{\left\| {{\bf{z}} - {{\bf{z}}^s}} \right\|_2^2}}} {\rm{ + }}a\lambda
\end{aligned}
 \end{equation}
Substitute these into (A.29) and (A.30), then we have:
for $i \in \Gamma _{\bf{y}}^s$,

\renewcommand{\theequation}{A.38}
\label{eqn_A.38}
 \begin{equation}
 \begin{aligned}
x_i^ *  = \frac{{{{\left\| {{{\bf{y}}^s}} \right\|}_2} - a\lambda }}{{{{\left\| {{{\bf{y}}^s}} \right\|}_2}}}\left( {1 + \frac{{a\lambda }}{{\sqrt {\left\| {{\bf{z}} - {{\bf{z}}^s}} \right\|_2^2 + {{\left( {{{\left\| {{{\bf{y}}^s}} \right\|}_2} - a\lambda } \right)}^2}} }}} \right){y_i}
\end{aligned}
 \end{equation}
 for $i \in \Gamma _{\bf{y}}^N\backslash \Gamma _{\bf{y}}^s$,

\renewcommand{\theequation}{A.39}
\label{eqn_A.39}
 \begin{equation}
 \begin{aligned}
x_i^ *  = \left( {1 + \frac{{a\lambda }}{{\sqrt {\left\| {{\bf{z}} - {{\bf{z}}^s}} \right\|_2^2 + {{\left( {{{\left\| {{{\bf{y}}^s}} \right\|}_2} - a\lambda } \right)}^2}} }}} \right){z_i}
\end{aligned}
 \end{equation}

2) If $\left| {{y}_{{{\pi }_{y}}\left( s+1 \right)}} \right|=\lambda $, for $i \in \Gamma _{\bf{y}}^N\backslash \Gamma _{\bf{y}}^s$, suppose that there are $k$ components of ${{y}_{i}}$ having the same amplitude of $\lambda $, i.e., $\left| {{y}_{{{\pi }_{y}}\left( s+1 \right)}} \right|=\cdots =\left| {{y}_{{{\pi }_{y}}\left( s+k \right)}} \right|=\lambda >\left| {{y}_{{{\pi }_{y}}\left( s+k+1 \right)}} \right|$.

For $i\in \left\{ {{\pi }_{y}}\left( s+k+1 \right),{{\pi }_{y}}\left( s+k+2 \right),\cdots ,{{\pi }_{y}}\left( N \right) \right\}$, we have $x_{i}^{*}=0$. Otherwise, for this $i$, if $0<{{y}_{i}}<\lambda $, then $x_{i}^{*}>0$ based on Proposition 5. Easily, we have ${{y}_{i}}-\lambda {{q}_{i}}<0$ and $1-\frac{a\lambda }{{{\left\| {{\mathbf{x}}^{*}} \right\|}_{2}}}<0$ from (A.30). Meanwhile, as $\left| {{y}_{{{\pi }_{y}}\left( s+1 \right)}} \right|=\lambda $, we have $\left| x_{{{\pi }_{y}}\left( s+1 \right)}^{*} \right|\ge \left| x_{i}^{*} \right|>0$, then ${{y}_{{{\pi }_{y}}\left( s+1 \right)}}-\lambda {{q}_{{{\pi }_{y}}\left( s+1 \right)}}=0$, and this contradicts to the equation $\left( 1-\frac{a\lambda }{{{\left\| {{\mathbf{x}}^{*}} \right\|}_{2}}} \right)x_{{{\pi }_{y}}\left( s+1 \right)}^{*}={{y}_{{{\pi }_{y}}\left( s+1 \right)}}-\lambda {{q}_{{{\pi }_{y}}\left( s+1 \right)}}$ in (A.30). If $-\lambda <{{y}_{i}}<0$, then $x_{i}^{*}<0$ based on Proposition 5, we have ${{y}_{i}}-\lambda {{q}_{i}}>0$ and $1-\frac{a\lambda }{{{\left\| {{\mathbf{x}}^{*}} \right\|}_{2}}}<0$ from (A.30). However, as ${{y}_{{{\pi }_{y}}\left( s+1 \right)}}-\lambda {{q}_{{{\pi }_{y}}\left( s+1 \right)}}=0$, this also contradicts to the equation $\left( 1-\frac{a\lambda }{{{\left\| {{\mathbf{x}}^{*}} \right\|}_{2}}} \right)x_{{{\pi }_{y}}\left( s+1 \right)}^{*}={{y}_{{{\pi }_{y}}\left( s+1 \right)}}-\lambda {{q}_{{{\pi }_{y}}\left( s+1 \right)}}$. If  ${{y}_{i}}=0$, we have $x_{i}^{*}=0$ based on (A.28). Then we obtain that $x_{i}^{*}=0$ for $i\in \left\{ {{\pi }_{y}}\left( s+k+1 \right),{{\pi }_{y}}\left( s+k+2 \right),\cdots ,{{\pi }_{y}}\left( N \right) \right\}$.

For $i\in \left\{ {{\pi }_{y}}\left( s+1 \right),{{\pi }_{y}}\left( s+2 \right),\cdots ,{{\pi }_{y}}\left( s+k \right) \right\}$, if there exits $x_{i}^{*}\ne 0$, for this $i$ we have ${{y}_{i}}-\lambda {{q}_{i}}=0$, then we obtain $1-\frac{a\lambda }{{{\left\| {{\mathbf{x}}^{*}} \right\|}_{2}}}=0$ and ${{\left\| {{\mathbf{x}}^{*}} \right\|}_{2}}=a\lambda $. Substitute this into (A.29), we have ${{\left\| {{\mathbf{y}}^{s}} \right\|}_{2}}=a\lambda $. As $\left| {{y}_{{{\pi }_{y}}\left( s+1 \right)}} \right|=\lambda $, then we have that there exits $x_{i}^{*}\ne 0$ if and only if the conditions of  $a=1$, $s=1$, $\left| {{y}_{{{\pi }_{y}}\left( 1 \right)}} \right|=\lambda $ and ${{\left\| {{\mathbf{x}}^{*}} \right\|}_{2}}=\lambda $ are all satisfied. In this case, there are infinite many solutions, and all these ${{\mathbf{x}}^{*}}$ should satisfy ${{\left\| {{\mathbf{x}}^{*}} \right\|}_{2}}=\lambda $, $x_{i}^{*}{{y}_{i}}\ge 0$ and $x_{i}^{*}=0$ when $i\in \left\{ {{\pi }_{y}}\left( k+2 \right),{{\pi }_{y}}\left( k+3 \right),\cdots ,{{\pi }_{y}}\left( N \right) \right\}$. For example,
\renewcommand{\theequation}{A.40}
\label{eqn_A.40}
 \begin{equation}
 \begin{aligned}
x_i^ *  = \left\{ {\begin{array}{*{20}{c}}
{{\mathop{\rm sign}} \left( {{y_{{\pi _y}\left( 1 \right)}}} \right)\lambda ,}&{i = {\pi _y}\left( 1 \right)}\\
{0,}&{i \in \left\{ {{\pi _y}\left( 2 \right),{\pi _y}\left( 3 \right), \cdots ,{\pi _y}\left( N \right)} \right\}}
\end{array}} \right.
\end{aligned}
 \end{equation}
 or
 \renewcommand{\theequation}{A.41}
\label{eqn_A.41}
 \begin{equation}
 \begin{aligned}
x_i^ *  = \left\{ {\begin{array}{*{20}{c}}
{\frac{{{\rm{sign}}\left( {{y_{{\pi _y}\left( i \right)}}} \right)\lambda }}{{\left( {k + 1} \right)}},}&{i \in \left\{ {{\pi _y}\left( 1 \right),{\pi _y}\left( 2 \right), \cdots ,{\pi _y}\left( {k + 1} \right)} \right\}}\\
{0,}&{i \in \left\{ {{\pi _y}\left( {k + 2} \right),{\pi _y}\left( {k + 3} \right), \cdots ,{\pi _y}\left( N \right)} \right\}}
\end{array}} \right.
\end{aligned}
 \end{equation}

When any of these conditions $a=1$, $s=1$, $\left| {{y}_{{{\pi }_{y}}\left( 1 \right)}} \right|=\lambda $ cannot be satisfied, we have $x_{i}^{*}=0$ for $i\in \left\{ {{\pi }_{y}}\left( s+1 \right),{{\pi }_{y}}\left( s+2 \right),\cdots ,{{\pi }_{y}}\left( s+k \right) \right\}$. Then we have ${{\mathbf{x}}^{*}}={{\mathbf{x}}^{*s}}$. Substitute this into (A.29), we have $x_{i}^{*}={{y}_{i}}$ for $i \in \Gamma _{\bf{y}}^s$. Then the solution ${{\mathbf{x}}^{*}}$ can be expressed as

 \renewcommand{\theequation}{A.42}
\label{eqn_A.42}
 \begin{equation}
 \begin{aligned}
x_i^ *  = \left\{ {\begin{array}{*{20}{c}}
{{y_i},}&{i \in \Gamma _{\bf{y}}^s}\\
{0,}&{i \in \Gamma _{\bf{y}}^N\backslash \Gamma _{\bf{y}}^s}
\end{array}} \right.
\end{aligned}
 \end{equation}

3) If $0<\left| {{y}_{{{\pi }_{y}}\left( s+1 \right)}} \right|<\lambda $, for $i \in \Gamma _{\bf{y}}^N\backslash \Gamma _{\bf{y}}^s$, suppose that there are $k$ components of ${{y}_{i}}$ having the same amplitude with ${{y}_{{{\pi }_{y}}\left( s+1 \right)}}$, i.e., $\left| {{y}_{{{\pi }_{y}}\left( s+1 \right)}} \right|=\cdots =\left| {{y}_{{{\pi }_{y}}\left( s+k \right)}} \right|>\left| {{y}_{{{\pi }_{y}}\left( s+k+1 \right)}} \right|$.

For $i\in \left\{ {{\pi }_{y}}\left( s+k+1 \right),{{\pi }_{y}}\left( s+k+2 \right),\cdots ,{{\pi }_{y}}\left( N \right) \right\}$, we have $x_{i}^{*}=0$. Otherwise, for this $i$, as $\left| {{y}_{{{\pi }_{y}}\left( s+1 \right)}} \right|>\left| {{y}_{i}} \right|$, we have $\left| x_{{{\pi }_{y}}\left( s+1 \right)}^{*} \right|\ge \left| x_{i}^{*} \right|>0$ based on (A.28). Then we obtain $1-\frac{a\lambda }{{{\left\| {{\mathbf{x}}^{*}} \right\|}_{2}}}<0$ from (A.30), and we have $\left( 1-\frac{a\lambda }{{{\left\| {{\mathbf{x}}^{*}} \right\|}_{2}}} \right)\left| x_{{{\pi }_{y}}\left( s+1 \right)}^{*} \right|\le \left( 1-\frac{a\lambda }{{{\left\| {{\mathbf{x}}^{*}} \right\|}_{2}}} \right)\left| x_{i}^{*} \right|$, which means that $\left| {{y}_{{{\pi }_{y}}\left( s+1 \right)}}-\lambda {{q}_{{{\pi }_{y}}\left( s+1 \right)}} \right|\ge \left| {{y}_{i}}-\lambda {{q}_{i}} \right|$ through (A.30). Since $\left| x_{{{\pi }_{y}}\left( s+1 \right)}^{*} \right|\ge \left| x_{i}^{*} \right|\ne 0$, then we can obtain ${{q}_{{{\pi }_{y}}\left( s+1 \right)}}=\text{sign}({{y}_{{{\pi }_{y}}\left( s+1 \right)}})$ based on Proposition 5. If ${{y}_{i}}\ne 0$, then we have ${{q}_{i}}=\text{sign}({{y}_{i}})$ and obtain $\left| \text{sign}\left( {{y}_{{{\pi }_{y}}\left( s+1 \right)}} \right) \right|\cdot \left| \left| {{y}_{{{\pi }_{y}}\left( s+1 \right)}} \right|-\lambda  \right|\ge \left| \text{sign}\left( {{y}_{i}} \right) \right|\cdot \left| \left| {{y}_{i}} \right|-\lambda  \right|$, which means that $\lambda -\left| {{y}_{{{\pi }_{y}}\left( s+1 \right)}} \right|\ge \lambda -\left| {{y}_{i}} \right|$. However, this contradicts $\left| {{y}_{{{\pi }_{y}}\left( s+1 \right)}} \right|>\left| {{y}_{i}} \right|$. If ${{y}_{i}}=0$, we have $\left| {{y}_{{{\pi }_{y}}\left( s+1 \right)}}-\lambda {{q}_{{{\pi }_{y}}\left( s+1 \right)}} \right|\ge \lambda $, then we can obtain $\left| \left| {{y}_{{{\pi }_{y}}\left( s+1 \right)}} \right|-\lambda  \right|\ge \lambda $, which contradicts $0<\left| {{y}_{{{\pi }_{y}}\left( s+1 \right)}} \right|<\lambda $. Then, we obtain that $x_{i}^{*}=0$ for $i\in \left\{ {{\pi }_{y}}\left( s+k+1 \right),{{\pi }_{y}}\left( s+k+2 \right),\cdots ,{{\pi }_{y}}\left( N \right) \right\}$.

For $i\in \left\{ {{\pi }_{y}}\left( s+1 \right),{{\pi }_{y}}\left( s+2 \right),\cdots ,{{\pi }_{y}}\left( s+k \right) \right\}$, if there exits $x_{i}^{*}\ne 0$, then we have $1-\frac{a\lambda }{{{\left\| {{\mathbf{x}}^{*}} \right\|}_{2}}}<0$ as the sign of ${{y}_{i}}-\lambda {{q}_{i}}$ and $x_{i}^{*}$ are opposite. For $i \in \Gamma _{\bf{y}}^s$, from (A.29), we have

 \renewcommand{\theequation}{A.43}
\label{eqn_A.43}
 \begin{equation}
 \begin{aligned}
{\left\| {{{\bf{x}}^{ * s}}} \right\|_2} = {{\left( {{{\left\| {{{\bf{y}}^s}} \right\|}_2} - a\lambda } \right)} \mathord{\left/
 {\vphantom {{\left( {{{\left\| {{{\bf{y}}^s}} \right\|}_2} - a\lambda } \right)} {\left( {1 - \frac{{a\lambda }}{{{{\left\| {{{\bf{x}}^ * }} \right\|}_2}}}} \right)}}} \right.
 \kern-\nulldelimiterspace} {\left( {1 - \frac{{a\lambda }}{{{{\left\| {{{\bf{x}}^ * }} \right\|}_2}}}} \right)}}
\end{aligned}
 \end{equation}
 If ${{\left\| {{\mathbf{y}}^{s}} \right\|}_{2}}\ge a\lambda $, we have ${{\left\| {{\mathbf{x}}^{*s}} \right\|}_{2}}\le 0$, which contradicts $x_{i}^{*}\ne 0$. So, when ${{\left\| {{\mathbf{y}}^{s}} \right\|}_{2}}\ge a\lambda $, we have $x_{i}^{*}=0$, and then the solution ${{\mathbf{x}}^{*}}$ is

 \renewcommand{\theequation}{A.44}
\label{eqn_A.44}
 \begin{equation}
 \begin{aligned}
x_i^ *  = \left\{ {\begin{array}{*{20}{c}}
{{y_i},}&{i \in \Gamma _{\bf{y}}^s}\\
{0,}&{i \in \Gamma _{\bf{y}}^N\backslash \Gamma _{\bf{y}}^s}
\end{array}} \right.
\end{aligned}
 \end{equation}
 If ${{\left\| {{\mathbf{y}}^{s}} \right\|}_{2}}<a\lambda $, for $i\in \left\{ {{\pi }_{y}}\left( s+1 \right),{{\pi }_{y}}\left( s+2 \right),\cdots ,{{\pi }_{y}}\left( s+k \right) \right\}$, suppose there are $c$ components of $x_{i}^{*}\ne 0$ and $c\le k$. From (A.30), we have ${{\left\| {{\mathbf{x}}^{*}}-{{\mathbf{x}}^{*s}} \right\|}_{2}}={\sqrt{c}\left( \left| {{y}_{\pi \left( s+1 \right)}} \right|-\lambda  \right)}/{\left( 1-\frac{a\lambda }{{{\left\| {{\mathbf{x}}^{*}} \right\|}_{2}}} \right)}\;$. Substitute this into ${{\left\| {{\mathbf{x}}^{*s}} \right\|}_{2}}={\left( {{\left\| {{\mathbf{y}}^{s}} \right\|}_{2}}-a\lambda  \right)}/{\left( 1-\frac{a\lambda }{{{\left\| \mathbf{x} \right\|}_{2}}} \right)}\;$ from (A.29), we have
  \renewcommand{\theequation}{A.45}
\label{eqn_A.45}
 \begin{equation}
 \begin{aligned}
{\left\| {{{\bf{x}}^ * }} \right\|_2} = a\lambda  - \sqrt {{{\left( {{{\left\| {{{\bf{y}}^s}} \right\|}_2} - a\lambda } \right)}^2} + c{{\left( {\left| {{y_{\pi \left( {s + 1} \right)}}} \right| - \lambda } \right)}^2}}
\end{aligned}
 \end{equation}
Reconsider the expression of $E\left( \mathbf{x} \right)$, and using the first-order optimality condition, we have
  \renewcommand{\theequation}{A.46}
\label{eqn_A.46}
 \begin{equation}
 \begin{aligned}
E\left( {{{\bf{x}}^ * }} \right) &= \frac{{\left\| {{{\bf{x}}^ * }} \right\|_2^2 + \left\| {\bf{y}} \right\|_2^2}}{{2\lambda }} - \left\langle {{{\bf{x}}^ * },\frac{{\bf{y}}}{\lambda }} \right\rangle \\
 &\quad + {\left\| {{{\bf{x}}^ * }} \right\|_1} - a{\left\| {{{\bf{x}}^ * }} \right\|_2} - {\left\| {{{\bf{x}}^ * }^s} \right\|_1} + a{\left\| {{{\bf{x}}^ * }^s} \right\|_2}\\
 &= \frac{{\left\| {{{\bf{x}}^ * }} \right\|_2^2 + \left\| {\bf{y}} \right\|_2^2}}{{2\lambda }} - \left\langle {{{\bf{x}}^ * }^s,\left( {\frac{1}{\lambda } - \frac{a}{{{{\left\| {{{\bf{x}}^ * }} \right\|}_2}}} + \frac{a}{{{{\left\| {{{\bf{x}}^ * }^s} \right\|}_2}}}} \right){{\bf{x}}^ * }^s} \right\rangle \\
 &\quad - \left\langle {{{\bf{x}}^ * } - {{\bf{x}}^ * }^s,q + \left( {\frac{1}{\lambda } - \frac{a}{{{{\left\| {{{\bf{x}}^ * }} \right\|}_2}}}} \right)\left( {{{\bf{x}}^ * } - {{\bf{x}}^ * }^s} \right)} \right\rangle \\
& \quad + {\left\| {{{\bf{x}}^ * }} \right\|_1} - a{\left\| {{{\bf{x}}^ * }} \right\|_2} - {\left\| {{{\bf{x}}^ * }^s} \right\|_1} + a{\left\| {{{\bf{x}}^ * }^s} \right\|_2}\\
 &= \frac{{\left\| {{{\bf{x}}^ * }} \right\|_2^2 + \left\| {\bf{y}} \right\|_2^2}}{{2\lambda }} - \frac{{\left\| {{{\bf{x}}^ * }^s} \right\|_2^2}}{\lambda } + \frac{{a\left\| {{{\bf{x}}^ * }^s} \right\|_2^2}}{{{{\left\| {{{\bf{x}}^ * }} \right\|}_2}}} - a{\left\| {{{\bf{x}}^ * }^s} \right\|_2}\\
 &\quad - {\left\| {{{\bf{x}}^ * } - {{\bf{x}}^ * }^s} \right\|_1} - \frac{{\left\| {{{\bf{x}}^ * } - {{\bf{x}}^ * }^s} \right\|_2^2}}{\lambda } + \frac{{a\left\| {{{\bf{x}}^ * } - {{\bf{x}}^ * }^s} \right\|_2^2}}{{{{\left\| {{{\bf{x}}^ * }} \right\|}_2}}}\\
 &\quad + {\left\| {{{\bf{x}}^ * }} \right\|_1} - a{\left\| {{{\bf{x}}^ * }} \right\|_2} - {\left\| {{{\bf{x}}^ * }^s} \right\|_1} + a{\left\| {{{\bf{x}}^ * }^s} \right\|_2}\\
 &=  - \frac{{\left\| {{{\bf{x}}^ * }} \right\|_2^2}}{{2\lambda }} + \frac{{\left\| {\bf{y}} \right\|_2^2}}{{2\lambda }}
\end{aligned}
 \end{equation}
 Then we have $E\left( {{\mathbf{x}}^{*}} \right)<E\left( \mathbf{0} \right)$, and we need to find the ${{\mathbf{x}}^{*}}$ with the largest norm among all ${{\mathbf{x}}^{*}}$ that satisfying (A.29) and (A.30). From this, we have that $c$ should be zero to make the largest $\left\| {{\mathbf{x}}^{*}} \right\|$ in (A.45). So, when ${{\left\| {{\mathbf{y}}^{s}} \right\|}_{2}}<a\lambda $, we have the solution ${{\mathbf{x}}^{*}}$ the same as (A.44).

 4) If ${{y}_{\pi \left( s+1 \right)}}=0$, for $i \in \Gamma _{\bf{y}}^N\backslash \Gamma _{\bf{y}}^s$, we have $x_{i}^{*}=0$. Otherwise, we can construct a vector $\mathbf{\tilde{x}}\in {{\mathbb{R}}^{N}}$, which is equal to ${{\mathbf{x}}^{*}}$ except setting these corresponding $\tilde{x}_{i}^{{}}$ to be zero. Then we can obtain a smaller objective value, which contradicts the optimality of ${{\mathbf{x}}^{*}}$. For $i \in \Gamma _{\bf{y}}^s$, we have $x_{i}^{*}={{y}_{i}}$. Then the solution ${{\mathbf{x}}^{*}}$ can be expressed as (A.44).

 Once again, if there exits one or more components of ${{y}_{i}}$, $i \notin \Gamma _{\bf{y}}^s$ having the same amplitude of ${{y}_{{{\pi }_{y}}\left( s \right)}}$, then we have a similar conclusion as Remark 6. This completes the proof.

\end{IEEEproof}

\section{Proof of Proposition 9}
\begin{IEEEproof}
Let ${{\mathbf{x}}^{[k+1]}}$ be the optimal solution of (\ref{eqn_13}) with $\mathbf{y}={{\mathbf{x}}^{[k]}}-\beta \nabla \phi \left( {{\mathbf{x}}^{[k]}} \right)$, then we have
  \renewcommand{\theequation}{A.47}
\label{eqn_A.47}
 \begin{equation}
 \begin{aligned}
&E\left( {{{\bf{x}}^{[k + 1]}}} \right) - E\left( {{{\bf{x}}^{[k]}}} \right)\\
& = \frac{{\left\| {{{\bf{x}}^{[k + 1]}} - {\bf{y}}} \right\|_2^2}}{{2\lambda }} + R\left( {{{\bf{x}}^{[k + 1]}}} \right) - R\left( {{{\bf{x}}^{[k + 1]}}^s} \right)\\
&\quad  - \frac{{\left\| {{{\bf{x}}^{[k]}} - {\bf{y}}} \right\|_2^2}}{{2\lambda }} - R\left( {{{\bf{x}}^{[k]}}} \right) + R\left( {{{\bf{x}}^{[k]}}^s} \right)\\
& =  - \frac{{\left\| {{{\bf{x}}^{[k + 1]}} - {{\bf{x}}^{[k]}}} \right\|_2^2}}{{2\lambda }} + \frac{{\left\langle {{{\bf{x}}^{[k + 1]}} - {{\bf{x}}^{[k]}},{{\bf{x}}^{[k + 1]}} - {\bf{y}}} \right\rangle }}{\lambda }\\
&\quad  + R\left( {{{\bf{x}}^{[k + 1]}}} \right) - R\left( {{{\bf{x}}^{[k + 1]}}^s} \right) - R\left( {{{\bf{x}}^{[k]}}} \right) + R\left( {{{\bf{x}}^{[k]}}^s} \right)\\
& =  - \frac{{\left\| {{{\bf{x}}^{[k + 1]}} - {{\bf{x}}^{[k]}}} \right\|_2^2}}{{2\lambda }} + \sum\limits_{i \in {\Lambda _{k + 1}}}^{} {\left( {x_i^{[k]} - x_i^{[k + 1]}} \right)\left( {\partial {r_i}\left( {x_i^{[k + 1]}} \right)} \right)} \\
&\quad  + \sum\limits_{i \in {\Lambda _{k + 1}}}^{} {{r_i}\left( {x_i^{[k + 1]}} \right)}  - \sum\limits_{i \in {\Lambda _k}}^{} {{r_i}\left( {x_i^{[k]}} \right)} \\
& \le  - \frac{{\left\| {{{\bf{x}}^{[k + 1]}} - {{\bf{x}}^{[k]}}} \right\|_2^2}}{{2\lambda }} + \sum\limits_{i \in {\Lambda _{k + 1}}}^{} {{r_i}\left( {x_i^{[k]}} \right)}  - \sum\limits_{i \in {\Lambda _k}}^{} {{r_i}\left( {x_i^{[k]}} \right)}
 \end{aligned}
 \end{equation}
 The third equation comes from Proposition 6, and the last inequality is based on the property of subgradient.
Then we have
  \renewcommand{\theequation}{A.48}
\label{eqn_A.48}
 \begin{equation}
 \begin{aligned}
E\left( {{{\bf{x}}^{[k + 1]}}} \right) - E\left( {{{\bf{x}}^{[k]}}} \right) \le \min \left\{ { - \frac{{\left\| {{{\bf{x}}^{[k + 1]}} - {{\bf{x}}^{[k]}}} \right\|_2^2}}{{2\lambda }} + {\Delta _k},0} \right\}
 \end{aligned}
 \end{equation}
where ${\Delta _k} = \sum\limits_{i \in {\Lambda _{k + 1}}}^{} {{r_i}\left( {x_i^{[k]}} \right)}  - \sum\limits_{i \in {\Lambda _k}}^{} {{r_i}\left( {x_i^{[k]}} \right)} $, ${\Lambda _{k + 1}} = \Gamma _{{{\bf{x}}^{[k + 1]}}}^N\backslash \Gamma _{{{\bf{x}}^{[k + 1]}}}^s$, and ${\Lambda _k} = \Gamma _{{{\bf{x}}^{[k]}}}^N\backslash \Gamma _{{{\bf{x}}^{[k]}}}^s$.
 Substitute this into (\ref{eqn_26}) then we have (\ref{eqn_32}). This completes the proof.
\end{IEEEproof}
\section*{Acknowledgment}

This work is partially supported by the National Natural Science Foundation of China (61701508).




\section*{References}

[1] E.J. Candès, J. Romberg, T. Tao. Robust uncertainty principles: Exact signal reconstruction from highly incomplete frequency information. IEEE Trans. Inf. Theory, 52(2): 489–509, 2006.

[2] Patel V M, Easley G R, Healy Jr D M, et al. Compressed synthetic aperture radar[J]. IEEE Journal of selected topics in signal processing, 2010, 4(2): 244-254.

[3] Yang J, Thompson J, Huang X, et al. Random-frequency SAR imaging based on compressed sensing[J]. IEEE Transactions on Geoscience and Remote Sensing, 2013, 51(2): 983-994.

[4] Berger C R, Wang Z, Huang J, et al. Application of compressive sensing to sparse channel estimation[J]. IEEE Communications Magazine, 2010, 48(11): 164-174.

[5] Chen Z, Jin X, Li L, et al. A limited-angle CT reconstruction method based on anisotropic TV minimization[J]. Physics in Medicine Biology, 2013, 58(7): 2119.

 [6] Lustig, Michael, David Donoho, and John M. Pauly. "Sparse MRI: The application of compressed sensing for rapid MR imaging." Magnetic Resonance in Medicine: An Official Journal of the International Society for Magnetic Resonance in Medicine 58.6 (2007): 1182-1195.

[7] Chartrand R, Staneva V. Restricted isometry properties and nonconvex compressive sensing[J]. Inverse Problems, 2008, 24(3): 035020.

[8] Candes E J, Wakin M B, Boyd S P. Enhancing sparsity by reweighted ℓ 1 minimization[J]. Journal of Fourier analysis and applications, 2008, 14(5-6): 877-905.

[9] Sun Y, Tao J. Few views image reconstruction using alternating direction method via‐norm minimization[J]. International Journal of Imaging Systems and Technology, 2014, 24(3): 215-223.

[10] Yu-Li S, Jin-Xu T. Image reconstruction from few views by ℓ0-norm optimization[J]. Chinese Physics B, 2014, 23(7): 078703.

[11] Chartrand R. Exact reconstruction of sparse signals via nonconvex minimization[J]. IEEE Signal Processing Letters, 2007, 14(10): 707-710.

[12] Chartrand R, Yin W. Iteratively reweighted algorithms for compressive sensing[C]//2008 IEEE International Conference on Acoustics, Speech and Signal Processing. IEEE, 2008: 3869-3872.

[13] Krishnan D, Fergus R. Fast image deconvolution using hyper-Laplacian priors[C]//Advances in neural information processing systems. 2009: 1033-1041.

 [14] Xu Z, Chang X, Xu F, et al. $ L_ {1/2} $ regularization: A thresholding representation theory and a fast solver[J]. IEEE Transactions on neural networks and learning systems, 2012, 23(7): 1013-1027.

[15] Lai M J, Xu Y, Yin W. Improved iteratively reweighted least squares for unconstrained smoothed $\ell_q$ minimization[J]. SIAM Journal on Numerical Analysis, 2013, 51(2): 927-957.

[16] Pant J K, Lu W S, Antoniou A. New Improved Algorithms for Compressive Sensing Based on $\ell_ {p}$ Norm[J]. IEEE Transactions on Circuits and Systems II: Express Briefs, 2014, 61(3): 198-202.

[17] Woodworth J, Chartrand R. Compressed sensing recovery via nonconvex shrinkage penalties[J]. Inverse Problems, 2016, 32(7): 075004.

[18] Wu L, Sun Z, Li D H. A Barzilai–Borwein-Like Iterative Half Thresholding Algorithm for the $ L_ {1/2} $ Regularized Problem[J]. Journal of Scientific Computing, 2016, 67(2): 581-601.

[19] Fengmin X, Shanhe W. A hybrid simulated annealing thresholding algorithm for compressed sensing[J]. Signal Processing, 2013, 93(6): 1577-1585.

[20] Miao C, Yu H. A General-Thresholding Solution for $ l_ {p}(0< p< 1) $ Regularized CT Reconstruction[J]. IEEE Transactions on Image Processing, 2015, 24(12): 5455-5468.

[21] Zhang T. Analysis of multi-stage convex relaxation for sparse regularization[J]. Journal of Machine Learning Research, 2010, 11(Mar): 1081-1107.

[22] Zhang T. Multi-stage convex relaxation for feature selection[J]. Bernoulli, 2013, 19(5B): 2277-2293.

[23] Lou Y, Yin P, Xin J. Point Source Super-resolution Via Non-convex $ L_1 $ Based Methods[J]. Journal of Scientific Computing, 2016, 68(3): 1082-1100.

[24] Z Zhang S, Xin J. Minimization of transformed $ L_1 $ penalty: theory, difference of convex function algorithm, and robust application in compressed sensing[J]. Mathematical Programming, 2018, 169(1): 307-336.

[25] Dinh T, Xin J. Convergence of a Relaxed Variable Splitting Method for Learning Sparse Neural Networks via $\ell_1,\ell_0 $, and transformed-$\ell_1 $ Penalties[J]. arXiv preprint arXiv:1812.05719, 2018.

[26] Lv J, Fan Y. A unified approach to model selection and sparse recovery using regularized least squares[J]. The Annals of Statistics, 2009, 37(6A): 3498-3528.

[27]Bogdan M, Berg E V D, Su W, et al. Statistical estimation and testing via the sorted L1 norm[J]. Statistics, 2013.

[28] Zeng X, Figueiredo M A T. Decreasing Weighted Sorted ${\ell_1} $ Regularization[J]. IEEE Signal Processing Letters, 2014, 21(10): 1240-1244.

[29] Lou Y, Yan M. Fast L1–L2 minimization via a proximal operator[J].
Journal of Scientific Computing, 2018, 74(2): 767-785.

[30]Lou, Y., Yin, P., He, Q., Xin, J.: Computing sparse representation in a highly coherent dictionary based on difference of l1 and l2. J. Sci. Comput. 64(1), 178–196 (2015)

[31] Yin P, Lou Y, He Q, et al. Minimization of L1-2 for compressed sensing[J]. SIAM Journal on Scientific Computing, 2015, 37(1): A536-A563.

[32] Fan J, Li R. Variable selection via nonconcave penalized likelihood and its oracle properties[J]. Journal of the American statistical Association, 2001, 96(456): 1348-1360.

[33] Mehranian A, Rad H S, Rahmim A, et al. Smoothly clipped absolute deviation (SCAD) regularization for compressed sensing MRI using an augmented Lagrangian scheme[J]. Magnetic resonance imaging, 2013, 31(8): 1399-1411.

[34] Zhang C H. Nearly unbiased variable selection under minimax concave penalty[J]. The Annals of statistics, 2010, 38(2): 894-942.

[35] Selesnick I. Sparse regularization via convex analysis[J]. IEEE Transactions on Signal Processing, 2017, 65(17): 4481-4494.

[36] Sun Y, Chen H, Tao J. Sparse signal recovery via minimax-concave penalty and $\ell _1 $-norm loss function[J]. IET Signal Processing, 2018, 12(9): 1091-1098.

[37] Blumensath T, Davies M E. Iterative thresholding for sparse approximations: The Journal of Fourier Analysis and Applications, 14, 629–654[J]. 2008.

[38] Blumensath T, Davies M E. Iterative hard thresholding for compressed sensing[J]. Applied and computational harmonic analysis, 2009, 27(3): 265-274.

[39] Blumensath T. Accelerated iterative hard thresholding[J]. Signal Processing, 2012, 92(3): 752-756.

[40] Lu Z. Iterative hard thresholding methods for $ \ell_0 $ regularized convex cone programming[J]. Mathematical Programming, 2014, 147(1-2): 125-154.

[41] Bao C, Dong B, Hou L, et al. Image restoration by minimizing zero norm of wavelet frame coefficients[J]. Inverse problems, 2016, 32(11): 115004.

[42] Zhang X, Zhang X. An accelerated proximal iterative hard thresholding method for $\ell_0 $ minimization[J]. arXiv preprint arXiv:1709.01668,
2017..

[43] Gotoh J, Takeda A, Tono K. DC formulations and algorithms for sparse optimization problems[J]. Mathematical Programming, 2018: 1-36.

[44] Tono K, Takeda A, Gotoh J. Efficient DC algorithm for constrained sparse optimization[J]. arXiv preprint arXiv:1701.08498, 2017.

[45] Tao P D, An L T H. Convex analysis approach to dc programming: Theory, algorithms and applications[J]. Acta mathematica vietnamica, 1997, 22(1): 289-355.

[46] Ahn M, Pang J S, Xin J. Difference-of-convex learning: directional stationarity, optimality, and sparsity[J]. SIAM Journal on Optimization, 2017, 27(3): 1637-1665.

[47] Yin P , Xin J . Iterative $\ell_1$ minimization for non-convex compressed sensing[J]. Journal of Computational Mathematics 2017, 35(4):439-451.

[48] Liu T, Pong T K, Takeda A. A successive difference-of-convex approximation method for a class of nonconvex nonsmooth optimization problems[J]. Mathematical Programming, 2017: 1-29.

[49] Yuille A L, Rangarajan A. The concave-convex procedure[J]. Neural computation, 2003, 15(4): 915-936.

[50] Artacho F J A, Fleming R M T, Vuong P T. Accelerating the DC algorithm for smooth functions[J]. Mathematical Programming, 2018, 169(1): 95-118.

[51] Wen B, Chen X, Pong T K. A proximal difference-of-convex algorithm with extrapolation[J]. Computational optimization and applications, 2018, 69(2): 297-324.

[52] Wen F, Pei L, Yang Y, et al. Efficient and robust recovery of sparse signal and image using generalized nonconvex regularization[J]. IEEE Transactions on Computational Imaging, 2017, 3(4): 566-579.

[53] Zhang J, Zhao C, Zhao D, et al. Image compressive sensing recovery using adaptively learned sparsifying basis via L0 minimization[J]. Signal Processing, 2014, 103: 114-126.

[54] Gong P, Zhang C, Lu Z, et al. A general iterative shrinkage and thresholding algorithm for non-convex regularized optimization problems[C]//International Conference on Machine Learning. 2013: 37-45.

[55] Li H, Lin Z. Accelerated proximal gradient methods for nonconvex programming[C]//Advances in neural information processing systems. 2015: 379-387.

[56] Beck A, Teboulle M. A fast iterative shrinkage-thresholding algorithm for linear inverse problems[J]. SIAM journal on imaging sciences, 2009, 2(1): 183-202.

[57] Dong B, Zhang Y. An efficient algorithm for ℓ 0 minimization in wavelet frame based image restoration[J]. Journal of Scientific Computing, 2013, 54(2-3): 350-368.

[58] Zhang Y, Dong B, Lu Z. ℓ₀ Minimization for wavelet frame based image restoration[J]. Mathematics of Computation, 2013, 82(282): 995-1015.

[59] Zhang X, Lu Y, Chan T. A novel sparsity reconstruction method from Poisson data for 3D bioluminescence tomography[J]. Journal of scientific computing, 2012, 50(3): 519-535.

[60] Trzasko J, Manduca A, Borisch E. Sparse MRI reconstruction via multiscale L0-continuation[C]//2007 IEEE/SP 14th Workshop on Statistical Signal Processing. IEEE, 2007: 176-180.

[61] Trzasko J, Manduca A. Highly Undersampled Magnetic Resonance Image Reconstruction via Homotopic $\ell_ {0} $-Minimization[J]. IEEE Transactions on Medical imaging, 2009, 28(1): 106-121.

[62] Pavlikov K, Uryasev S. CVaR norm and applications in optimization[J]. Optimization Letters, 2014, 8(7): 1999-2020.

[63] Gotoh J, Uryasev S. Two pairs of families of polyhedral norms versus $\ell _p $-norms: proximity and applications in optimization[J]. Mathematical Programming, 2016, 156(1-2): 391-431.

[64] Sun Y, Chen H, Tao J, et al. Computed tomography image reconstruction from few views via Log-norm total variation minimization[J]. Digital Signal Processing, Volume 88, Pages 172-181, May 2019.

[65] Wen F, Liu P, Liu Y, et al. Robust Sparse Recovery in Impulsive Noise via $\ell _p $-$\ell _1 $ Optimization[J]. IEEE Transactions on Signal Processing, 2017, 65(1): 105-118.

[66] Combettes P L, Pesquet J C. Proximal splitting methods in signal processing[M]//Fixed-point algorithms for inverse problems in science and engineering. Springer, New York, NY, 2011: 185-212.

[67] Lu Z. Sequential convex programming methods for a class of structured nonlinear programming[J]. arXiv preprint arXiv:1210.3039, 2012.

[68] Barzilai J, Borwein J M. Two-point step size gradient methods[J]. IMA journal of numerical analysis, 1988, 8(1): 141-148.

[69] Sidky E Y, Chartrand R, Pan X. Image reconstruction from few views by non-convex optimization[C]//2007 IEEE Nuclear Science Symposium Conference Record. IEEE, 2007, 5: 3526-3530.

[70] Rahimi Y, Wang C, Dong H, et al. A Scale Invariant Approach for Sparse Signal Recovery[J]. arXiv preprint arXiv:1812.08852, 2018.

[71]. Nocedal, Jorge, Wright, Stephen J.: Numerical Optimization 2nd. Springer, Berlin (2006)

[72] Boyd S, Parikh N, Chu E, et al. Distributed optimization and statistical learning via the alternating direction method of multipliers[J]. Foundations and Trends® in Machine learning, 2011, 3(1): 1-122.

\end{document}